%% file: forward-model-speed-and-accuracy.tex
\documentclass[a4paper,11pt]{article}
\pdfoutput=1 
\usepackage{jcappub} 
\usepackage[T1]{fontenc}

\usepackage{amsmath, amssymb}
\usepackage{bm}
\usepackage{graphicx,color}
\usepackage{soul}
\usepackage{layouts}
\usepackage{multirow}
\usepackage{makecell}
\usepackage{mathtools}
\usepackage{bbold}
\usepackage{hyperref}

\usepackage{tikz}
\usetikzlibrary{arrows.meta}

\renewcommand{\vec}[1]{\boldsymbol{\bm{#1}}}
\newcommand{\dg}{\delta_\mathrm{g}}
\newcommand{\dgd}{\delta_\mathrm{g,det}}

\newcommand{\op}{\mathcal{O}}
\newcommand{\ee}{\mathrm{e}}
\newcommand{\dirac}{\mathrm{D}}
\newcommand{\lagrangian}{\mathrm{L}}
\newcommand{\eulerian}{\mathrm{E}}
\newcommand{\nyquist}{\mathrm{Ny}}
\newcommand{\grid}{\mathrm{G}}
\newcommand{\alias}{\mathrm{alias}}
\newcommand{\lin}{^{(1)}}
\newcommand{\lpt}{\mathrm{LPT}}
\newcommand{\ini}{\mathrm{G,in}}
\newcommand{\fwd}{\mathrm{G,fwd}}
\newcommand{\eul}{\mathrm{G,Eul}}
\newcommand{\fin}{\mathrm{G,final}}
\newcommand{\lh}{\mathrm{G,LH}}
\newcommand{\mpc}{\mathrm{Mpc}}
\newcommand{\nn}{\nonumber}
\newcommand{\LO}{\mathrm{L.O.}}
\newcommand{\bias}{\mathrm{bias}}
\newcommand{\leftfield}{\texttt{LEFTfield}}
\newcommand{\filterfwd}{\Lambda_\mathrm{fwd}}
\newcommand{\filterbias}{\Lambda_\mathrm{bias}}

\newlength{\figuresize}
\title{Fast, Accurate and Perturbative Forward Modeling of Galaxy Clustering \\[.5\baselineskip]
Part I: Galaxies in the Restframe}

\author[a,b]{Julia Stadler,}
\author[a,b]{Fabian Schmidt,}
\author[a]{Martin Reinecke}
\affiliation[a]{Max-Planck-Institut für Astrophysik, Karl-Schwarzschild-Str. 1, 85748 Garching, Germany}
\affiliation[b]{Excellence Cluster ORIGINS, Boltzmannstr. 2, 85748 Garching, Germany}
\emailAdd{jstadler@mpa-garching.mpg.de}
\emailAdd{fabians@mpa-garching.mpg.de}
\emailAdd{martin@mpa-garching.mpg.de}

\abstract{Forward models of the galaxy density field enable simulation based inference as well as field level inference of galaxy clustering. However, these analysis techniques require forward models that are both computationally fast and robust to modeling uncertainties in the relation between galaxies and matter. Both requirements can be addressed with the Effective Field Theory of Large Scale Structure. Here, we focus on the physical and numerical convergence of the \leftfield~model. Based on the perturbative nature of the forward model, we derive an analytic understanding of the leading numerical errors, and we compare our estimates to high-resolution and N-body references. This allows us to derive a set of best-practice recommendations for the numerical accuracy parameters, which are completely specified by the desired order of the perturbative solution and the cut-off scale. We verify these recommendations by an extended set of parameter recovery tests from fully nonlinear mock data and find very consistent results. A single evaluation of the forward model takes seconds, making cosmological analyses of galaxy clustering data based on forward models computationally feasible.}

\begin{document}
\maketitle
\flushbottom

\section{Introduction}
\label{sec:intro}

The new generation of stage-IV spectroscopic galaxy surveys, including DESI \cite{DESI:2016fyo}, Euclid \cite{Amendola:2016saw}, PFS \cite{2014PASJ...66R...1T} or SPHEREx \cite{Dore:2014cca}, will measure three-dimensional galaxy clustering at an unprecedented volume and depth. They were designed to investigate some long-standing open questions in cosmology regarding the nature of dark energy, inflation and gravity. Traditionally, the analysis of galaxy surveys has focused on Baryon Acoustic Oscillations (BAO) and redshift space distortions (RSD) \cite{2dFGRS:2005yhx,SDSS:2005xqv}, two features in the two-point correlation function that are particularly robust with respect to galaxy bias \cite{Desjacques:2016bnm}. Alternatively, the Effective Field Theory of Large Scale Structure (EFT of LSS) enables rigorous marginalization over theoretical uncertainties; within its framework, the power spectrum \cite{DAmico:2019fhj, Ivanov:2019pdj} and the bispectrum \cite{DAmico:2022osl, Ivanov:2023qzb} have been considered. Presently, it is unclear how quickly the information in the hierarchy of N-point functions converges \cite{Carron:2015hha,Cabass:2023nyo}, and there are indications for significant information beyond the bispectrum \cite{Nguyen:2024yth}. With the simultaneous advance in data quality and the theoretical understanding of the dark matter-galaxy bias relation on quasi-linear scales, the question of how cosmological information can be optimally extracted from galaxy surveys gains urgency.

One avenue to access more information contained in the galaxy distribution is the design of novel summary statistics. They include one-point statistics \cite{Uhlemann:2019gni}, weighted skew-spectra \cite{MoradinezhadDizgah:2019xun}, marked power spectra \cite{Massara:2020pli}, cosmic voids \cite{2019BAAS...51c..40P}, wavelet-based summaries \cite{Cheng:2020qbx, Allys:2020vld}, nearest neighbor distributions \cite{Banerjee:2020umh} and machine-learned compressions \cite{Alsing:2017var, Heavens:2020spq, Philcox:2020zyp}; see also \cite{beyond2pt} for a recent overview and comparison. Since analytical model predictions and covariances are difficult for many of these, simulation based inference (SBI, \cite{2020PNAS..11730055C}) is gaining importance in LSS cosmology (see e.g. \cite{Hahn:2022zxa,Hahn:2023udg} for recent analyses of SDSS/BOSS). SBI requires a simulator or forward model to generate data predictions given a set of cosmological and nuisance parameters, the samples then implicitly define the likelihood of the data vector. For the high-dimensional data vectors in typical cosmological analyses, a fast simulator is crucial to generate the large number of samples required for the convergence of the algorithm.

The compression into summaries can be forgone entirely by analyzing the three-dimensional galaxy distribution at the level of the (density) field. By definition, such an approach extracts all available information. The cosmological parameters, which determine the gravitational evolution and the distribution of initial conditions, are then inferred simultaneously with the initial conditions realization. This huge parameter space can be explored by sampling algorithms such as Hamiltonian Monte Carlo (HMC, \cite{2011hmcm.book..113N}). Field-level analysis has been applied at fixed cosmological parameters to recover initial conditions from mock observations \cite{2013MNRAS.432..894J, 2013MNRAS.429L..84K, 2013ApJ...772...63W, Wang:2014hia, Modi:2018cfi, Shallue:2022mhf, Chen:2023uup, Modi:2022pzm, Dai:2022dso, Qin:2023dew, Jindal:2023qew, Charnock:2019rbk, Doeser:2023yzv} and from observed galaxy distributions \cite{Jasche:2018oym, Lavaux:2019fjr}, where the gravity description ranged from Lagrangian Perturbation Theory (LPT) over full particle-mesh simulations to neural networks. Several works have explored the joint recovery of initial conditions and cosmological parameters from perturbative mocks \cite{Kostic:2022vok}, the non-linear dark matter field \cite{Bayer:2023rmj}, N-body halos \cite{Nguyen:2024yth} and HOD-based mocks \cite{beyond2pt}. As for SBI, an accurate and efficient forward model is crucial for field-level analysis. The impact of the forward model on the recovered initial conditions and parameters has been investigated for power-law bias descriptions \cite{Nguyen:2020hxe}, while \cite{Schmittfull:2018yuk, Schmittfull:2020trd} studied the residuals of a LPT gravity model with EFT bias expansion, at fixed initial conditions, with respect to N-body halos.

Perturbative forward models have two great advantages for SBI and field level analysis: first, they are fast and hence facilitate analyses over cosmological volumes; second, the EFT treatment of galaxy bias allows for robust marginalization over a major source of theoretical uncertainty. In this work, we focus on the \leftfield~code \cite{Schmidt:2020ovm} (Lagrangian EFT-based forward model at the field level), which models the gravitational evolution of (nearly-)Gaussian initial conditions based on Lagrangian perturbation theory and includes the complete set of EFT bias terms in either the Lagrangian or Eulerian expansion. From a detailed understanding of all relevant numerical effects, we derive a set of best-practice recommendations that leave no tunable parameters, once the perturbative order and the cut-off scale have been specified. We validate these guidelines by inferring the amplitude of initial density fluctuations $\sigma_8$ from halos in N-body simulations.  In these tests, we fix the initial conditions, specifically the normalized Gaussian random field, to the ground truth. This speeds up the analysis considerably, and it allows for scanning a wide range of configurations while removing a considerable source of uncertainty. Previous works on the \leftfield\ forward model accuracy \cite{Schmidt:2020ovm} did not systematically optimize all numerical parameters. Rather, they considered conservative settings at the sacrifice of computational runtime. Their fixed-phase analyses \cite{Schmidt:2020viy, Schmidt:2020tao, Babic:2022dws} used a profile likelihood, while we here perform fully Bayesian parameter inferences. The best practice recommendations are an important reference for field-level analyses and SBI already under way \cite{Kostic:2022vok, Nguyen:2024yth, Tucci:2023bag, Babic:2024wph}. In a companion paper \cite{Stadler:2024}, we extend this work to the \leftfield~redshift space modeling \cite{Stadler:2023hea}.

This work is structured as follows. We describe the forward model in section \ref{sec:forward-model}, investigate the numerical accuracy in section \ref{sec:restframe-accuracy}, provide timing information in section \ref{sec:timing} and results on the $\sigma_8$ inference in section \ref{sec:results}. We conclude in section \ref{sec:conclusions}.

\section{Summary of the restframe forward model}
\label{sec:forward-model}

\subsection{Perturbative modeling of the large-scale galaxy density}
\label{sec:forward-model-physics}

The forward model predicts the large-scale galaxy density for a given realization of the initial conditions, cosmological and bias parameters. We are interested in quasi-linear scales exclusively, where the EFT bias description is valid, and where perturbative methods are very accurate and fast. \leftfield~\cite{Schmidt:2020ovm}, in particular, is based on Lagrangian Perturbation Theory (LPT) for the gravitational evolution, coupled with either a Lagrangian or Eulerian bias expansion.

To only keep modes under strict perturbative control, the computation starts from $\delta\lin_\Lambda$, the linear density field filtered at a cut-off scale $\Lambda$ \cite{Schmidt:2020viy} with a Fourier space top-hat. LPT describes the gravitational evolution by a displacement field $\vec{s}$, that relates particles' initial $\vec{q}$ and final positions $\vec{x}$,
\begin{equation}
\vec{x}\left(\vec{q}, \tau\right) = \vec{q} + \sum_{n=1}^{n_\lpt} \vec{s}^{(n)}\left(\vec{q}, \tau\right)\,.
\label{eq:forward-model__lagrangian-to-eulerian-position}
\end{equation}
The displacement is determined by the continuity and Poisson equations, which are solved perturbatively to the order $n_\lpt$. To that end, it is useful to decompose $\vec{s}$ into a curl-free longitudinal, and a curl or transverse component,
\begin{equation}
\vec{s}^{(n)} = \frac{\vec{\nabla}}{\nabla^2} \sigma^{(n)} \left(\vec{q}, \tau\right) - \frac{1}{\nabla^2} \vec{\nabla} \times \vec{t}^{(n)} \left(\vec{q}, \tau\right)\,.
\label{sec:forward-model__shift-decomposition}
\end{equation}
At linear order, $\sigma^{\lin} = -\delta^{\lin}$, only the longitudinal component contributes; the transverse term starts at third order. At any order, the temporal and the spatial dependencies of the LPT solution factorize, and they can be found iteratively from a set of recursion relations \cite{Rampf:2012up, Zheligovsky:2013eca, Matsubara:2015ipa, Rampf:2015mza, Schmidt:2020ovm}. This factorization also holds for a generic expansion history, however, we here approximate the higher-order evolution by the Einstein-de Sitter (EdS) solution, for which
\begin{equation}
\sigma^{(n)}\left(\vec{q},\lambda\right) = \ee^{n\lambda}\, \sigma^{(n)}\left(\vec{q},\lambda=0\right)\,,
\end{equation}
where we have changed the time coordinate from conformal time $\tau$ to $\lambda = \ln D$, the logarithm of the growth factor $D(\tau)$, computed in the given (non-EdS) cosmology. The EdS approximation has been shown to be accurate to the level of $0.1-0.2\%$ (at $z=0$) in predicting the power spectrum \cite{Schmidt:2020ovm}, while significantly simplifying the numerical computations. By choosing the Lagrangian coordinates such that each infinitesimal volume element $d^3\vec{q}$ contains equal mass, the evolved matter density contrast is related to the Jacobian of the coordinate transformation between $\vec{q}$ and $\vec{x}$ as
\begin{equation}
1 + \delta_\mathrm{fwd} \left(\vec{x}\left(\vec{q}\right),\lambda\right) 
=\left|\frac{\partial\vec{x}}{\partial\vec{q}}\right|^{-1}_{\vec{x}=\vec{x}\left(\vec{q},\lambda \right)} = \left| \bm{1} + \bm{H}\left(\vec{q},\lambda\right)\right|^{-1}\,,
\label{eq:forward-model__deltafwd}
\end{equation} 
where the Lagrangian distortion tensor is
\begin{equation}
H_{ij} = \frac{\partial}{\partial q_i}\,s_j \left(\vec{q},\lambda\right) \,.
\label{eq:forward-model__Lagrangian-distortion-tensor}
\end{equation}
In general, $H_{ij}$ has a symmetric component, denoted by $M_{ij}$ in the following, and an anti-symmetric component, where the latter is only sourced by the transverse part of the displacement. Note that eq.~\eqref{eq:forward-model__deltafwd} assumes there is only a single solution for $\vec{q}$ given $\vec{x}$ (single-stream), which is appropriate here as conventional perturbation theory breaks down beyond stream crossing. That is, while multistreaming is technically included in the forward model, since the displacement step is implemented by moving pseudo-particles as described below, we do not expect the forward model to be accurate in that regime.

For the bias expansion there are two options; either Eulerian bias with operators constructed from $\delta_\mathrm{fwd}$ (using nonlocal derivative operators), or Lagrangian bias where the operators are given by all scalar invariants of the symmetric part $M_{ij}$ of the Lagrangian distortion tensor. In the latter case, it is important to note that each order $n$ in $M_{ij}$ has a different time dependence due to gravitational evolution, and hence each contribution has to be considered independently in the bias expansion. However, the fields $\sigma^{(n)} = \mathrm{tr} M^{(n)}$ with $n > 1$ can be re-expressed in terms of lower-order operators through the LPT recursion relations; hence they are redundant and can be excluded from the bias expansion. Here, we will follow previous \leftfield\ works, and replace $\sigma^{(1)}$ by $\sigma = \sum_n \sigma^{(n)}$ in the bias expansion. This choice was initially made for a more convenient implementation. Note that the two variants of the bias expansion only differ by terms higher than the maximum order $o_\bias$ of the bias expansion, and in appendix~\ref{sec:additional-tests__lagrangian-bias-inference} we compare parameter inference results obtained for the Lagrangian bias expansions in terms of $\sigma^{(1)}$ and $\sigma$. The Lagrangian bias operators need to be transformed to the Eulerian frame. Applying the coordinate transformation yields
\begin{equation}
\left[1 + \delta_\mathrm{fwd}\left(\vec{x},\lambda\right)\right]\, \op_{\lagrangian}\left(\vec{x},\lambda\right)
 = \int d^3\vec{q} ~ \op_{\lagrangian}\left(\vec{q}\right)\, \delta_\mathrm{D}\left[ \vec{x} - \vec{q} - \vec{s}\left(\vec{q}, \lambda\right)\right]
\,.
\label{eq:forward-model__opL}
\end{equation}
At leading order, these correspond to the desired Lagrangian bias operators $\op_{\lagrangian}\left(\vec{q}\left(\vec{x}\right)\right)$, and at higher order, for a complete basis of bias operators, the prefactor $1+\delta$ is absorbed by a redefinition of the bias coefficients. Lagrangian and Eulerian bias operator to any order $n$ form a complete basis and can be transformed into each other; that is, they differ only by terms that are higher order than $o_\bias$. However, both alternatives have different implications for the numerical implementation, as discussed below. The bias operators relevant to this study are detailed in appendix~\ref{sec:analysis-settings}.

Once a basis of bias operator $\left\{\op \right\}$ has been constructed to order $o_\bias$, the deterministic large-scale galaxy density simply follows as
\begin{equation}
\dgd\left(\vec{x}\right) = \sum\nolimits_{\left\{\op \right\}}{}{} b_\op \, \op\left(\vec{x}\right) \,,
\end{equation}
where the bias coefficients $b_\op$ are unknown a priori and need to be inferred from the data. Here and in the following, we will always assume a fixed time (or redshift) at which the model is evaluated, and hence drop the explicit time argument. One also expects a stochastic contribution to the observed galaxy density, arising from the superposition of many small-scale modes that are not included into the perturbative solution. At leading order, the expected noise is Gaussian with a white spectrum, while at higher orders scale-, density-dependent and non-Gaussian noise terms arise. For the fixed-initial-conditions analyses presented here, the assumed properties of the noise are of secondary importance. Hence, we restrict the simplest case of Gaussian noise with the subleading scale-dependent correction. A more detailed investigation of the noise model is deferred to future work. In this case, the EFT-likelihood of the field-level galaxy density in Fourier space is \cite{Schmidt:2018bkr, Cabass:2019lqx}
\begin{equation}
- \ln \mathcal{P}\left(\dg | \dgd \right) = \frac{1}{2} \sum_{k\neq 0}^{k_\mathrm{max}} \frac{1}{\sigma^2\left(k\right)} \left| \dg\left(\vec{k}\right) - \dgd \left[\delta_\Lambda^{\lin}, \left\{\theta\right\}, \left\{b_\op\right\}\right]\left(\vec{k}\right) \right|^2 + \ln \sigma^2\left(k\right)\,
\label{eq:forward-model__likelihood}
\end{equation}
where we have written explicitly the dependence of $\dgd$ on the initial conditions, bias coefficients and cosmological parameters $\left\{\theta\right\}$. As with the bias coefficients, the noise level is a priori unknown and needs to be inferred from the data. We parametrize the assumed Gaussian noise with the subleading scale-dependent correction as
\begin{equation}
\sigma^2\left(k\right) = \sigma_0^2 \left[1 + \sigma_{\epsilon,2} k^2\right]^2\,.
\label{eq:forward-model__noise-parametrization}
\end{equation}
The likelihood cut-off, $k_\mathrm{max} \leq \Lambda$, ensures that only modes under perturbative control enter the analysis (see \cite{Schmidt:2020tao} for an implementation in real space).

\subsection{Numerical implementation of the forward model}
\label{sec:forward-model-numerics}

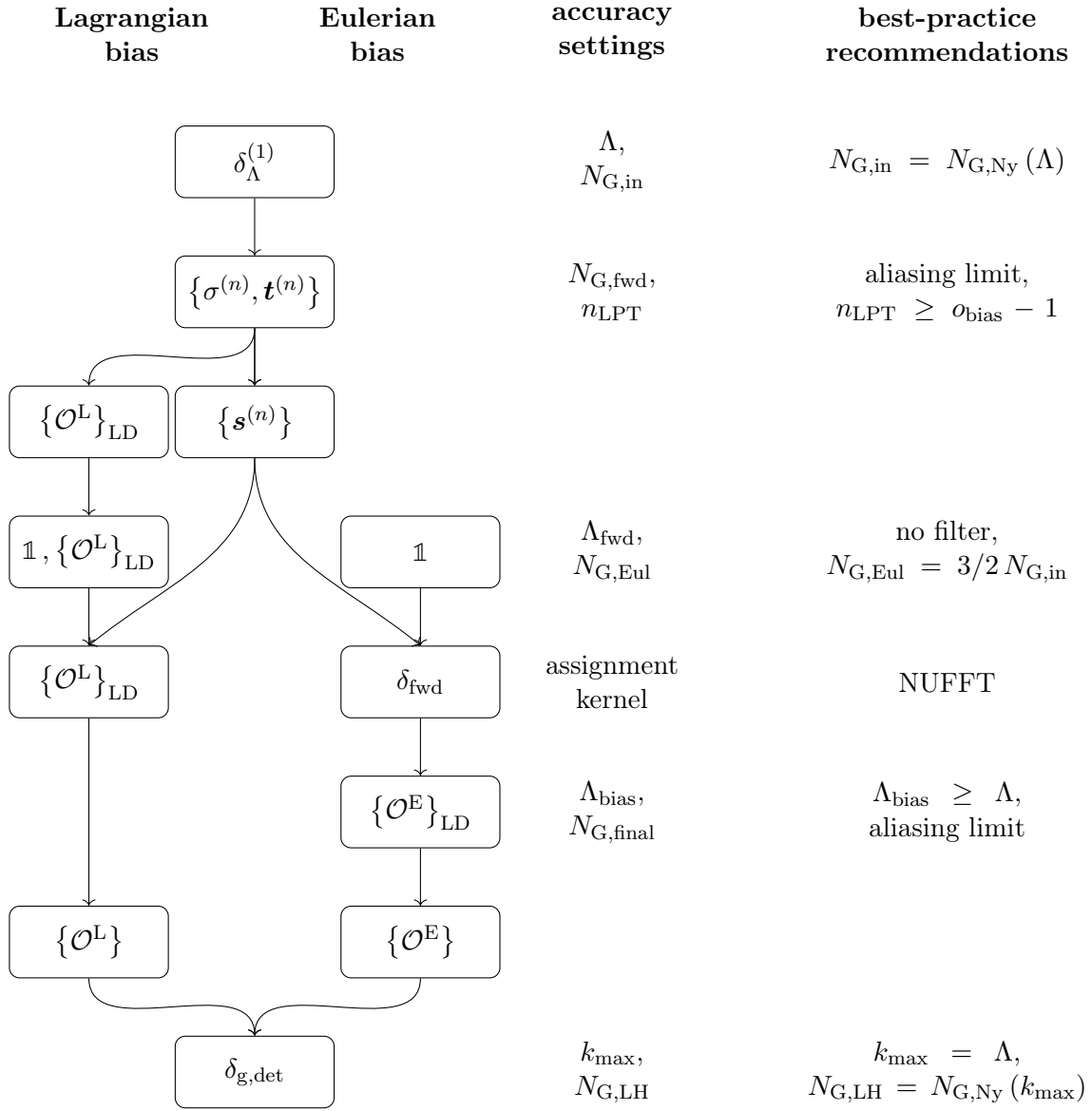
\begin{figure}
\centering
\input{./graphics/flowchart.tex}
\caption{Forward model for the large-scale galaxy density $\dgd$ in the restframe. Two variants, with a Lagrangian (left) or an Eulerian space (right) bias expansion are considered. In the middle column, we list filters, grid sizes, and further parameters which control the numerical accuracy. These parameters and the computational steps are further detailed in section \ref{sec:forward-model-numerics}. Our best-practice recommendations, based on the results of section \ref{sec:restframe-accuracy}, are listed in the rightmost column.}
\label{fig:forward-model}
\end{figure}

The nLPT forward model with Lagrangian and Eulerian bias is implemented in \leftfield. To this end, eqs.~(\ref{eq:forward-model__lagrangian-to-eulerian-position}) to (\ref{eq:forward-model__noise-parametrization}) are solved on three dimensional grids, whose sizes govern the numerical accuracy and the computational demands. Due to the perturbative nature of the forward model and the initial cut-off at $\Lambda$, it is often possible to exactly predict the highest wavenumber with non-zero excitation. Ideally, the grid size $N_\mathrm{G}$ can be chosen such that the Nyquist frequency,
\begin{equation}
k_\nyquist = N_\mathrm{G}\,\frac{\pi}{L_\mathrm{box}}\,.
\label{eq:forward-model__kny}
\end{equation}
exceeds this highest wavenumber, where $L_\mathrm{box}$ is the side length of a cubic simulation box. We denote the minimum grid size that fulfills this condition for a given scale $\Lambda$ as
\begin{equation}
N_{\grid,\nyquist}(\Lambda) = \left\lceil \frac{\Lambda L_\mathrm{box}}{\pi} \right\rceil
\label{eq:NGNy}
\end{equation}
In this case, no numerical resolution effects will impact the model prediction. In the following, we summarize the main computational steps and their relevant frequency limits and accuracy settings (see also figure \ref{fig:forward-model}).
\begin{enumerate}
\item The initial conditions are filtered by a Fourier-space top-hat that only passes modes below the cut-off frequency $\Lambda$, and $\delta_\Lambda^{\lin}$ is represented on a grid of size $N_\ini = N_{\grid,\nyquist}\left(\Lambda\right)$. 
\item The initial conditions are zero-padded in Fourier space to $N_\fwd$, before computing $\sigma$ and $\vec{t}$ to desired order $n_\lpt$. In principle, the higher-order LPT terms populate modes up to $n_\lpt \Lambda$. However, we derive a less demanding criterion for $N_\fwd$ from aliasing constraints in eq.~(\ref{eq:restframe-accuracy_resolution_ngfwd-rule}).
\item The LPT displacement and (if required) Lagrangian bias operators $\{ \op^\lagrangian\}_\mathrm{LD}$ are constructed on grids of size $N_\fwd$. We only construct leading-order in derivatives (LD) operators in the Lagrangian space and add higher-order derivatives later, see point 5. For the displacement, the grid size follows obviously from the fact that eq.~(\ref{sec:forward-model__shift-decomposition}) is a linear operation. For the Lagrangian bias operators, we discuss the impact of $N_\fwd$ in section~\ref{sec:restframe-accuracy__lagrangian-bias}. We also consider the option of applying an intermediate filter at $\filterfwd$ to $\vec{s}$ and $\{ \op^\lagrangian\}_\mathrm{LD}$.
\item The coordinate transformation from Lagrangian to Eulerian frame is implemented by generating an ensemble of equally spaced particles, shifting each particle by the LPT displacement, and re-assigning the particle density to a grid. Particles with unit mass yield $\delta_\mathrm{fwd}$ (eq.~\eqref{eq:forward-model__deltafwd}). By weighting the particles with the operator values eq.~(\ref{eq:forward-model__opL}) is implemented. Before the displacement, we resize the grid to $N_\eul$ by zero padding or removing modes in Fourier space (for details on this operation, see appendix~\ref{sec:fourier-resize}). The number of particles is $N_\eul^3$ and they are assigned to a grid of equal size. Since the displacement is a fully nonlinear operation, the choice of $N_\eul$, and also of the assignment kernel, is not self-evident. We discuss them in section~\ref{sec:restframe-accuracy__density-assignment} and \ref{sec:restframe-accuracy__resolution}, respectively.
\item For Lagrangian bias operators, we construct higher-order derivatives in the Eulerian frame to save the computational costs of additional operator displacements. The difference between Lagrangian and Eulerian derivatives can be absorbed by a redefinition of the bias coefficients for a complete basis of operators (which also includes higher-derivative operators at higher order in perturbations, see Sec.~2.6 in \cite{Desjacques:2016bnm}). We here restrict the set of higher-derivative operators to those obtained by (repeated) application of the Laplace operator \cite{Schmidt:2020ovm}. This is a linear operation and does not require resizing or filtering the grids.
\item In the case of Eulerian bias operators, we filter the evolved density field with a Fourier-space top-hat at $\filterbias$ and resize the grid to $N_\fin$. The filtering is necessary to avoid uncontrolled aliasing from higher-order modes which have been populated by the nonlinear displacement. Its choice is not obvious and discussed in section~\ref{sec:restframe-accuracy__eulerian-bias}. Once $\filterbias$ has been specified, $N_\fin$ follows from aliasing constraints and the bias order.
\item The final grid size $N_\lh$, on which the evolved galaxy density is constructed, is determined via eq.~\eqref{eq:NGNy} using the likelihood cutoff $k_\mathrm{max}$ such that the highest mode in the likelihood is represented.
\end{enumerate}
In the code itself, we replace the minimum grid size calculated using the above guidelines with the smallest equal or larger grid size whose maximum prime factor is 11. This is to ensure numerical efficiency of the Fast Fourier Transform, which becomes inefficient in case of grid sizes with large prime factors.

\section{Accuracy of the restframe forward model}
\label{sec:restframe-accuracy}

The \leftfield~forward model achieves a remarkable accuracy in predicting the gravitational evolution of the matter. The recursive implementation allows to request any LPT order; here we focus on $n_\lpt \leq 4$. In this regime, it is possible to optimize the numerical settings (filters and grid sizes) such that the error of the forward model is dominated by higher-order perturbative terms. The most impactful choices to this regard are the assignment kernel and the resolution of the displacement step $N_\eul$. We start this section by introducing our reference simulations against which we evaluate the numerical accuracy (subsection \ref{sec:restframe-accuracy_reference-sims}), then turn to the optimization of assignment kernel and $N_\eul$ in the subsections \ref{sec:restframe-accuracy__density-assignment} and \ref{sec:restframe-accuracy__resolution}, respectively. The hasty reader can directly jump to subsection \ref{sec:restframe-accuracy__lpterror}, where we discuss the overall gravitational accuracy of \leftfield. Finally, numerical effects in the bias operators are discussed in subsections \ref{sec:restframe-accuracy__lagrangian-bias} and \ref{sec:restframe-accuracy__eulerian-bias} for Lagrangian and Eulerian bias, respectively.

\subsection{Reference simulations to evaluate the perturbative accuracy}
\label{sec:restframe-accuracy_reference-sims}
As reference to evaluate the \leftfield~accuracy, we consider N-body simulations whose initial conditions have an identical cut-off as in the forward model. Essentially, the simulations contain all LPT terms to arbitrarily high order, and their comparison with the forward model can evaluate the LPT implementation as well as its convergence (see also \cite{1993MNRAS.260..765C}).

The simulations were run with the GADGET-2 code \cite{Springel:2005mi} in a box of size $L_\mathrm{box} = 2000 \, \mpc/h$ using $1536^3$ particles of mass $1.8\times 10^{11}\, M_\odot/h$. Initial conditions were set at $z_\mathrm{ini}=24$ with particle displacements generated by second order LPT to minimize transients \cite{Michaux:2020yis}. As fiducial cosmology we adopt throughout this work
\begin{equation}
\Omega_\mathrm{m}=0.30\,,\quad \Omega_\Lambda=0.70\,,\quad h=0.70\,, \quad \sigma_8 = 0.84\,,\quad n_\mathrm{s} = 0.967\,.
\end{equation}
There are two realizations available which start from the same initial conditions but use different cut-offs, $\Lambda = 0.1\,h/\mpc$ and $\Lambda = 0.2\,h/\mpc$, and for each realization snapshots at $z = 1.0\,,0.5\,,0.0$. The simulations were already considered in previous works \cite{Schmidt:2020ovm, Stadler:2023hea}.

\subsection{Density assignment kernel}
\label{sec:restframe-accuracy__density-assignment}
There are two instances of density assignment in a field-level analysis. One in the forward model, to compute the gravitational displacement of matter particles and Lagrangian bias operators. The other in the generation of the data vector, where a set of discrete tracer positions is assigned to a density grid. If the same kernels are used in either operation, one can expect the kernel to cancel to some extent. On the other hand, data and forward model contain different particle numbers and small scales are populated differently in respective cases, rendering the cancellation incomplete. 

We consider two options for the density assignment, a cloud-in-cell (CIC) kernel, as it is frequently used in cosmological simulations and the NUFFT scheme with an ``exponential of semicircle'' kernel \cite{2019SJSC...41C.479B} which derives from algorithms developed for non-uniform Fourier transforms. Similar techniques have also been considered to reduce discreteness effects in N-body simulations \cite{List:2023kbb}. In the CIC scheme, the mass fraction of a particle at position $\vec{x}_p$ assigned to a grid node at position $\vec{x}$ is given by
\begin{equation}
W\left(\vec{x}_p - \vec{x}\right) = \prod_{i=1}^3 \int dx'_i S\left(x_{p,i} - x'_i\right)\,,
\end{equation}
with the assignment kernel,
\begin{equation}
S(x) = \frac{1}{\Delta x}
\begin{cases}
1 & \mathrm{if}~ |x|<\Delta x/2\,, \\
0 & \mathrm{else}
\end{cases}
\end{equation}
and the resulting density at the node position $\vec{x}$ is obtained from summing up all particles weights
\begin{equation}
\rho\left(\vec{x}\right) = \sum_{p=1}^{N_p} m_p \, W\left(\vec{x}_p - \vec{x}\right)\,.
\end{equation}
The kernel size $\Delta x = L_\mathrm{box}/N_\eul$ is set by the grid spacing, and $N_\eul$ governs the resolution of the density assignment for a fixed box size $L_\mathrm{box}$. The CIC assignment ensures mass conservation at machine precision and prevents the generation of spurious large-scale noise.

In the NUFFT scheme, particles are assigned to a super-sampled grid, $N_\mathrm{G,NUFFT} > N_\eul$. The ``exponential of semicircle'' kernel \cite{2019SJSC...41C.479B, 2021A&A...646A..58A},
\begin{equation}
S_\beta\left(z\right) = \begin{cases}
\exp\left[ \beta\left(\left( 1-z^2 \right)^\gamma -1\right)\right] & \mathrm{if}~ |z|<1\,, \\
0 & \mathrm{else}\,,
\end{cases}
\end{equation}
strikes a compromise between two mutually exclusive criteria \cite{97598}: a kernel that is well-localized in real space reduces the computational cost, while a well-localized kernel in Fourier space allows for accurate deconvolution. The algorithm\footnote{we use the \texttt{ducc} implementation, \url{https://gitlab.mpcdf.mpg.de/mtr/ducc}} proceeds to perform a Fast Fourier Transform (FFT) on the super-sampled grid, deconvolves the kernel, and resizes to the target grid size $N_\eul$. This method is approximate, but its accuracy can be tuned to user-specific requirements by the selection of $N_\mathrm{G,NUFFT}$, $\beta$ and $\gamma$. Typically, $N_\mathrm{G,NUFFT} = \left(1.2 - 2.0\right)\,N_\eul$ and the kernel has support over $4-16$ cells. For this work, we require a precision of $10^{-7}$.

\begin{figure}[]
\centering
\includegraphics[width=\figuresize]{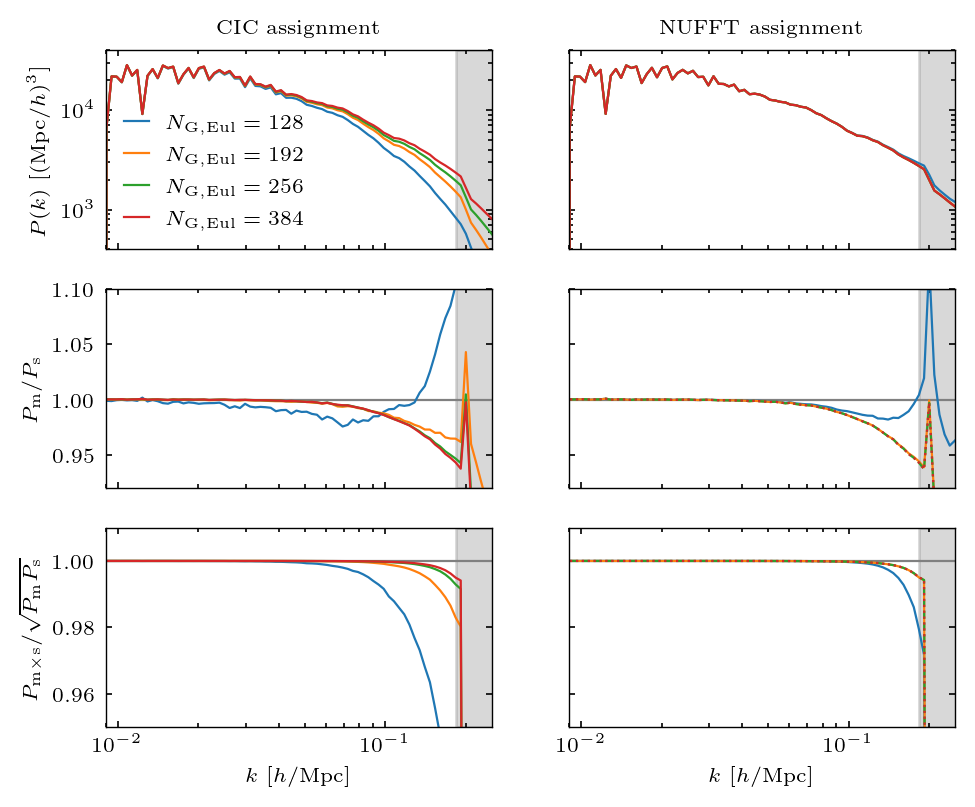}
\caption{Impact of the displacement/assignment step on the accuracy of the forward model. We compute this step at different numerical resolutions, using a CIC (left) and a NUFFT (right) assignment, and compare the 3LPT prediction (subscript ``m'') to reference N-body simulation (subscript ``s'') at $z=0$. Model and simulations are always compared at the same $N_\eul$, and they are computed from identical initial conditions which are filtered at $\Lambda = 0.20\,h/\mpc$. Some level of disagreement is expected due to the perturbative error. The residuals between model and simulation converge much faster at increasing resolution $N_\eul$ for the NUFFT scheme than for the CIC kernel, and we adopt the former for the reminder of this work.}
\label{fig:restframe-accuracy__density-assignment}
\end{figure}

To investigate the effect of the assignment kernel on the agreement between forward model and data, we compare reference simulations and the forward model for $\Lambda=0.20\,h/\mpc$ (figure~\ref{fig:restframe-accuracy__density-assignment}). Due to perturbative errors, we do not expect a perfect agreement between model and simulations, rather we search for convergence with increasing $N_\eul$. Even for a relatively high CIC resolution (large $N_\eul$) the power spectrum ratio and the cross-correlation between forward model and simulation keeps evolving. In the NUFFT scheme, in contrast, the residuals converge at $N_\eul \geq 192 = 3/2\, N_\ini$. Therefore, even though the NUFFT scheme is computationally more expensive at fixed $N_\eul$, it allows for smaller grid sizes and provides a more effective forward model. It allows to speed up the forward model by more than a factor four, as we show in section \ref{sec:timing}. We therefore focus exclusively on the NUFFT assignment for the reminder of this work.

\subsection{Resolution effects in the particle displacement}
\label{sec:restframe-accuracy__resolution}

Given the choice of the NUFFT assignment scheme, we seek to further optimize the resolution. The goal is to keep $N_\eul$ as low as possible for numerical efficiency while sufficiently suppressing numerical errors. To develop a better understanding for the latter, we express the generic displacement to the Eulerian frame in eq.~(\ref{eq:forward-model__opL}) as
\begin{align}
\label{eq:eulerian-frame-accuracy-displaced-operators}
\op_{\eulerian}(\vec{k}) &= \int d^3\vec{q} \, \op_{\lagrangian}(\vec{q}) \,\ee^{-i\vec{k}\cdot\left[\vec{q} + \vec{s}(\vec{q})\right]} \\
&= \int \frac{d^3\vec{k'}}{(2\pi)^3}\, \op_{\lagrangian}(\vec{k'}) \, \int d^3\vec{q}\, \ee^{-i(\vec{k}-\vec{k'})\cdot\vec{q}} ~ \ee^{-i\vec{k}\cdot\vec{s}(\vec{q})} \nn\\
&= \int \frac{d^3\vec{k'}}{(2\pi)^3}\, \op_{\lagrangian}(\vec{k'}) \, \int d^3\vec{q}\, \left[1 - i\vec{k}\cdot\vec{s}(\vec{q}) - \frac{1}{2} k_l k_m s^l(\vec{q}) s^m(\vec{q}) + \mathcal{O}\left((\vec{k}\cdot\vec{s})^3\right)\right] \ee^{-i(\vec{k}-\vec{k'})\cdot\vec{q}} \nn\,.
\end{align}
The matter density follows from the displacement of a unit field or equivalently in Fourier space $\op_{\lagrangian}(\vec{k}) = (2\pi)^3 \delta^\dirac(\vec{k})$, and hence is given by
\begin{align}
\delta_{\rm fwd}(\vec{k})  = - i\vec{k}\cdot\vec{s}(\vec{k}) - \frac{1}{2} k_l k_m {\left(s^l s^m\right)}(\vec{k}) + \mathcal{O}\left((\vec{k}\cdot\vec{s})^3\right)\,,
\label{eq:restframe-accuracy_resolution_displaced-delta}
\end{align}
where ${\left(s^l s^m\right)}(\vec{k})$ indicates the Fourier transform of the product $s^l(\vec{q}) s^m(\vec{q})$. The forward model aims to predict galaxy clustering on quasi-linear scales, where the expansion in powers of the displacement converges. Eq.~(\ref{eq:restframe-accuracy_resolution_displaced-delta}) allows to investigate the leading sources for numerical errors. The finite grid spacing can give rise to the following effects:
\begin{itemize}
\item \textbf{Truncation errors}: if we choose $N_\ini < N_\eul < N_\fwd$ we filter out modes from $\vec{s}$ between $\Lambda$ and $n_\lpt\,\Lambda$ that have been populated by higher-order LPT contributions. Since the displacement couples modes at different scales, the truncation can impact the low-$k$ modes we are interested in.
\item \textbf{Aliasing}: since eq.~(\ref{eq:eulerian-frame-accuracy-displaced-operators}) represents a fully nonlinear coupling it can in principle populate arbitrarily high modes. Aliasing from these high modes to lower ones can affect predictions on the scales of interest.
\end{itemize}

\subsubsection{Truncation errors}

\paragraph{Theoretical expectation}
For an analytical intuition about truncation errors, we decompose the shift vector $\vec{s}$ into a long-scale and a short-scale contribution $\vec{s} = \vec{s}_{\mathrm{l,fwd}} + \vec{s}_{\mathrm{s,fwd}}$, where the split is determined by the grid size $N_\eul$. The long-short decomposition is applied to the (evolved) $n_\lpt$-order shift vector and represents the effect of down-sizing the grid in Fourier space after the perturbative solution has been obtained. The leading-order term in eq.~(\ref{eq:restframe-accuracy_resolution_displaced-delta}) does not introduce any mode-coupling for $N_\eul \geq N_\ini$. The second-order contribution decomposes as
\begin{align}
\left(s^l s^m\right)(\vec{k}) 
&= \int \frac{d^3\vec{k'}}{(2\pi)^3} \left[ 
s^l_{\mathrm{l,fwd}}(\vec{k'}) s^m_{\mathrm{l,fwd}}(\vec{k}-\vec{k'}) 
+ 2\, s^l_{\mathrm{l,fwd}}(\vec{k'}) s^m_{\mathrm{s,fwd}}(\vec{k}-\vec{k'}) \right. \nn \\
&\left.\hspace{1.6cm}
+\, s^l_{\mathrm{s,fwd}}(\vec{k'}) s^m_{\mathrm{s,fwd}}(\vec{k}-\vec{k'}) \right] \,,
\label{eq:restframe-accuracy__leading-order-resolution-effects}
\end{align}
and the leading contribution neglected by truncation errors is
\begin{align}
\Delta\delta_\mathrm{trunc.}^\LO = 
- \frac{1}{2} k_l k_m  \int \frac{d^3\vec{k'}}{(2\pi)^3} \left[ 
2 s^l_{\mathrm{l,fwd}}(\vec{k'}) s^m_{\mathrm{s,fwd}}(\vec{k}-\vec{k'})
+ s^l_{\mathrm{s,fwd}}(\vec{k'}) s^m_{\mathrm{s,fwd}}(\vec{k}-\vec{k'}) \right]\,.
\label{eq:restframe-accuracy_resolution_truncation-error}
\end{align}
The easiest scenario to discuss truncation effects is for $k_\nyquist\left(N_\eul\right)=\Lambda$. In this case, small-scale modes are only populated by the nonlinear evolution,
\begin{align}
\vec{s}_{\mathrm{l,fwd}}(\vec{k}) &= \vec{s}^{\lin}_{\mathrm{l,fwd}}(\vec{k}) + \vec{s}^{(2)}_{\mathrm{l,fwd}}(\vec{k}) + \vec{s}^{(3)}_{\mathrm{l,fwd}}(\vec{k})  + \ldots \nn\\
\vec{s}_{\mathrm{s,fwd}}(\vec{k}) &= \vec{s}^{(2)}_{\mathrm{s,fwd}}(\vec{k}) + \vec{s}^{(3)}_{\mathrm{s,fwd}}(\vec{k}) + \dots\,.
\end{align}
The leading truncation error arises at third order from the combination $\vec{s}^{\lin}_{\mathrm{l,fwd}} \vec{s}^{(2)}_{\mathrm{s,fwd}}$. Its scaling with $k$ is determined by the convolution of a low-pass and a high-pass sharp-k filter (see appendix~\ref{sec:filters}), and we expect
\begin{equation}
\left\langle \left(\Delta\delta_\mathrm{trunc.}^\LO\right)^2 \right\rangle \propto k^6\,.
\label{eq:restframe-accuracy_resolution_truncation-residuals}
\end{equation} 
The truncation error correlates with the full high-resolution solution (see appendix~\ref{sec:analytic-residuals_truncation}) as
\begin{equation}
\left\langle \Delta\delta_\mathrm{trunc.}^\mathrm{L.O.}\, \delta \right\rangle \propto k^3\,P(k)\,.
\label{eq:restframe-accuracy_resolution_truncation-correlation}
\end{equation}

\paragraph{Numerical experiments}
\begin{figure}[h]
\centering
\includegraphics[width=\figuresize]{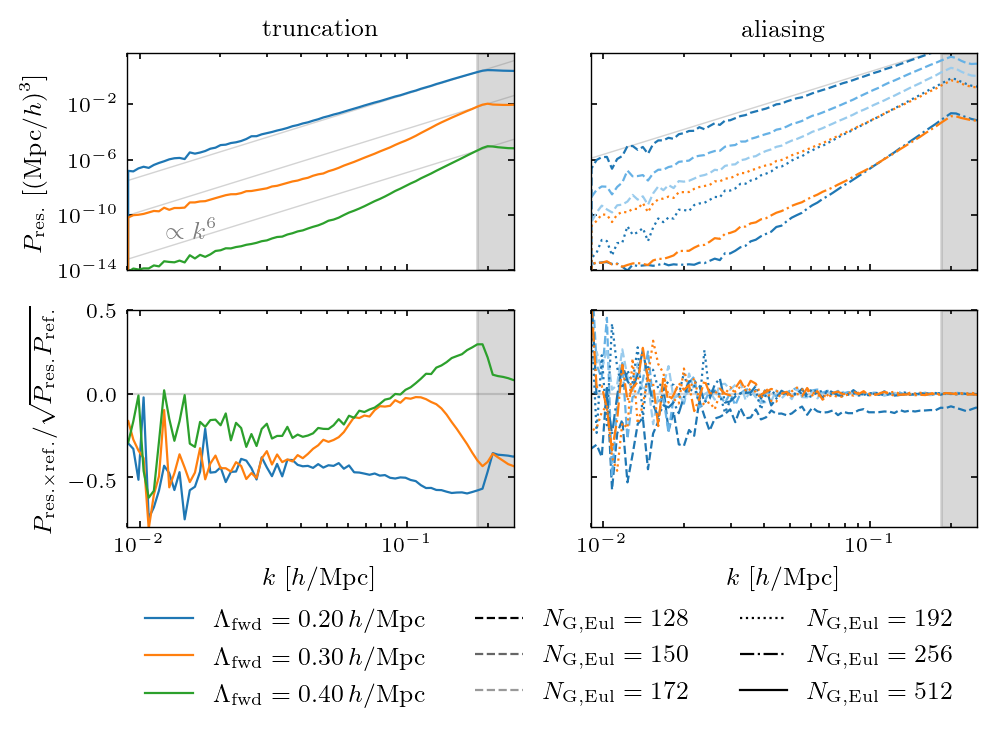}
\caption{Impact of truncation and aliasing effects in the displacement step on the evolved matter density at $z=0$. For \textbf{truncation effect  (left)}, we apply different filters $\filterfwd$ to the displacement vector before the actual displacement operation and compute the latter at high resolution, $N_\eul=512$. The reference model has no filter. For \textbf{aliasing effects (right)}, we compare densities obtained with identical filters in the forward model at different displacement resolutions. The reference model has the highest resolution, $N_\eul=512$. In either case we depict the power spectrum of the residuals (top) and the cross-correlation between residuals and reference (bottom). We here specifically consider $\Lambda=0.2\,h/\mpc$ and the 3LPT model; in figure~\ref{fig:restframe-accuracy_resolution_truncation-lpt} we explore different LPT orders. The numerical results reproduce the $k^6$-scaling (gray lines) expected for $\filterfwd=\Lambda$ and $N_\eul=N_\ini$. We also predict the white ($k$-independent) residual correlations at low-$k$ observed for truncation effects. Resolution effects can be effectively suppressed by choosing $N_\eul = 3/2\,N_\ini$ without an intermediate filter $\filterfwd$.}
\label{fig:restframe-accuracy_resolution_truncation}
\end{figure}

To investigate truncation effects numerically, we choose $N_\fwd = 3\,N_\ini$ and $N_\eul > n_\lpt\, N_\ini$, so the displacement grid can in principle represent all modes, and we apply a cubic filter at $\filterfwd$ before we compute the displacement. This filter mimics truncation effects from resizing the grid to $N_\eul < N_\fwd$, and it allows to isolate them from aliasing contributions. The densities obtained from filtered grids are compared to an unfiltered reference in figure~\ref{fig:restframe-accuracy_resolution_truncation} (left).

Indeed we observe the expected $k^6$ scaling in the top-right panel, in particular for $\filterfwd = k_\nyquist\left(\Lambda\right)$ (the same truncation error would be implied by $N_\eul = N_\ini$). Truncation errors lead to a sizable correlation between the residuals and the unfiltered reference. At low-$k$ this correlation is white, and indeed from eq.~(\ref{eq:restframe-accuracy_resolution_truncation-correlation}) we would expect a $\sqrt{P(k)}$ dependence. Importantly, truncation effects decrease considerably as $\filterfwd$ (or correspondingly $N_\eul$) increases; going from $\filterfwd = \Lambda$ ($N_\eul = N_\ini$) to $\filterfwd = 3/2\, \Lambda$ ($N_\eul=3/2\, N_\ini$) the residual power spectrum has diminished by a factor $2\times 10^{-4}$. If we apply a spherical filter instead of the cubic one chosen for figure ~\ref{fig:restframe-accuracy_resolution_truncation}, the residuals increase slightly, but the qualitative behavior remains unchanged. This result can be expected, since the spherical filter cuts modes in the ``corners'' of the cube, which would be kept otherwise.

\subsubsection{Aliasing errors}
\paragraph{Theoretical expectation}
Going back to eq.~(\ref{eq:restframe-accuracy_resolution_displaced-delta}), it is clear that the leading term which can give rise to aliasing is $\propto s^{(1)} s^{(1)}$. Due to the initial cut-off, $s^{(1)}$ only contains modes up to $\Lambda$, and the leading-order aliasing contribution populates modes up to $2\Lambda$. These high-k modes can be aliased to modes above the aliasing frequency (``Orszag rule'')
\begin{equation}
k_\alias = 2 k_\nyquist\left(N_\eul\right) - 2\Lambda\,.
\end{equation}
Analytical predictions for the aliasing error are easiest for $N_\eul=N_\ini$, and we expect (see appendix~\ref{sec:analytic-residuals_aliasing})
\begin{equation}
\left\langle \left( \Delta\delta_\alias^\LO \right)^2\right\rangle \propto k^6\,.
\end{equation}
On the other hand, we can eliminate the leading-order aliasing effect for modes of interest by requiring $k_\alias \geq \Lambda$. This imposes
\begin{equation}
N_\eul \geq \frac{3}{2}\,\frac{L_\mathrm{box} \Lambda}{\pi} = \frac{3}{2} N_\ini \,,
\label{eq:restframe-accuracy_resolution_ngeul-rule}
\end{equation}
and is referred to as the ``3/2-rule'' \cite{1971JAtS...28.1074O}.

\paragraph{Numerical experiments} We investigate aliasing errors numerically by computing the displacement at different grid sizes, and by comparing those results with a high-resolution reference. The results are summarized in the right panel of figure~\ref{fig:restframe-accuracy_resolution_truncation}. To control which modes are present and can contribute to aliasing, we in addition apply filters of different sizes $\filterfwd$ before the displacement. The grids are always chosen large enough so all modes that pass the filter can be represented, and the comparison always is between identical filters to eliminate truncation errors. In figure~\ref{fig:restframe-accuracy_resolution_truncation-lpt}, we compare the aliasing error at different LPT orders and find a qualitatively similar behavior.

For the coarsest grid with $k_\nyquist\left(N_\eul\right)=\Lambda$, we recover the $\propto k^6$ scaling that we estimated analytically. By comparing different filter sizes at identical resolution, we can estimate the impact of small-scale modes with $k > \Lambda$. It appears to be minor and only becomes recognizable at low-$k$, where the residuals are considerably suppressed. This matches the qualitative expectation that, to alias into very low $k$, one needs to generate a high-k mode first, and these are perturbatively suppressed due to the initial cut-off. The correlation between residuals and the high-resolution reference decays very quickly and can only be noted for the coarsest grid.

\subsubsection{Best-practice to minimize resolution effects}
For computationally practical displacement resolutions $N_\eul$, both aliasing and truncation errors will be present in the forward model. However, we can suppress aliasing at the cost of increasing truncation by applying a low-pass filter at $\filterfwd$ to the shift vector $\vec{\hat{s}}$ before computing the displacement. From the results of this subsection, such a low-pass filter seems not advisable  as the nonlinearly populated modes, which would be filtered, only have a subdominant contribution to the aliasing error. A more restrictive filtering, on the other hand, quickly increase the truncation error and the correlated component from the residuals. For example, figure \ref{fig:restframe-accuracy_resolution_truncation} demonstrates that the truncation error dominates over any aliasing benefits if we were to chose $\filterfwd=\Lambda$ for $N_\eul = 3/2\, N_\ini$. We therefore advise to set $N_\eul = 3/2\, N_\ini$ as in eq.~(\ref{eq:restframe-accuracy_resolution_ngeul-rule}) and to not apply a filter before the displacement. With these settings, resolution effects are suppressed well below the perturbative error (see section~\ref{sec:restframe-accuracy__lpterror}) while still achieving good computational speed (see section~\ref{sec:timing}).

\subsection{Perturbative Accuracy}
\label{sec:restframe-accuracy__lpterror}

To compute the nLPT solution in the previous section, we use the ``no-mode-left-behind'' scheme, i.e. we chose the grids large enough that all higher-order modes generated could be represented, $N_\fwd = N_{\grid,\nyquist}\left(n_\lpt\Lambda\right)$. Before the displacement, the grids are resized to $N_\eul$ and we have established $N_\eul = 3/2\, N_\ini$ as a good criterion to suppress aliasing errors in the previous section. Modes above $k_\nyquist\left(N_\eul\right)$ will not affect the final evolved density grid, so it is actually sufficient to suppress aliasing below this threshold by imposing
\begin{equation}
N_\fwd \geq \frac{1}{2}\left( N_\eul + n_\lpt N_\ini\right) = \frac{1}{2} \left(\frac{3}{2} + n_\lpt\right) N_\ini\,,
\label{eq:restframe-accuracy_resolution_ngfwd-rule}
\end{equation}
where for the second equality, we assumed the $3/2$-rule. Still one might worry that some back-coupling in the nLPT solution of high-k modes to modes below $k_\nyquist\left(N_\eul\right)$ is missed. In appendix~\ref{sec:optimizing-ngfwd}, we show that this cannot be the case and verify explicitly, that the less demanding criterion for $N_\fwd$ leaves the gravitational prediction unaffected at the level of machine precision. We therefore advise to select $N_\fwd$ by eq.~(\ref{eq:restframe-accuracy_resolution_ngfwd-rule}).

\paragraph{Theoretical expectations}
In the implementation of the gravitational evolution, we solve perturbatively for the displacement vector, but the coordinate transformation itself is completely nonlinear. While there is a one-to-one correspondence between Lagrangian and Eulerian perturbation theory at fixed order \cite{Rampf:2012xa}, the nonlinear displacement generates shift terms, which are protected by symmetries, at arbitrarily high orders. They somewhat complicate the residuals arising from truncating the perturbative solution at finite order. We here use Eulerian perturbation theory, to provide an intuitive understanding of the residuals, and in appendix \ref{sec:analytic-residuals_lpt}, we derive identical scaling relations explicitly for the first and second order LPT solution.

The leading-order error of the n-th order perturbative solution is given by 
\begin{equation}
\Delta\delta_{n_\lpt}^\LO = \delta^{(n_\lpt+1)}\,,
\end{equation}
and correspondingly, in the low-$k$ limit the residual power spectrum is
\begin{equation}
P_\mathrm{res.}^\LO(k) = P^{(n_\lpt+1, n_\lpt+1)},
\end{equation}
which scales as $k^4$ (or even higher power) for $k\rightarrow 0$.
On the other hand, the residuals correlate with the all-order solution as
\begin{equation}
\left\langle \Delta\delta_{n\lpt}\, \delta \right\rangle_\LO =
\begin{cases}
\left\langle \delta^{(n_\lpt+1)} \delta^{(1)} \right\rangle &\mathrm{if}~n~\mathrm{even} \\
\left\langle \delta^{(n_\lpt+1)} \delta^{(2)} \right\rangle &\mathrm{if}~n~\mathrm{odd}\,.
\end{cases}
\end{equation}
The lowest odd term with $n_\lpt=1$ scales as $\left\langle \delta^{(2)}\delta^{(2)} \right\rangle \propto k^4$ and the lowest even term, $n_\lpt=2$, $\left\langle \delta^{(3)}\delta^{(1)} \right\rangle \propto k^2 P(k)$ for $k\rightarrow 0$. At higher orders, the correlation between LPT error and the all-order solution is limited from above by the respective even/odd lowest order scalings. In the cross-correlation \emph{coefficient} of the residuals, this leads to an approximately constant plateau at low-$k$ for even LPT orders, and trend toward zero for odd LPT orders. Again, the residual power spectrum as well as its cross power spectrum with the model go to zero on large scales in all cases.

\paragraph{Numerical experiments}

\begin{figure}
\centering
\includegraphics[width=\figuresize]{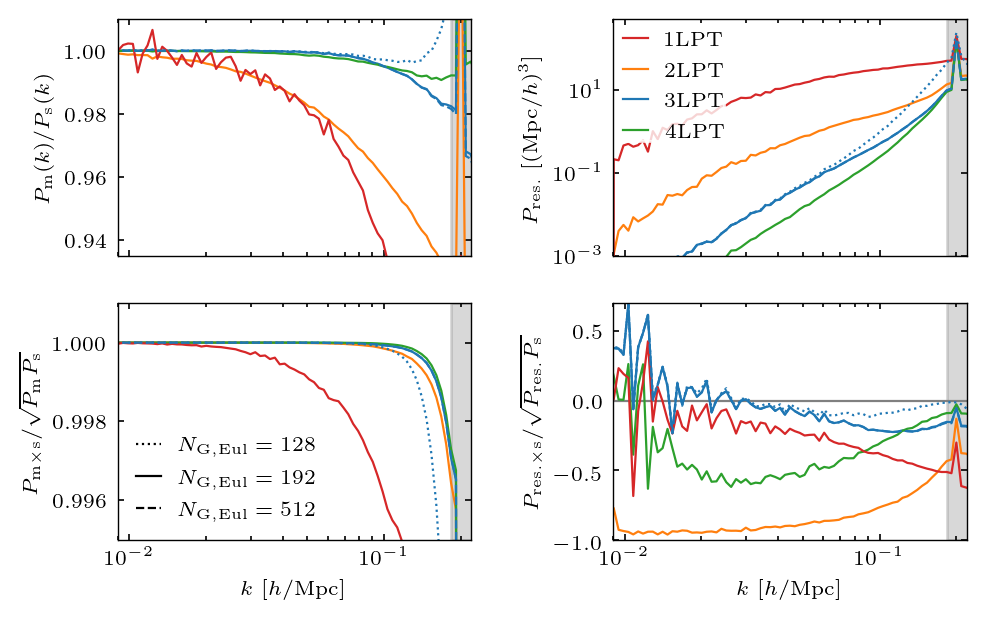}
\caption{Accuracy of the perturbative forward model at $z=0.5$ for different LPT orders. We compare the forward model to a reference N-body simulation run from identical initial conditions filtered at $\Lambda=0.2\,h/\mpc$ (see section~\ref{sec:restframe-accuracy_reference-sims}). The comparison is shown, going clockwise from the top left, in terms of power spectrum ratios, the residual power spectrum, the correlation between model and reference and the correlation between residuals and reference. For the third order solution, we additionally vary the resolution of the displacement, $N_\eul$, and we neglect transverse terms in the 3LPT and 4LPT solution, which are shown to be negligible in figure~\ref{fig:restframe-accuracy_lpt_curl}. For higher LPT orders, the forward model achieves remarkable accuracy at a level of $\sim 2\%$ down to the smallest scales tested. The accuracy improves further for smaller cut-off values $\Lambda$.}
\label{fig:restframe-accuracy_lpt_lpterror}
\end{figure}

We compare predictions from the forward model to the reference simulations (section~\ref{sec:restframe-accuracy_reference-sims}) in figure~\ref{fig:restframe-accuracy_lpt_lpterror}. Indeed, the residuals decrease systematically with LPT order, and their cross-correlation exhibits the expected alternating behavior between even and odd LPT orders. The overall accuracy of our forward model is at the level of $\sim 2\%$ up to modes of $0.2\,h/\mpc$ at redshift $z=0.5$. At $z=0$, it still would be at the level of $4\%$. The accuracy further improves for lower cut-off values and for $\Lambda=0.10\,h/\mpc$ we find sub-percent agreements at $z=0.5$ (figure~\ref{fig:restframe-accuracy_lpt_lpterror_l010}). For some selected scenarios in figure~\ref{fig:restframe-accuracy_lpt_lpterror}, we also test higher and lower displacement-resolutions, $N_\eul$. While a too-low resolution has a visible impact on the accuracy, there is no discernible improvement once the 3/2-rule in eq.~(\ref{eq:restframe-accuracy_resolution_ngeul-rule}) has been satisfied. This indicates that the forward model accuracy indeed is dominated by the perturbative error rather than numerical effects. The results of figure~\ref{fig:restframe-accuracy_lpt_lpterror} have been obtained while neglecting the transverse contribution to the displacement in eq.~(\ref{sec:forward-model__shift-decomposition}). This contribution starts at third order, and we assess its impact explicitly in figure \ref{fig:restframe-accuracy_lpt_curl}, finding it to be negligible.

\begin{figure}
\centering
\includegraphics[width=\figuresize]{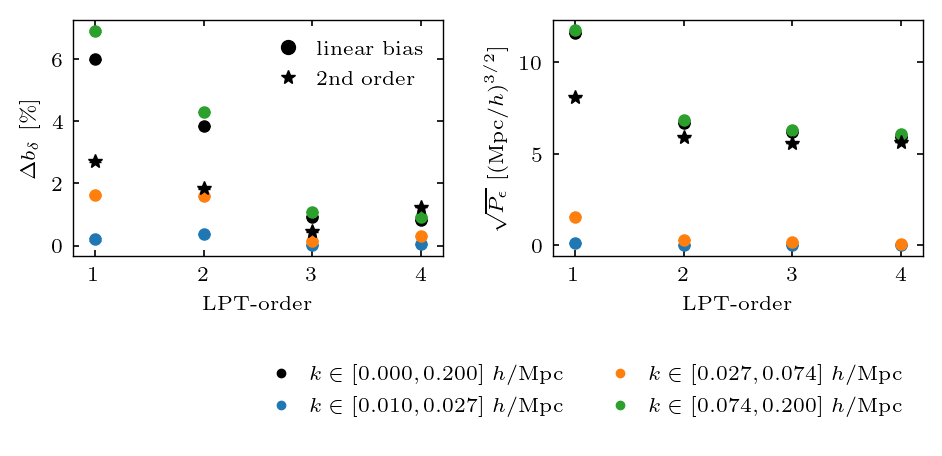}
\caption{Linear bias parameter and noise power spectrum inferred from the reference simulations at $z=0$ with cut-off $\Lambda=0.20\,h/\mpc$. Only the wavenumber-interval indicated in the legend is considered in the likelihood and we assume white Gaussian noise. In a realistic analysis, tracer noise is likely to dominate over the LPT error and the parameter shifts are not indicative for those scenarios. Rather, they demonstrate how the forward model accuracy increases with LPT order and that higher-order bias operators can (partly) absorb the correlated component of the LPT residuals.}
\label{fig:restframe-accuracy_lpt_inference}
\end{figure}

To illustrate how the LPT error can impact an analysis, we infer the linear bias parameter $b_\delta$ from the simulations with cut-off, which should be recovered to be consistent with 1. The only additional free parameter is the noise amplitude squared, $P_\epsilon$. As figure~\ref{fig:restframe-accuracy_lpt_inference} shows, the noise $P_\epsilon$ and the mismatch in $b_\delta$ systematically decrease as the LPT order increases and the model becomes more accurate. The shift in $b_\delta$ from the ground truth is mostly driven by the highest $k$-modes; if we restrict the modes considered in the likelihood to a spherical shell over some $k$-range, only the shell with the highest wavenumbers infers $b_\delta$ significantly offset from the ground truth.

In the analysis of a biased tracer, there are noise contributions from integrating out small scales and discreteness. They can easily dominate over the LPT error. The correlated component of the LPT residuals, on the other hand, can partly be absorbed by higher-order bias operators. Indeed, the results in figure~\ref{fig:restframe-accuracy_lpt_inference} show that we infer $b_\delta$ closer to the true value by allowing for a second-order bias expansion. The precision and accuracy found here, however, does not translate immediately to a biased tracer population. We investigate the latter explicitly in section \ref{sec:results}. Before doing so, we turn to the accuracy at which higher-order bias operators are computed in the forward model.

\subsection{Lagrangian bias and weighted displacements}
\label{sec:restframe-accuracy__lagrangian-bias}

The numerical effects in Lagrangian bias operators are very similar to the matter density field. We first focus on the resolution of the displacement step, $N_\eul$. Subsequently, we optimize $N_\fwd$, the size of the Lagrangian-space grid on which the bias operators are constructed.

\paragraph{Theoretical Expectations}
For an intuition of resolution effects, we again expand the displacement perturbatively. For a generic weight field $\op\left(\vec{q}\right)$, eq.~(\ref{eq:eulerian-frame-accuracy-displaced-operators}) becomes
\begin{equation}
\op^{\eulerian}(\vec{k}) = \op^{\lagrangian}(\vec{k}) - i k_i \int \frac{d^3\vec{k'}}{(2\pi)^3} \op^{\lagrangian}(\vec{k'}) s^i(\vec{k}-\vec{k'}) -\frac{1}{2} k_l k_m \int \frac{d^3\vec{k'}}{(2\pi)^3} \op^{\lagrangian}(\vec{k'}) {\left(s^l s^m\right)}(\vec{k}-\vec{k'}) + \ldots
\label{eq:eulerian-frame-accuracy-displaced-operator-2}
\end{equation}
We again split $\vec{s}$ and $\op^{\lagrangian}$ into a long-wave and a short-wave contribution and for simplicity assume $N_\eul = N_\ini$. The leading-order truncation effect originates from the first integral in eq.~(\ref{eq:eulerian-frame-accuracy-displaced-operator-2}), where terms of the type
\begin{equation}
k_i \op^{\lagrangian}_{\mathrm{l,fwd}} s^i_\mathrm{s,fwd}\,,\quad
k_i \op^{\lagrangian}_{\mathrm{s,fwd}} s^i_\mathrm{l,fwd}\,,\quad
k_i \op^{\lagrangian}_{\mathrm{s,fwd}} s^i_\mathrm{s,fwd}\,,
\end{equation}
are neglected. With the $k$-scaling from the of low-pass and high-pass or two high-pass filters (see appendix~\ref{sec:filters}), we expect the residuals due to the truncation error to scale as
\begin{equation}
\left\langle \left(\Delta\op^\eulerian_{\mathrm{trunc.}}\right)^2 \right\rangle_\LO \propto k^4\,.
\label{eq:eulerian-frame-accuracy__lagrangianop-truncation-residuals}
\end{equation}
The leading-order aliasing error arises from the term $k_i \op^\lagrangian_{\rm l,fwd} s^i_{\rm l,fwd}$. Under the assumption $k_\nyquist\left(N_\eul\right)=\Lambda$ it follows analogously to eq.~(\ref{eq:analytic-residuals__aliasing-scaling}) that
\begin{equation}
\left\langle \left(\Delta\op^\eulerian_{\alias} \right)^2 \right\rangle_\LO \propto k^4\,.
\label{eq:eulerian-frame-accuracy__lagrangianop-aliasing-residuals}
\end{equation}

\begin{figure}
\centering
\includegraphics[width=\figuresize]{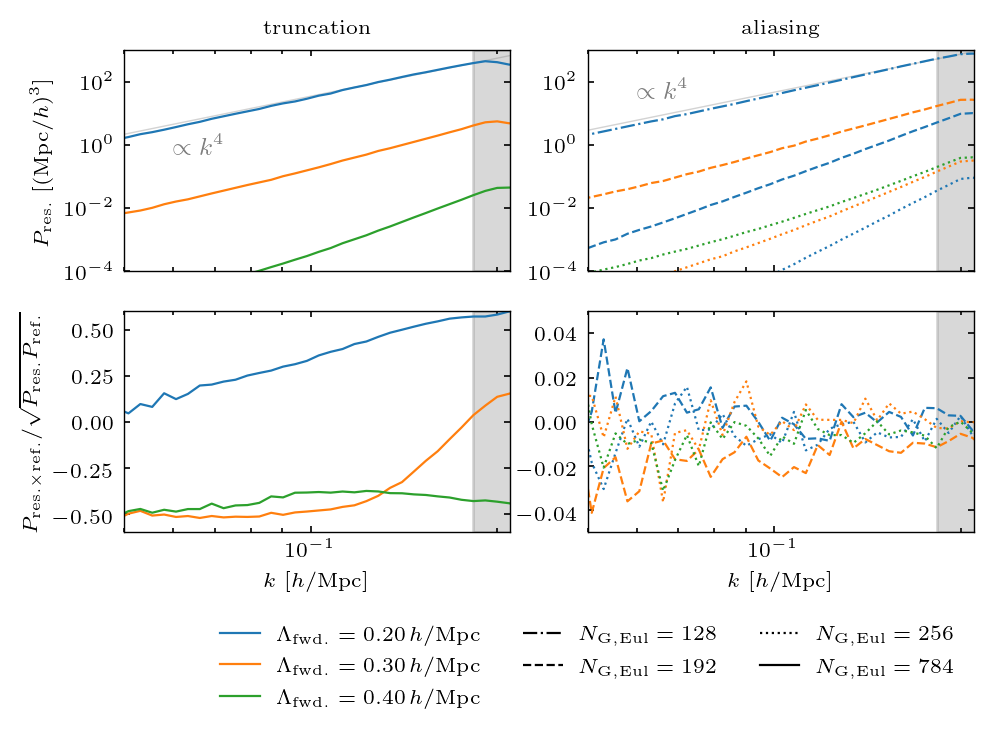}
\caption{Impact of truncation and resolution effects on higher-order Lagrangian bias operators, after they have been transformed to the Eulerian frame; as representative example we consider $\sigma^2$, where $\sigma=\sum\sigma^{(n)}$ is the divergence of the displacement field. The forward model computes the gravitational evolution to $z=0$ at third LPT order for a cut-off $\Lambda=0.2\,h/\mpc$. For \textbf{truncation effect  (left)}, we apply different filters $\filterfwd$ to the displacement vector before the actual displacement operation and compute the latter at high resolution, $N_\eul=768$. The reference model has no filter. For \textbf{aliasing effects (right)}, we compare densities obtained with identical filters in the forward model at different displacement resolutions. The reference model has the highest resolution, $N_\eul=768$. The numerical tests recover the expected $k^4$ scaling in the residuals. If $N_\eul \geq 3/2\,N_\ini$ resolution effects in the bias operators are suppressed below the perturbative accuracy.}
\label{fig:restframe-accuracy_bias_lagrangian-truncation-resolution}
\end{figure}

\paragraph{Numerical experiments} We test resolution effects in Lagrangian bias operators for the example of $\sigma^2$, where $\sigma = \boldsymbol{\nabla}\cdot\vec{s}$ is the divergence of the displacement field. The operator contains modes up to $2\times n_\lpt$. To isolate aliasing and truncation errors in the displacement step, we choose $N_\fwd=2 n_\lpt N_\ini$, that is we construct $\sigma^2$ in the Lagrangian frame on a grid that is large enough to represent all nonlinearly populated modes. For truncation effects, we compare results on a large reference grid, $N_\eul=784$, with and without applying a filter before the displacement. For aliasing effects, we compare results at identical filters but different displacement resolutions. The results are summarized in figure~\ref{fig:restframe-accuracy_bias_lagrangian-truncation-resolution}.

The numerical results indeed recover the leading-order scaling predicted in eqs.~(\ref{eq:eulerian-frame-accuracy__lagrangianop-truncation-residuals}) and (\ref{eq:eulerian-frame-accuracy__lagrangianop-aliasing-residuals}). For a filter size of $\filterfwd=\Lambda=0.20\,h/\mpc$, truncation and aliasing effects become very severe. However, if the grid and filter sizes obey the 3/2-rule, the residual power spectrum for either type of resolution effect falls below the corresponding LPT error. As a reference, we here consider the 3LPT model at $z=0.5$ and identical cut-off $\Lambda = 0.20\,h/\mpc$. We therefore conclude that resolution effects in the displacement are well controlled by the 3/2-rule also for higher-order bias operators, and we expect that the overall accuracy of the forward model is dominated by the order of the LPT solution.

\begin{figure}
\centering
\includegraphics[width=\figuresize]{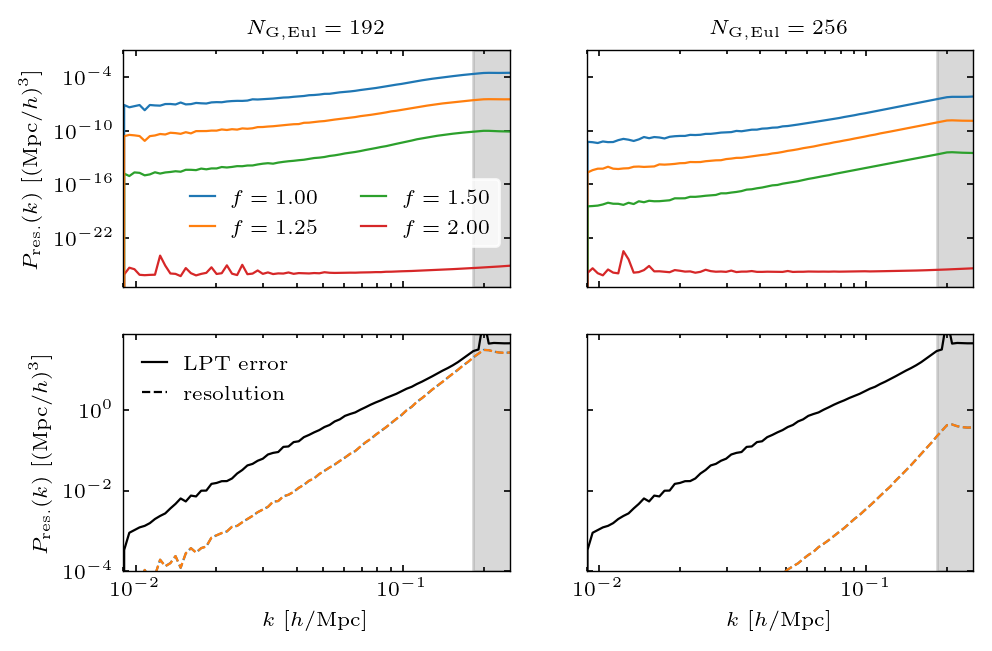}
\caption{Impact of $N_\fwd$ on higher-order Lagrangian bias operators after they have been transformed to the Eulerian frame. We consider $\sigma^{(n)}$ as representative example, $n_\lpt=3$, $\Lambda=0.2\,h/\mpc$ and $z=0$. \textbf{Top:} we compare $\sigma^2$ computed for different grid sizes $N_\fwd = 0.5 \times \left(N_\eul + f\, n_\lpt N_\ini \right)$ to a high-resolution version with $N_\fwd=786$. The latter allows to represent all nonlinearly populated modes, while $f=2$ is required to fulfill the aliasing bound of eq.~(\ref{eq:ngfwd-avoid-aliasing-sigma2}). \textbf{Bottom:} we compare $\sigma^2$ to a high-resolution reference with $N_\fwd=N_\eul=786$. Results are shown for $f=1.0$ and $f=1.25$, however, both are indistinguishable. For comparison, the black line gives the LPT error as found in figure~\ref{fig:restframe-accuracy_lpt_lpterror}. Resolution effects in the higher-order bias operators are well below the perturbative error and the use of $N_\fwd$ as required by the gravitational solution ($f=1$) does not introduce significant residuals for the Lagrangian bias operators.}
\label{fig:restframe-accuracy_bias_lagrangian-full}
\end{figure}

For the tests above, we choose $N_\fwd$ large enough so all modes in $\sigma^2$ could be represented. Smaller grid sizes will inevitably introduce some aliasing during the computation of the nLPT solution. However, if this effect is small relative to the overall model accuracy, choosing smaller $N_\fwd$ might allow to further optimize the computational time and memory, requirements. In the case of $\sigma^2$, the generalized Orszag rule to avoid aliasing into modes below $k_\nyquist\left(N_\eul\right)$ becomes
\begin{equation}
N_\fwd \geq \frac{N_\eul}{2} + n_\lpt \times N_\ini\,.
\label{eq:ngfwd-avoid-aliasing-sigma2}
\end{equation}
where the second term has picked up an additional factor two with respect to the pure matter case. In figure~\ref{fig:restframe-accuracy_bias_lagrangian-full} we test the effect of lowering $N_\fwd$ by comparing $\sigma^2$ to a high-resolution reference with $N_\fwd=2 n_\lpt N_\ini$. If the aliasing rule in eq.~(\ref{eq:ngfwd-avoid-aliasing-sigma2}) is fulfilled, residuals are very small, at the level of the machine precision. But even if $N_\fwd$ is chosen according to the less-demanding matter criterion in eq.~(\ref{eq:restframe-accuracy_resolution_ngfwd-rule}) ($f=1$ in figure~\ref{fig:restframe-accuracy_bias_lagrangian-full}), the numerical accuracy effects are small in comparison to the overall amplitude of the operator, and below the level of the LPT error. For Lagrangian forward models, we therefore use these smaller values for $N_\fwd$. In figure~\ref{fig:restframe-results__lagrangianbias__var-ngfwd}, we verify for one scenario explicitly that this choice does not affect the accuracy at which cosmological parameters are inferred. 

We have focused on the $\sigma^2$ operator here, which contains contributions up to order $2n_\lpt$. Typically, the order of the bias expansion is chosen smaller, $o_\bias = n_\lpt+1$. Hence operators containing powers of $\sigma$ are the most susceptible to resolution effects. Of all further bias operators, only the highest-order ones are subject to aliasing if we use the minimal requirement $n_\lpt = o_\bias-1$ and choose $N_\fwd$ by the aliasing criterion for matter in eq.~(\ref{eq:restframe-accuracy_resolution_ngfwd-rule}).

\subsection{Eulerian bias}
\label{sec:restframe-accuracy__eulerian-bias}

Performing the bias expansion in Eulerian frame, only a single field has to be displaced, which is computationally efficient as the density assignment is usually the costliest step of the forward model. But the study of numerical effects is complicated by the fact that the evolved density field has support to arbitrarily high wavenumbers. We therefore explore effects in the Eulerian-bias forward model numerically, focusing on $\delta^2$.

\begin{figure}
\centering
\includegraphics[width=\figuresize]{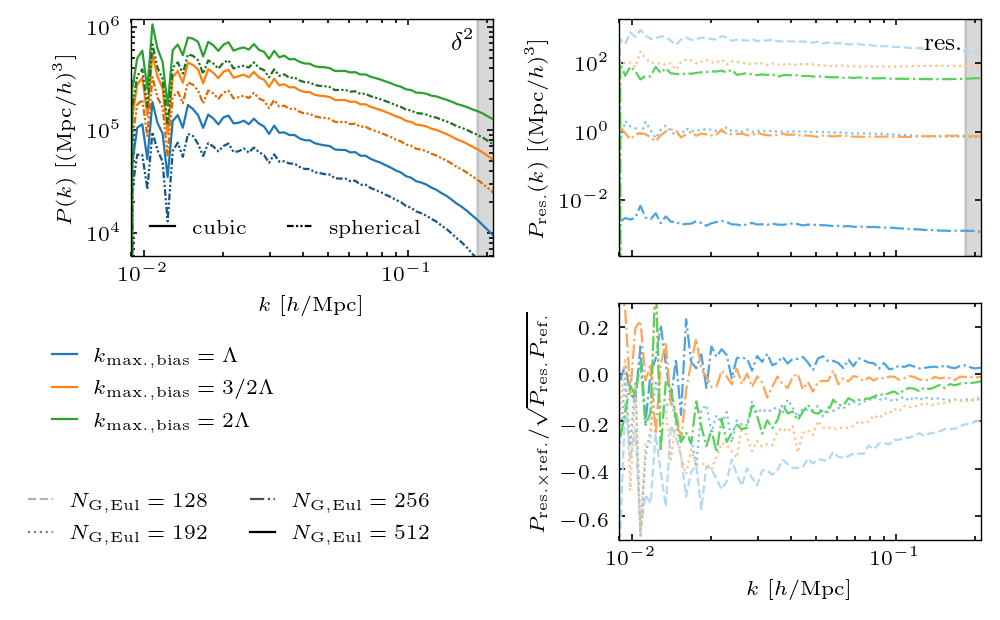}
\caption{Numerical accuracy effects for the Eulerian bias operator $\delta^2$. \textbf{Left:} power spectrum for different choices of the filter $\filterbias$ which is applied to the evolved density field before constructing the bias operators from it. \textbf{Right:} Impact of the numerical resolution in the displacement step on the bias operator; the reference grid was computed for $N_\eul=512$. All results are shown at $z=0$ for a 3LPT forward model and $\Lambda=k_\mathrm{max}=0.2\,h/\mpc$. Nonlinearly populated modes have a higher impact on Eulerian bias operators which makes their numerical convergence more tricky and necessitates filtering the evolved density field before constructing the bias operators. The impact of this filter on the inference of cosmological parameters is investigated in section~\ref{sec:results-eulerian}.}
\label{fig:displacement-accuracy__delta2-eulerian-filter-resolution}
\end{figure}

For higher-order Eulerian bias operators, $N_\eul$ has a significant impact even at low-$k$. Small-scale modes, populated by the nonlinear displacement, couple back into larger ones. The filtering implied by $N_\eul$ controls which modes can contribute to the effect. To avoid uncontrolled aliasing, we always apply a sharp-k filter at  $\filterbias$ to $\delta_\mathrm{fwd}$ before constructing the bias operators. In principle, the effect of such a filter can be studied analytically as well, but this is challenging due to the presence of two filters (one in the initial conditions and one on the final density field). Thus, the results of e.g. \cite{Rubira:2023vzw, Rubira:2024tea} are not directly applicable. In the top left panel of figure \ref{fig:displacement-accuracy__delta2-eulerian-filter-resolution}, we directly investigate the effect of $\filterbias$. To that end we choose a large $N_\eul=512$ and construct the operators on grids which can represent all modes, i.e. $N_\fin = N_{\grid,\nyquist}\left(2 \filterbias \right)$ for the case of second order bias. In fact, if $N_\fin$ is lowered to obey the aliasing constraint eq.~(\ref{eq:restframe-accuracy_resolution_ngfwd-rule}) this has a negligible impact, and we switch to the more efficient constraint for the remainder of this work. It is obvious in figure~\ref{fig:displacement-accuracy__delta2-eulerian-filter-resolution} that the choice of the filter has a sizeable impact on higher-order bias operators at all scales. We expect however that the bulk of this dependence can be absorbed by a corresponding change in the bias coefficients (including higher-derivative coefficients).

Finally, we investigate the impact of the resolution of the displacement step $N_\eul$ on the bias parameters. Since Eulerian operators are significantly impacted by small-scale modes, we can expect them to be more sensitive to $N_\eul$. The right side of figure~\ref{fig:displacement-accuracy__delta2-eulerian-filter-resolution} shows the residuals for difference choices of $N_\eul$ with a high-resolution reference at identical filtering $\filterbias$. In contrast to the $k^4$-scaling found for Lagrangian bias operators, these residuals are now white. The resolution has a stronger effect for filters that pass higher-k modes. However, for $N_\eul = 3/2\, N_\ini$ and $\filterbias=\Lambda$ the maximum amplitude of the residuals due to $N_\eul$ in $\delta^2$ and the Lagrangian operator $\sigma^2$ is comparable. In section~\ref{sec:results-eulerian}, we test how the different numerical behavior of Eulerian operators affects the inference of cosmological parameters.

\section{Computing time}
\label{sec:timing}

The execution time of the forward model is crucial to facilitate field-level analysis but also SBI over cosmological volumes. The perturbative treatment in \leftfield~allows to focus exclusively on modes that impact the scales of interest and for very efficient predictions.

\begin{figure}
\centering
\includegraphics[width=\figuresize]{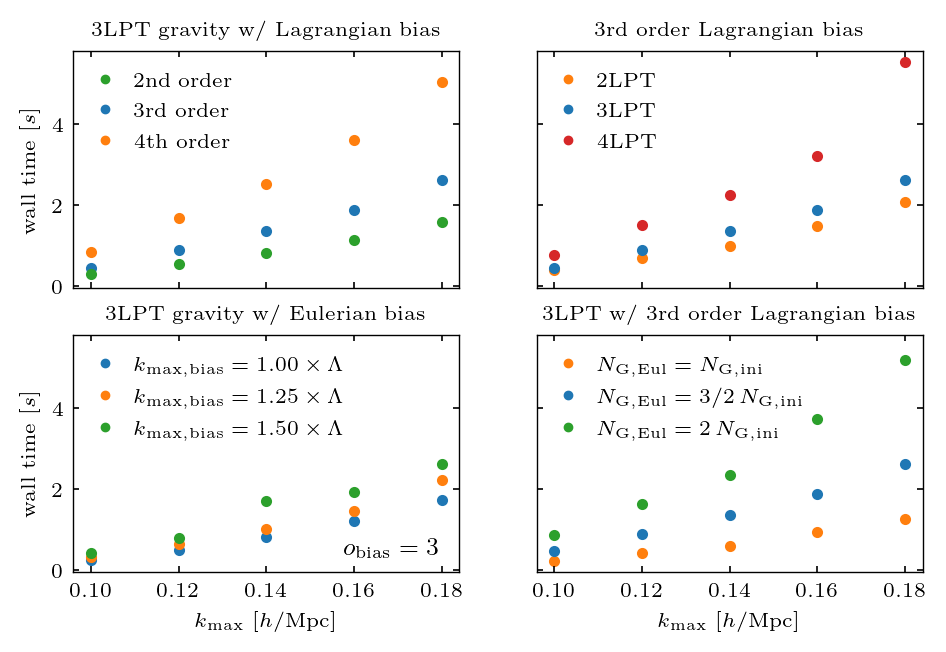}
\caption{Average wall time over ten evaluations of the forward model run on a Intel(R) Xeon(R) Gold 6138 CPU @ 2.00GHz with 20 cores. The box size is $8\,h^{-3}\mathrm{Gpc}^3$, and we compare different cut-off values and forward model choices. Precision settings not explicitly specified are set to the best-practice recommendations.}
\label{fig:runtime__walltime}
\end{figure}

We test the execution time of the \leftfield~forward model on a Intel(R) Xeon(R) Gold 6138 CPU @ 2.00GHz with 20 cores. These CPUs were launched in 2017, and they represent a decent node on a computing cluster but not the best available resources to date. Our results, summarized in figure~\ref{fig:runtime__walltime}, show that it is possible to compute the large-scale galaxy density field in an $8\,h^{-3}\mathrm{Gpc}^3$ box within a fraction of a second to a few seconds.

Not surprisingly, the computing time increases for higher cut-off values $\Lambda$, which require finer grid resolutions at all steps of the forward model. The Eulerian bias model is computationally the most efficient, since it requires the displacement of only a single field from Lagrangian to Eulerian frame. In figure~\ref{fig:runtime__walltime}, we show the runtime of a 3LPT Eulerian bias model; the corresponding 2LPT models have a computational cost that is lower by a factor of $0.65 - 0.85$. For Lagrangian bias models, there is a clear increase of the computational costs with bias order, because each new bias operator requires the displacement of an additional field from Lagrangian to Eulerian frame.

The run time of most forward models is dominated by the displacement step, and thus directly increases for higher $N_\eul$ as the bottom right panel of figure~\ref{fig:runtime__walltime} shows. The situation, however, changes with increasing LPT order. The three models in the top right corner of figure~\ref{fig:runtime__walltime} all have identical $N_\eul$ but differ in $N_\fwd$ according to eq.~(\ref{eq:restframe-accuracy_resolution_ngfwd-rule}). At 3LPT the displacement dominates the overall run-time and takes up about half of it, while the computation of the LPT solution contributes $\sim 20 \%$. Larger grids (by a factor $1.2$) and additional operations required to find the 4th order term reverse the situation; at 4LPT the displacement only contributes $\sim 30\%$ of the total runtime, which is dominated by the LPT solution with $40-50\%$. This indicates that increasing the LPT order is ``cheap'' in terms of runtime until a threshold between 3LPT and 4LPT is reached.

Finally, we investigate the impact of the assignment kernel on the execution time of the forward model. To this end, we focus on the 2LPT model with linear bias only and set $\Lambda=0.20\,h/\mpc$, $z=0.0$. We compare CIC assignments at different resolutions $N_\eul$ to the baseline NUFFT model with $N_\eul = 3/2\, N_\ini = 192$. In section \ref{sec:restframe-accuracy__resolution} we established that the density assignment with the CIC kernel is not converged for grid sizes as large as $N_\eul=384$. However, the wall time in this configuration is already more than four times higher than for the baseline NUFFT scheme; for the CIC kernel with $N_\eul=512$ the factor increases to more than eight. We note however that our CIC implementation could also be optimized further.

\section{Parameter inference at the field level}
\label{sec:results}

In section~\ref{sec:restframe-accuracy}, we studied the residuals of the forward model with respect to high-resolution tests and reference N-body simulations. However, it is not straightforward to relate the residuals to the accuracy at which cosmological parameters can be inferred. To address the latter point, we here present a set of field-level parameter inference tests. As data we consider halos in N-body simulations, which are a non-trivially biased tracer of the fully nonlinear matter density in the simulation. We have two simulation realizations available for which we consider halos in two disjoint mass bins at three redshift snapshots; further details are given in appendix~\ref{sec:halo-sample}.

The analysis of simulated data has the additional benefit that the initial conditions are known, and for our tests we fix them to the ground truth, up to an overall scaling parameter $\alpha$ (see below). This removes uncertainty due to cosmic variance from our parameter estimates to the largest possible extent, and it also reduces the time for an inference considerably, allowing to scan through different model configurations. We focus on the amplitude of matter fluctuations $\sigma_8$, which is parametrized as 
\begin{equation}
\alpha = \frac{\sigma_8}{\sigma_{8,\mathrm{fid}}}\,,
\label{eq:results-alpha-parametrization}
\end{equation}
and is implemented by rescaling the linear density field $\delta^{(1)}$.
At linear order, $\sigma_8$ is completely degenerate with $b_\delta$. Its inference allows to probe directly the extraction of information from features beyond linear order and should be very sensitive to any shortcomings of the forward model in higher-order terms. Of the model parameters, the linear bias $b_\delta$ and the amplitudes $\sigma_{\epsilon,0}$ and $\sigma_{\epsilon,2}$ of the white and $k^2$-dependent noise contributions (see eq.~\eqref{eq:forward-model__noise-parametrization}) are inferred explicitly and marginalized numerically. Bias parameters enter the likelihood (eq.~\eqref{eq:forward-model__likelihood}) quadratically, and it is possible to marginalize over them analytically \cite{Elsner:2019rql}. We choose this option for all higher-order bias parameters, as it has been shown to reduce the correlation length of the sampling chains \cite{Kostic:2022vok}. We depict parameter constraints in terms of mode-centered 68\% credible intervals; details on sampling algorithm, stopping criteria and convergence tests are given in appendix~\ref{sec:sampling}.

Parameter constraints at fixed initial conditions provide a stringent test of the forward model due to the smallness of the error bars. Based on the EFT of LSS, we expect convergence to the ground truth for smaller cut-offs or higher expansion order, and correspondingly our tests scan through a range of values for $k_\mathrm{max}$. In interpreting the results, it is important to note that the two simulations are statistically independent, but there are correlations between model choices and cut-offs. The tests presented are designed to unveil systematic biases introduced by the forward model; but two simulations are insufficient to perform a full coverage test of the error bars. Further, the highest wavenumber up to which an analysis can be applied will in general depend on the tracer population and redshift. Unbiased parameter constraints at fixed initial conditions are an important prerequisite for the full field-level analysis, but they do not automatically guarantee consistent results when the initial conditions are inferred simultaneously. The full field-level analysis of initial conditions and cosmological parameters has been demonstrated recently \cite{beyond2pt, Nguyen:2024yth, Babic:2024wph}. These studies alo present detailed convergence tests of the inferred error bars. Ref. \cite{beyond2pt} in addition shows the posterior consistency across ten independent data sets.

\subsection{Lagrangian bias results}
\label{sec:results-lagrangina-bias}
\begin{figure}
\centering
\includegraphics[width=\figuresize]{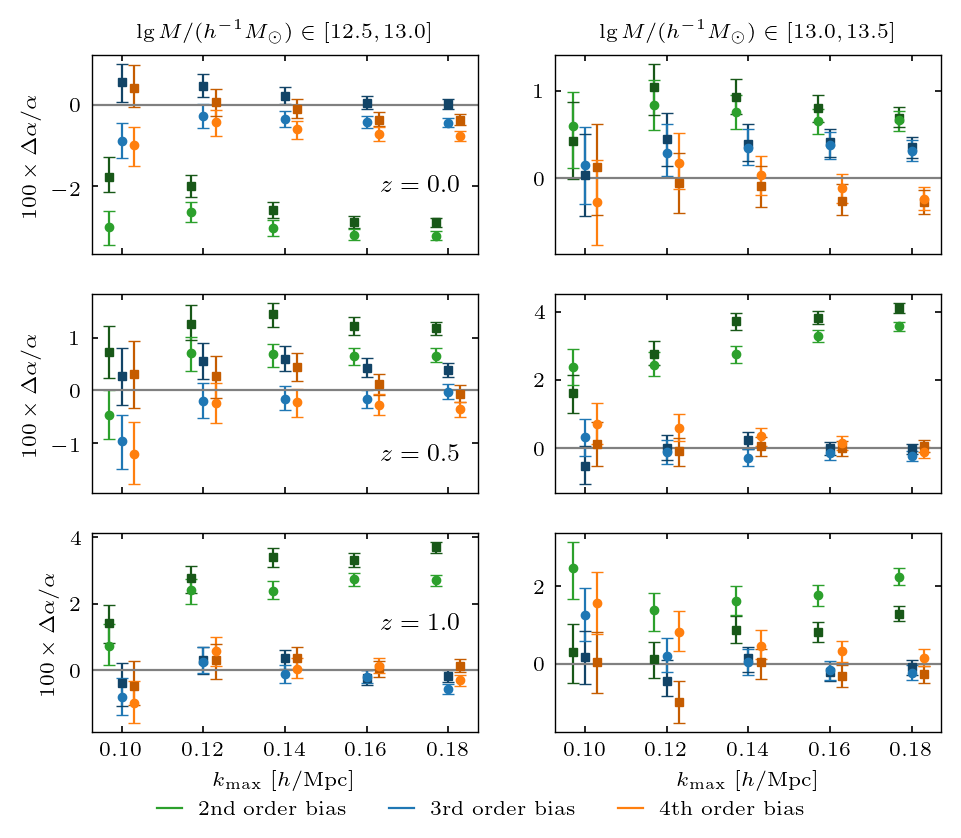}
\caption{Inference of the scaled primordial fluctuation amplitude $\alpha=\sigma_8/\sigma_{8,\mathrm{fid}}$ from N-body halos. Constraints are shown in terms of the mode-centered 68\% credible intervals. We compare results between different orders of the Lagrangian bias expansion for a 3LPT forward model and use $\Lambda=k_\mathrm{max}$. Different symbols indicate two realizations of the initial conditions that are analyzed independently. The convergence of the results with bias order and cut-off scale is precisely as expected for an EFT analysis of galaxy clustering.}
\label{fig:restframe__results-3lpt-lagrangian__var-bias-order}
\end{figure}

The main Lagrangian bias results are summarized in figure \ref{fig:restframe__results-3lpt-lagrangian__var-bias-order}, where we compare results at third order in the LPT solution and different orders in the bias expansion. 
We explore two mass bins, three redshift slices, and two realizations of the simulation for each sample. 
Results with second order bias are offset from the ground truth by a small amount, $\sim 2.5\%$, which nevertheless is significant due to the high accuracy of the measurement. As the bias order increases, the results converge to the ground truth. This is precisely the convergence behavior expected for an EFT analysis.

For the lower mass bin, one can note a discrepancy between the two realizations at the level of $\sim2\sigma$. This discrepancy is very persistent across different analysis choices. Previous works, which investigated parameter inference at fixed phase of the initial conditions with a profile likelihood from the same simulations, also show some discrepancies in the lowest mass bin around $\Lambda\sim 0.10\,h/\mpc$, although at an apparently lower level of statistical significance (see e.g. figure~2 of \cite{Schmidt:2020tao}). One possible explanation for the difference between the two simulations is simply sample variance. Prior effects from the marginalization over nuisance parameters might play a role, and would not be present in a profile likelihood analysis such as that of \cite{Schmidt:2020tao}. On the other hand, this effect would be expected to affect both mass bins.

\begin{figure}
\centering
\includegraphics[width=\figuresize]{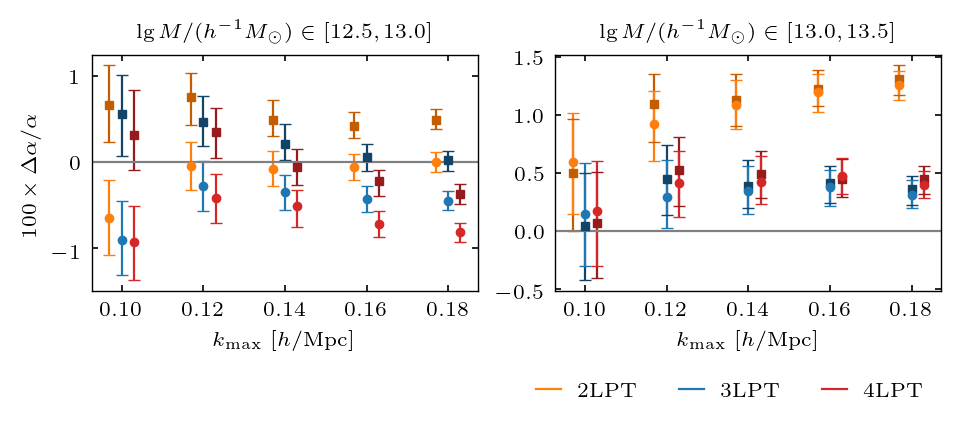}
\caption{Same as figure~\ref{fig:restframe__results-3lpt-lagrangian__var-bias-order} at $z=0$, but using a third order Lagrangian bias expansion and different LPT orders for the gravity forward model. There appears to be a small benefit in using a higher-order LPT solution than what is strictly required for the bias expansion, in particular in the higher mass bin. The effect, however, is much less significant than pushing to higher bias orders in figure~\ref{fig:restframe__results-3lpt-lagrangian__var-bias-order}.}
\label{fig:restframe__lagrangianbias_var-lpt-order}
\end{figure}

We repeat the inference now at fixed third order of the bias expansion for different orders of the LPT solution in figure~\ref{fig:restframe__lagrangianbias_var-lpt-order}. The tests start at 2LPT, the lowest order from which a complete 3rd order bias expansion can be constructed. In the lower mass bin, using 3LPT or 4LPT lowers the inferred value of $\alpha$ by $\lesssim 0.2\%$, which is much smaller than the difference between the two realizations. The difference between 2LPT and the higher order becomes significant (in the sense of exceeding one standard deviation) only for modes $k\geq 0.16\,h/\mpc$. In the higher mass bin, deviations from the ground truth are highest for 2LPT. Going to higher LPT orders significantly shifts the results and makes them more consistent with the ground truth. The difference between 2LPT and 3LPT is significant already at small cut-offs, $k \geq 0.12\,h/\mpc$, while between 3LPT and 4LPT it remains below  $0.5\,\sigma$ up to the highest $k_\mathrm{max}$. This seems to indicate a slight benefit of extending the order of the perturbative solution beyond what is strictly needed for the bias expansion. The greater benefit, however, comes from extending the bias expansion to higher orders. As figure~\ref{fig:restframe__results-3lpt-lagrangian__var-bias-order} shows, at 3LPT and fourth order in the bias expansion, the results in the second mass bin are consistent with the ground truth up to the highest wavenumbers tested.

We present a set of additional test in appendix~\ref{sec:additional-tests__lagrangian-bias-inference}. The results in brief are:
\begin{itemize}
\item We test the effect of $N_\eul$ in figure~\ref{fig:restframe__lagrangianbias_var-ngeul}. For small values, $N_\eul=N_\ini$ the inferred clustering amplitude is significantly offset from the ground truth and we cannot recover the expected convergence towards small $k_\mathrm{max}$. Increasing $N_\eul$ beyond $3/2\, N_\ini$, on the other hand, has a much weaker effect and does not improve the convergence. We find that the difference between $3/2\, N_\ini$ and $2 N_\ini$ in the lower mass bin is at a significance $\leq 0.5\sigma$ for $k \leq 0.14\,h/\mpc$ and below the scatter between the two realizations for all cut-offs; in the higher mass bin, it is always $\leq 0.2\sigma$. Thus the chosen resolution of the displacement step indeed sits at the sweet spot between numerical accuracy and
computational efficiency.
\item We investigate the effect of increasing $N_\fwd$ in figure~\ref{fig:restframe-results__lagrangianbias__var-ngfwd}. We concluded in section~\ref{sec:restframe-accuracy__lagrangian-bias} that choosing $N_\fwd$ by the aliasing constraint of the gravitational solution was sufficient, even though there might be some residual effects on the highest-order bias operators. Indeed, we find that effect of increasing $N_\fwd$ is below $0.1\sigma$ significance in both mass bins and for all cut-offs. Therefore, our results indicate that this potential aliasing in the bias parameters does not affect the analysis of $\sigma_8$ in any discernible manner.
\item We test the effect of allowing for additional, higher-order in derivatives bias operators (figure~\ref{fig:sampling-results__lagrangianbias__highderiv}) and of choosing $\Lambda > k_\mathrm{max}$ (figure~\ref{fig:sampling-results__lagrangianbias__highlambda}). None of these impact the convergence behavior, and the shift in the inferred parameters always is small (less than $0.4\,\sigma$ for all cases explored). We conclude that back-reactions from small-scale modes are sufficiently absorbed by the leading derivative operator $\nabla^2\delta$, in case of the Lagrangian bias expansion and inference at fixed initial conditions.
\item We test replacing $\sigma = \sum_n \sigma^{(n)}$ by $\sigma^{(1)}$ in the construction of the Lagrangian bias operators in figure~\ref{fig:crosscheck-sigma-constraints}. Both versions are equivalent up to terms of order higher than $o_\bias$. Indeed, we only see a small shift in the higher mass bin for large values of $\Lambda$, where we would expect higher-order bias operators to have the strongest effect. More concretely, the shift exceeds $0.5\,\sigma$ significance in the higher mass bin for $k \geq 0.14\,h/\mpc$, where results with $\sigma^{(1)}$ are more consistent with the ground truth. There also is a small shift in the inferred value of $b_\delta$, pointing to a degeneracy between linear and higher-order bias parameters, but the qualitative correlation structure between the inferred parameters remains the same (see figure~\ref{fig:crosscheck-sigma-contours}).
\end{itemize}

\subsection{Eulerian bias results}
\label{sec:results-eulerian}

\begin{figure}
\centering
\includegraphics[width=\figuresize]{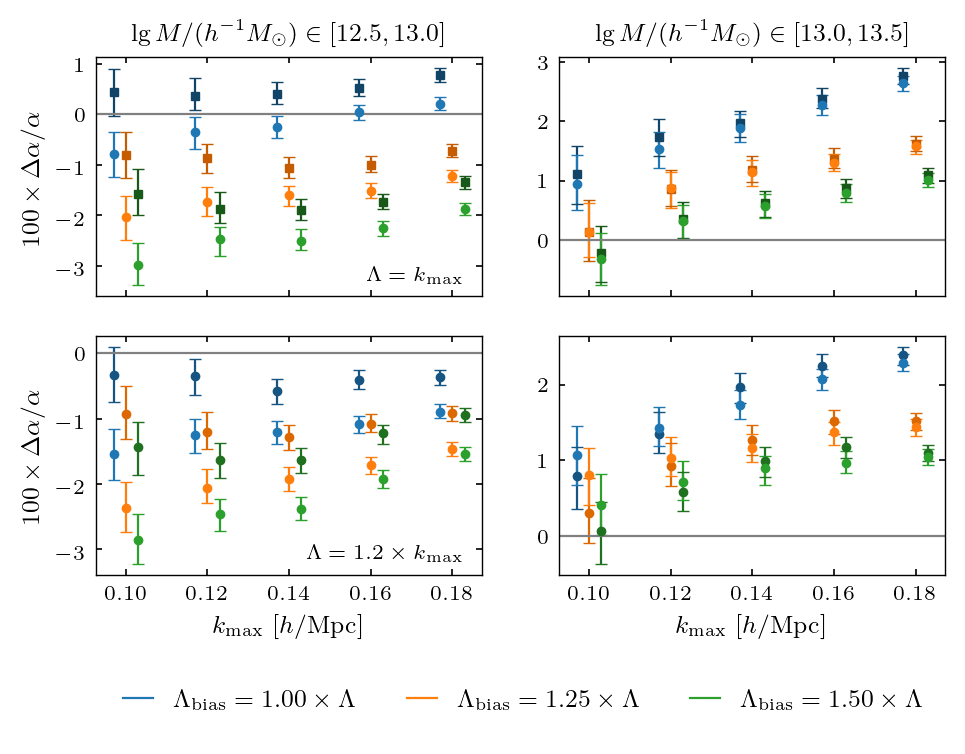}
\caption{Inference of the scaled primordial fluctuation amplitude $\alpha=\sigma_8/\sigma_{8,\mathrm{fid}}$ from N-body halos at $z=0$ using a third order Eulerian bias expansion and the 2LPT gravity model. We compare different spherical filters, applied to the evolved density before the construction of the bias operators. In the upper panels, we set $\Lambda=k_\mathrm{max}$, i.e. the cut-off in the initial conditions agrees with the largest wavenumber considered in the likelihood, while the lower panels explore increasing $\Lambda$ to $\Lambda = 1.2 \, k_\mathrm{max}$. Choosing $k_{\rm max} < \Lambda$ reduces the dependency on $\filterbias$.}
\label{fig:restframe-results__eulerianbias__2LPT-varLambda-varKmaxBias}
\end{figure}

Our baseline results for the scaled primordial fluctuation amplitude $\alpha=\sigma_8/\sigma_{8,\mathrm{fid}}$ with third order Eulerian bias operators and a 2LPT gravity model are summarized in figure~\ref{fig:restframe-results__eulerianbias__2LPT-varLambda-varKmaxBias}. We focus on two mass bins and two realizations of the simulation in the $z=0$ snapshot. In the upper panel of figure~\ref{fig:restframe-results__eulerianbias__2LPT-varLambda-varKmaxBias}, we test different values of $\filterbias$, i.e. the filter which is applied to the evolved density field before constructing the bias parameters. The inferred parameter values are shifted with respect to the ground truth at the level of $1-2 \%$. A systematic trend of lower inferred $\alpha$ with increasing $\filterbias$ is evident, with no clear evidence of convergence. In the lower panel of figure~\ref{fig:restframe-results__eulerianbias__2LPT-varLambda-varKmaxBias}, we test if choosing $\Lambda>k_{\max}$ can compensate for additional higher-derivative terms that would be necessitated by the bias running. The sensitivity of the inferred value of $\alpha$ to variations in $\filterbias$ tends to be smaller when $k_{\rm max} < \Lambda$.

\begin{figure}
\centering
\includegraphics[width=\figuresize]{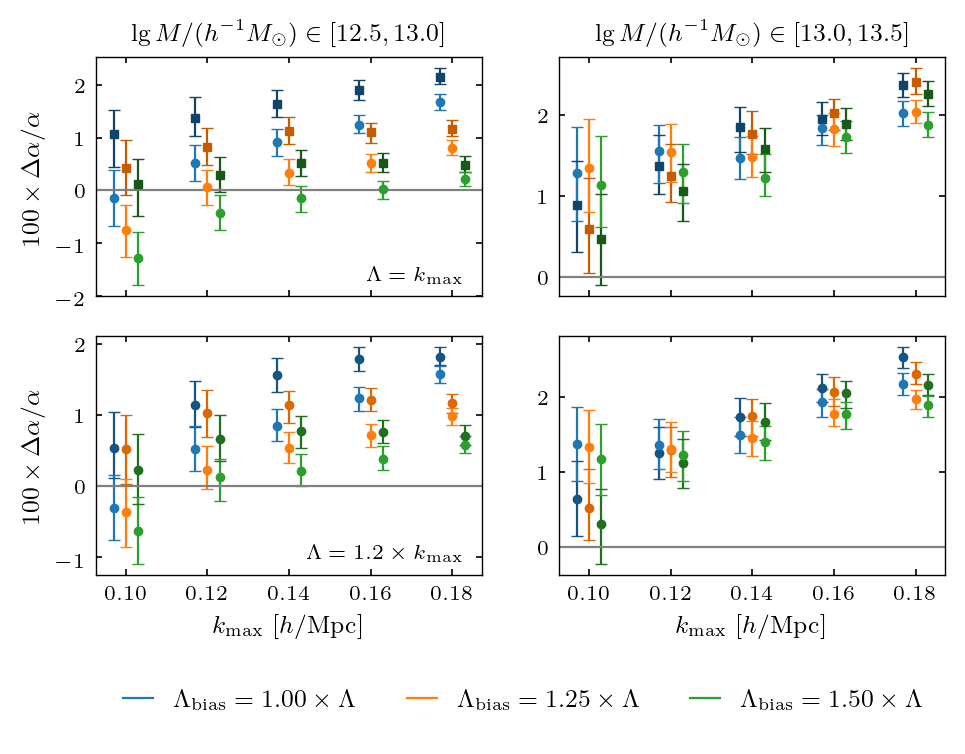}
\caption{Same as figure~\ref{fig:restframe-results__eulerianbias__2LPT-varLambda-varKmaxBias}, but for N-body halos at $z=0.5$. The consistency with the ground truth improves at higher redshifts but convergence in $\filterbias$ remains unclear.}
\label{fig:restframe-results__eulerianbias__2LPT-varLambda-varKmaxBias_z050}
\end{figure}

We repeat the baseline analysis for the halo snapshots at $z=0.5$, shown in figure~\ref{fig:restframe-results__eulerianbias__2LPT-varLambda-varKmaxBias_z050}. While the same qualitative trends as for $z=0$ are found, the effect of varying $\filterbias$ and $k_{\rm max}/\Lambda$ is evidently weaker at higher redshifts, especially for the higher mass bin. Again, choosing $k_{\rm max} < \Lambda$ somewhat reduces the dependence of the results on $\filterbias$.

We present additional test in appendix~\ref{sec:additional-tests__eulerian-bias-inference} and find the following results:
\begin{itemize}
\item We repeat the analysis with the 3LPT forward model, still considering third order Eulerian bias in figure~\ref{fig:restframe-results__eulerianbias__3LPT-varLambda-varKmaxBias}. For the Lagrangian bias expansion, we could see an improved convergence as we push to higher orders in the LPT expansion. For Eulerian bias this is not the case. Although the parameter shifts in the second mass bin reduce slightly, they still are at a high level of significance given the small error bar.
\item We test the choice of filters in figure~\ref{fig:restframe-results__eulerianbias__filtertypes}. The analyses presented in this section previously assumed spherical filters in the initial conditions and for the evolved density field, and they did not filter the displacement field before the actual displacement operation. Here, we switch to cubic filters (top row), and investigate additionally filtering the displacement vector at $\filterfwd=\Lambda$ (bottom). The cubic filter in the initial conditions leads to lower inferred $\alpha$  in comparison to the spherical one. In the lower mass bin, the results become less consistent with the ground truth, while in the higher mass bin the consistency improves. Filtering the displacement vector itself, on the other hand, has very little impact.
\item Finally, we investigate if higher-order derivative operators are able to absorb the bias running from filtering the evolved density field before the construction of the Eulerian bias parameters (figure~\ref{fig:restframe-results__eulerianbias__NLOderivatives}). The impact of these operators is always much smaller than that of $\filterbias$ and only becomes significant in the higher mass bin for $\filterbias=\Lambda$. For all other scenarios in figure~\ref{fig:restframe-results__eulerianbias__NLOderivatives}, higher-order derivative bias impacts the results at $\leq 0.5\sigma$ significance.
\end{itemize}

In summary, we find a shift at the level of up to $\sim 3\%$ in the inferred $\sigma_8$ values if we use a forward model with Eulerian bias expansion. This shift is significant given the high precision of the field-level analysis at fixed initial conditions. The dependence of the inferred $\sigma_8$ on the choice of the filter $\filterbias$ applied to $\delta_{\rm fwd}$ suggests that this operation is at least partly responsible. The effect of this filter should be reduced by choosing $k_{\rm max} < \Lambda$, and while we find some evidence for this, it does not eliminate this sensitivity. Alternatively, the filter effects are expected to be absorbed by higher-derivative operators; however, at least the inclusion of the subset of subleading higher-derivative terms considered in figure~\ref{fig:restframe-results__eulerianbias__NLOderivatives} does not seem sufficient either. In the future, it would be interesting to explore this by introducing a significant hierarchy between $k_{\rm max}$ and $\filterbias$, and/or by including a complete set of higher-derivative bias terms.

\section{Conclusions}
\label{sec:conclusions}

We have presented the Lagrangian perturbative forward model \leftfield\ for galaxy clustering (see figure~\ref{fig:forward-model}), and discussed in detail the physical (perturbative) and numerical sources of error in the model, as well as their mitigation. The perturbative description starts from initial conditions filtered at the cut-off scale $\Lambda$, allowing to identify and focus only on those modes which will impact the final cosmological results, which makes the forward model very efficient. We use the perturbativity of the forward model to identify the leading-order contributions to the numerical error from finite-resolution effects, and we derive a set of best-practice recommendations for those parameters controlling the numerical precision (see figure~\ref{fig:forward-model}). These analytical estimates are confirmed by a wide range of numerical tests in which we compare the forward model to high-resolution references and to N-body simulations that have an identical cut-off in the initial conditions.

The \leftfield~forward model is very accurate in predicting the gravitational evolution of matter; at LPT order $n_\lpt \geq 3$, $z=0.5$ and for a cut-off $\Lambda=0.20\,h/\mpc$ its accuracy is at the level of $2\%$ in the overall matter clustering amplitude, and it further improves for smaller cut-offs to sub-percent level accuracy at $\Lambda=0.10\,h/\mpc$ (see figures~\ref{fig:restframe-accuracy_lpt_lpterror} and~\ref{fig:restframe-accuracy_lpt_lpterror_l010}, respectively). \leftfield~allows for two different bias descriptions with operators either constructed in Lagrangian or Eulerian frame. While numerically more efficient, the latter is more challenging to keep under control; as the nonlinear bias operators are obtained from the evolved density the back coupling of higher-wavenumber modes can become significant and is hard to control numerically. This necessitates the introduction of a filter $\filterbias$, which is applied to the evolved density. After introduction of this filter, and in general for the Lagrangian bias expansion, we show that numerical effects in the bias operators are well below the level of the perturbative accuracy for LPT orders ($n_\lpt \leq 4$) that we considered here. One evaluation of the forward model takes of order seconds, and therefore is sufficiently fast to enable field-level analysis and SBI.

We test the forward model by inferring cosmological parameters, and in particular the fluctuation amplitude $\sigma_8$ from halos in N-body simulations. The halos provide a non-trivial tracer of the fully nonlinear density field and $\sigma_8$ is particularly interesting due to its linear-order degeneracy with the bias parameter $b_\delta$; it allows to directly investigate the information extracted from nonlinear scales. We fix the initial conditions {(up to an overall scaling parameter) to the known ground truth. On the one hand, this removes a considerable amount of uncertainty from the final parameter constraints and thus enables a very stringent test of the forward model. On the other hand, it makes the inference fast enough so we can scan through a broad range of analysis configurations and cut-off scales.

In the case of Lagrangian bias, we precisely recover the convergence behavior expected from the EFT of LSS, where parameter shifts with respect to the ground truth shrink both for smaller cut-off scales and higher bias expansions. As figure~\ref{fig:restframe__results-3lpt-lagrangian__var-bias-order} demonstrates, we recover $\sigma_8$ consistent with the ground truth for both mass bins considered in three redshift snapshots, $z=0.0,\,0.5,\,1.0$ up to the highest cut-off tested at $\Lambda = k_\mathrm{max} = 0.18\,h/\mpc$.

Results with the Eulerian bias expansion, on the other hand, exhibit a small shift at the level of a few percent in the inferred parameter values away from the ground truth. We could not establish convergence for the filter scale $\filterbias$; some results show a better convergence for $\filterbias=\Lambda$, others for $\filterbias > \Lambda$. There are some indications that choosing $k_{\rm max} < \Lambda$ reduces the unwanted dependence on this filter scale, but it does not remove it. This should be studied more systematically in the future, in particular by also including the full set of subleading higher-derivative operators.

With the optimal numerical choices for the Lagrangian bias model established in this paper, the only user-defined parameters in \leftfield\ are the order of the LPT solution, the bias order and the cut-off scale $\Lambda$, i.e. precisely those parameters that are varied in an EFT analysis to study the convergence of the results. We will extend the study to the modeling of redshift space distortions in \leftfield~\cite{Stadler:2023hea} in a future publication \cite{Stadler:2024}. A similar convergence study as presented here should also be performed including joint inference of the initial conditions \cite{Kostic:2022vok, Nguyen:2024yth,Babic:2024wph}. However, such a study also needs to consider a more realistic model of the stochastic contributions. Here we focused on contributions in Fourier space and included a flat and a sub-leading scale-dependent noise term, but we neglected higher-order effects that are are more conveniently described in real-space such as density-dependent noise. The explicit modeling of the noise field, which is the subject of ongoing work, will allow to combine different sources of stochasticity at higher orders, and it provides a real-space formulation of the likelihood which makes the inclusion of survey effects such as masking more straightforward (for a real-space formulation of the likelihood which neglects scale-dependent noise, see \cite{Schmidt:2020tao}). In the meantime, the results presented here should already make \leftfield~ready for SBI applications \cite{Tucci:2023bag}.

\acknowledgments{
We thank Ivana Babi\'c, Safak Celik, Ivana Nikolac, Minh Nguyen, Moritz Singhartinger and Beatriz Tucci for useful discussion and Cornelius Rampf for insightful comments on the manuscript. This work  was funded by the Deutsche Forschungsgemeinschaft (DFG, German Research Foundation) under Germany´s Excellence Strategy – EXC 2094 – 390783311”.}

\appendix

\section{List of bias parameters}
\label{sec:analysis-settings}

All Lagrangian bias expansions considered in this work contain the leading-order bias operator $\delta$ and the leading higher-derivative operator $\nabla^2\delta$. In addition, at increasing order the following operators are added:
\begin{itemize}
\item 2nd order: $\sigma^2$, $\mathrm{tr}\left[ M^{(1)}\,M^{(1)}\right]$ \,,
\item 3rd order: $\sigma^3$, $\sigma\,\mathrm{tr}\left[M^{(1)}\,M^{(1)}\right]$, $\mathrm{tr}\left[M^{(1)}\,M^{(1)}\,M^{(1)}\right]$,  $\mathrm{tr}\left[M^{(1)}\,M^{(2)}\right]$ \,,
\item 4th order: $\sigma^4$, $\sigma^2\,\mathrm{tr}\left[M^{(1)}\,M^{(1)}\right]$, $\sigma\,\mathrm{tr}\left[M^{(1)}\,M^{(1)}\,M^{(1)}\right]$, $\sigma\,\mathrm{tr}\left[M^{(1)}\,M^{(2)}\right]$, $\mathrm{tr}\left[M^{(1)}\,M^{(1)}\,M^{(2)}\right]$,\newline\phantom{4th order:} $\left(\mathrm{tr}\left[ M^{(1)}\,M^{(1)}\right]\right)^2$,  $\mathrm{tr}\left[M^{(1)}\,M^{(3)}\right]$,  $\mathrm{tr}\left[M^{(2)}\,M^{(2)}\right]$\,.
\end{itemize}
where $M^{(n)}$ is the symmetric component Lagrangian distortion tensor (eq.~\eqref{eq:forward-model__Lagrangian-distortion-tensor}) at nth order and $\sigma = \sum_{n} \sigma^{(n)}$ the divergence of the displacement vector (eq.~\eqref{sec:forward-model__shift-decomposition}), see also section~\ref{sec:forward-model-physics}. We explore the effect of replacing $\sigma$ by $\sigma^{(1)}$ in figures~\ref{fig:crosscheck-sigma-constraints} and \ref{fig:crosscheck-sigma-contours}.

In the Eulerian bias expansion, we equally have the leading-order operators $\delta$ and $\nabla^2\delta$. At increasing order, the following operators are added
\begin{itemize}
\item 2nd order: $\delta^2$, $K^2$ \,,
\item 3rd order: $\delta^3$, $K^3$, $\delta K^2$, $O_\mathrm{td}$ \,.
\end{itemize}
The definition of these operators and their relation to other bias conventions is discussed in appendix C of \cite{Desjacques:2016bnm}.

\section{Resizing of arrays in Fourier space}
\label{sec:fourier-resize}

On several occasions, the \leftfield\ forward model requires the increasing or decreasing of the resolution of a discretized field, which is performed in the usual fashion by zero-padding/truncating the Fourier representation of the field.
However, since the fields in question are real-valued, special care must be taken to preserve this property during the resizing operation.

Without loss of generality we can consider a one-dimensional case with a small grid of size $n$ and a large grid of size $N$. 
The operation is straightforward when $n$ is odd. In this case, the highest frequency component of the small field is stored in two distinct array locations $(n-1)/2$ and $(n+1)/2$ (using the most common indexing convention for Fourier-space data). Since $N>n$, the larger array always has two corresponding distinct entries, and the values are simply copied over, both in the downsampling and upsampling direction.

When $n$ is even, there is by definition only a single array entry for the component at the Nyquist frequency in the smaller array, located at index $n/2$, while there are two distinct corresponding entries in the large array, at indices $n/2$ and $N-n/2$.
During upsampling the Nyquist value from the small array must therefore be split, by entering half its value into the two locations in the larger array:
\begin{equation}
T_{n/2} = T_{(N-n)/2} = \frac{S_{n/2}}{2}\text{,}
\end{equation}
with $S$ and $T$ being the source and target arrays, respectively.
Correspondingly, when downsampling, the two values from the large array must be added together and written into the single Nyquist entry of the small array:
\begin{equation}
T_{n/2} = S_{n/2} + S_{(N-n)/2}\text{.}
\end{equation}

As described, it is evident that, other than in the odd-$n$ case, the up- and downsampling operations for the even-$n$ case are not adjoint to each other.
Therefore special care must be taken in situations where the actual adjoint (in contrast to the ``pseudo-inverse'' operation described above) is needed.
In particular this is the case when back-propagating the gradient through a resize operation.
In such a case, the Nyquist value is simply copied to both target locations (i.e.\ without the factor 0.5) during upsampling:
\begin{equation}
T_{n/2} = T_{(N-n)/2} = S_{n/2}\text{,}
\end{equation}
and averaging is performed during downsampling:
\begin{equation}
T_{n/2} = \frac{1}{2}(S_{n/2} + S_{(N-n)/2})\text{.}
\end{equation}

The described 1D scenario can be extended analogously to multidimensional arrays; for the three-dimensional arrays in \leftfield\ we perform the above operations for all elements of the $yz$, $xz$, and $xy$ Nyquist planes successively. The edges (vertices) of the Nyquist cube correspondingly involve splitting/averaging over four (eight) modes.

In addition to the specific treatment of entries at the Nyquist frequencies, the resampling of arrays may also include a rescaling of the Fourier coefficients, depending on the desired convention. In the case of \leftfield\ we perform a rescaling that ensures that, for a band-limited field, statistics such as the variance and the power spectrum computed on the resized grid remain the same. In our discrete Fourier convention, this implies a rescaling factor of $(n_\text{tgt}/n_\text{src})^3$, where $n_\text{src}$ and  $n_\text{tgt}$ are the numbers of grid cells per dimension of the source and target arrays.

\section{Filter convolution}
\label{sec:filters}
The convolution of two top-hat filters contributes significantly to the $k$-scaling of numerical resolution effects, see section \ref{sec:restframe-accuracy__resolution} and \ref{sec:restframe-accuracy__lagrangian-bias}. We here compute these convolutions explicitly.

\subsection{One-dimensional top-hat}
The one-dimensional top-hat filter is defined as
\begin{equation}
W_\Lambda(k) = 
\begin{cases}
1\,,& \text{if~} |k| < \Lambda \\
1/2\,,& \text{if~} |k| = \Lambda \\
0\,, & \text{else} \,.
\end{cases}
\end{equation}
The convolution of two top-hats with each other gives
\begin{equation}
\int dk'~ W_{\Lambda}\left(k'\right)  W_{\Lambda}\left(k-k'\right) =
\begin{cases}
2\Lambda - |k| &\text{if~} |k| \leq 2\Lambda \\
0 &\text{else}
\end{cases}\,.
\end{equation}
For the convolution of a high-pass and a low-pass filter we obtain
\begin{align}
\int dk'~ W_{\Lambda}\left(k'\right) \left[1- W_{\Lambda}\left(k-k'\right)\right] = |k| \quad \mathrm{if} \quad |k| \leq 2\Lambda\,.
\end{align}

\subsection{Three-dimensional cubic filter}
We define the three-dimensional cubic top-hat as
\begin{equation}
W^\mathrm{C}_\Lambda(\vec{k}) = \prod_{i=1}^3 W_\Lambda(k_i)\,,
\end{equation}
The integral in the convolution of two filters factorizes,
\begin{align}
\int d^3\vec{k'}~ W^\mathrm{C}_\Lambda(\vec{k'}) W^\mathrm{C}_\Lambda(\vec{k}-\vec{k'})
&= \prod_{i=1}^3 \int dk'_i~ W_\Lambda(k'_i)\,W_\Lambda(k_i) W_\Lambda(k_i-k'_i)
\nn \\
&= \prod_{i=1}^3 \left( 2\Lambda - |k_i| \right)\,.
\end{align}
Therefore, the convolution of a cubic low-pass and a high-pass filter gives
\begin{equation}
\int d^3\vec{k'}~ W^\mathrm{C}_\Lambda(\vec{k}-\vec{k'}) \left[1 - W^\mathrm{C}_\Lambda(\vec{k'})\right]
= \left(2\Lambda\right)^2 \sum_{i=1}^3\,|k_i| - 4\Lambda \left(|k_1 k_2| + |k_1 k_3| + |k_2 k_3|\right) + \prod_{i=1}^3 |k_i|
\end{equation}
We can cast the fields we are interested in in the form
\begin{equation}
f\left(\vec{k}\right) = \int d^3\vec{k'} ~ W^\mathrm{C}_\Lambda(\vec{k}-\vec{k'}) \left[1 - W^\mathrm{C}_\Lambda(\vec{k'})\right] g\left(\vec{k'}\right)\,,
\end{equation}
and for the cross-correlation with a field $h$,
\begin{align}
\left\langle f\left(\vec{k}_1\right) h\left(\vec{k}_2\right) \right\rangle = \int d^3\vec{k'} ~ W^\mathrm{C}_\Lambda(\vec{k}-\vec{k'}) \left[1 - W^\mathrm{C}_\Lambda(\vec{k'})\right] \left\langle g\left(\vec{k'}\right) h\left(\vec{k}_2\right) \right\rangle\,.
\end{align}
If $\langle g^2 \rangle$ and $\langle gh \rangle$ are sufficiently smooth, the leading $k$-scaling is dominated by the filter convolution. In the power spectrum, we average over all angular directions. Since
\begin{equation}
\left(4\pi\right)^{-1}\int d\Omega \sum_i |k_i| = k\, \frac{\pi}{2}\left(\sqrt{2} + 1\right)\\
\end{equation}
we get for the leading $k$-scaling
\begin{equation}
P_{f^2} \propto k^2 
\quad \mathrm{and} \quad
P_{fh} \propto k \,.
\end{equation}

\subsection{Three-dimensional spherical filter}
We define the three-dimensional spherical top-hat filter as
\begin{equation}
W^\mathrm{S}_\Lambda(\vec{k}) = W_\Lambda(|\vec{k}|)\,.
\end{equation}
And perform the convolution in spherical coordinates, writing
\begin{equation}
\left|\vec{k}-\vec{k'}\right| = k^2 + k'^2 + 2 \vec{k}\cdot \vec{k'} =  k^2 + k'^2 + 2k k' \cos \theta\,.
\end{equation}
In the second line, we have rotated the integration variables $\vec{k'}$ such that their z-axis aligns with $\vec{k}$ and subsequently identified the angle between $\vec{k}$ and $\vec{k'}$ with the $\theta$ coordinate. 
\begin{align}
&\int d^3\vec{k'}~ W^\mathrm{S}_\Lambda(\vec{k'}) W^\mathrm{S}_\Lambda(\vec{k}-\vec{k'}) = \nn\\
&= \int_{0}^{\infty} dk' k'^2~ \int_{-1}^{1} dx~ W_\Lambda(k')\, W_{\Lambda^2}(k^2+k'^2-2kk'x) \int_{0}^{2\pi} d\varphi = \nn \\
&= 2\pi \int_{-1}^{1} dx \int_{0}^{\min(\Lambda, kx+\sqrt{k^2(x^2-1)+\Lambda^2})} dk'~ k'^2 = \nn\\
&= \frac{2\pi}{3} \int_{-1}^{k/2\Lambda} dx \left[kx+\sqrt{k^2(x^2-1)+\Lambda^2}\right]^3 + \frac{2\pi}{3} \int_{k/2\Lambda}^{1} \Lambda^3 = \nn \\
&= \frac{2\pi}{3} \frac{1}{8k} \left[\left(\sqrt{k^2\left(x^2-1\right)+\Lambda^2} + kx\right)^2 \left(2kx \sqrt{k^2\left(x^2-1\right)+\Lambda^2} + k^2\left(2 x^2 - 3\right) + 3\Lambda^2 \right)\right]_{-1} ^ {k/2\Lambda} \nn \\
&+ \frac{2\pi}{3} \left[1 - \frac{k}{2\Lambda}\right] 
=\frac{2\pi}{3} \left[2\Lambda^3 - \frac{3}{2}k\Lambda^2 + \frac{1}{8} k^3\right]\,.
\end{align}
In the second line we used $x=\cos\theta$ and $W_\Lambda(x)=W_{\Lambda^2}(x^2)$. In the third line, the first filter limits $k'\leq\Lambda$ and the second $k' \leq kx + \sqrt{k^2(x^2-1)+\Lambda^2}$; the transition determining which of the limits is more constraining occurs at $x=k/2\Lambda$, which we use to split the x-integral in the third line. For the convolution of a low-pass and a high-pass filter, we then obtain
\begin{equation}
\int d^3\vec{k'}~ W^\mathrm{S}_\Lambda(\vec{k'}) \left[1 - W^\mathrm{S}_\Lambda(\vec{k}-\vec{k'}) \right] = \pi\Lambda^2 k - \frac{\pi}{12} k^3\,.
\end{equation}

\section{Analytic prediction of the residuals}

\subsection{Truncation errors}
\label{sec:analytic-residuals_truncation}
We here calculate the cross-correlation of the leading-order truncation error (eq.~\eqref{eq:restframe-accuracy_resolution_truncation-error}) with the full solution under the assumption that $k_\nyquist\left(N_\eul\right)=\Lambda$, and for simplicity of notation we evaluate all results at $z=0$; the redshift dependence can be very easily re-instated. The truncation error arises at third order
\begin{align}
\Delta\delta_\mathrm{trunc.}^\LO\left(\vec{k}\right) &= - k^l k^m \int \frac{d^3\vec{p}}{(2\pi)^3}~ 
W_\Lambda\left(\vec{k}-\vec{p}\right)\, \left[1 - W_\Lambda\left(\vec{p}\right)\right]~
s^{(1)}_l\left(\vec{k}-\vec{p}\right) s^{(2)}_m\left(\vec{p}\right) \nn\\
&= - \frac{k^l k^m}{2} \int \frac{d^3\vec{p}_1}{(2\pi)^3}\, \frac{d^3\vec{p}_2}{(2\pi)^3}~
W_\Lambda\left(\vec{k}-\vec{p}_1\right)\, \left[1 - W_\Lambda\left(\vec{p}_1\right)\right] 
\nn \\
&\times
 L_l^{(1)}\left(\vec{k}-\vec{p}_1\right) L_m^{(2)}\left(\vec{p}_2, \vec{p}_1-\vec{p}_2 \right)
\delta^{(1)}\left(\vec{k}-\vec{p}_1\right)
\delta^{(1)}\left(\vec{p}_1-\vec{p}_2\right)
\delta^{(1)}\left(\vec{p}_2\right)\,,
\end{align}
where we have written $\vec{s}_{\mathrm{s},i}=\left[1 - W_\Lambda\right]\,s_i$ and $\vec{s}_{\mathrm{l},i} = W_\Lambda\,s_i$. The LPT kernels $L^{(n)}$ are
\begin{equation}
\vec{L}^{(1)}\left(\vec{p}\right) = \frac{i \vec{p}}{p^2}\,,\quad \mathrm{and} \quad 
\vec{L}^{(2)}\left(\vec{p}_1, \vec{p}_2\right) = \frac{3}{7}\,\frac{i \vec{p}_{12}}{p_{12}^2} \left[1 - \left(\frac{\vec{p}_1 \cdot \vec{p}_2}{p_1\,p_2}\right)^2\right]\,,
\label{eq:analytic-residuals__lpt-kernels}
\end{equation}
where $\vec{p}_{12} \equiv \vec{p}_1 + \vec{p}_2$. The leading-order contribution to the density is
\begin{equation}
\delta\lin\left(\vec{k}\right) 
= - i\,\vec{k}\cdot \vec{s}\lin\left(\vec{k}\right) =  - i\, k^i L_i^{(1)}\left(\vec{k}\right)\,\delta\lin\left(\vec{k}\right)
\end{equation}
After contracting the two, one term is set to zero by the sharp-k filters and the remaining expression becomes
\begin{align}
\left\langle \Delta\delta^\LO_\mathrm{trunc.}\left(\vec{k}\right) \delta\lin\left(\vec{k'}\right) \right\rangle
&= \left(2\pi\right)^3 \delta^\dirac\left(\vec{k}+\vec{k'}\right) P(k) ~ k_l k_m \int \frac{d^3\vec{p}}{\left(2\pi\right)^3}\, W_\Lambda\left(\vec{p}\right)\left[1 - W_\Lambda\left(\vec{k} - \vec{p}\right)\right] \nn\\
&\times L\lin_l\left(\vec{p}\right) L_m^{(2)}\left(-\vec{p}, \vec{k}\right) P(p)\,.
\end{align}
The low-$k$ scaling of this integral is dominated by the convolution of the two sharp-k filters so we eventually obtain
\begin{equation}
\left\langle \Delta\delta^\LO_\mathrm{trunc.}\left(\vec{k}\right) \delta\lin\left(\vec{k'}\right) \right\rangle \propto k^3\,P(k) \,.
\end{equation}

\subsection{Aliasing errors}
\label{sec:analytic-residuals_aliasing}

Consider a finite one-dimensional grid with Nyquist frequency $k_\nyquist$. When represented on this grid, a mode $k$ is indistinguishable from modes at $k + 2nk_\nyquist$, where $n\in \mathbb{Z}$ (the factor $2$ arises because by definition  grids cover wavenumbers in the range $\pm k_\nyquist$). In the case of the forward model, the leading-order aliasing contribution arises from the coupling of two linear, large-scale modes which populates wavenumbers up to $2\Lambda$. Generalizing to three dimensions, if $k_\nyquist\left(N_\eul\right)=\Lambda$, the aliasing contribution to a mode with $|\vec{k}| < \Lambda$ is
\begin{equation}
\Delta \delta_\mathrm{alias}^\LO\left(\vec{k}\right) \propto k^2 \int \frac{d^3\vec{k'}}{(2\pi)^3}~ s\lin_\Lambda\left(\vec{k'}\right)\, s\lin_\Lambda\left(\vec{k}-\vec{k'} \pm 2\Lambda\right) \propto k^3\,,
\label{eq:analytic-residuals__aliasing-scaling}
\end{equation}
where the notation $\vec{k}\pm2\Lambda$ indicates that $2\Lambda$ is added/subtracted to each Cartesian component of $\vec{k}$ separately. The leading $k$-scaling arises from the convolution of two low-pass sharp-k filters, where one of them is additionally shifted by $2\Lambda$.

\subsection{LPT errors}
\label{sec:analytic-residuals_lpt}
From eq.~(\ref{eq:restframe-accuracy_resolution_displaced-delta}), we can write the Zeldovich (1LPT) density field and its leading-order error as
\begin{align}
\delta_{1\lpt}\left(\vec{k}\right) &= - i\vec{k}\cdot\vec{s}^{(1)}\left(\vec{k}\right) - 
\frac{1}{2} k^l k^m \left(s^{(1)}_l\,s^{(1)}_m\right)\left(\vec{k}\right) + \ldots \\
\Delta\delta_{1\lpt}\left(\vec{k}\right) &= - i\vec{k}\cdot\vec{s}^{(2)}\left(\vec{k}\right) - 
\frac{1}{2} k^l k^m \left(s^{(2)}_l\,s^{(1)}_m\right)\left(\vec{k}\right) + \ldots \,.
\end{align}
The cross-correlation between the residual and the full solution then has three contributions
\begin{align}
\left\langle \delta\left(\vec{k}_1\right) \Delta\delta_{1\lpt}\left(\vec{k}_2\right) \right\rangle_\LO 
&=
- k_1^l k_2^m \left\langle s_l^{(2)}\left(\vec{k}_1\right)\,s_m^{(2)}\left(\vec{k}_2\right) \right\rangle
+
\frac{i}{2}\, k_1^k k_2^l k_2^m \left\langle s_j^{(2)}\left(\vec{k}_1\right) \left(s_l^{(1)} s_m^{(1)}\right)\left(\vec{k}_2\right)\right\rangle \nn\\
&+
\frac{i}{2}\, k_1^k k_2^l k_2^m \left\langle s_j^{(1)}\left(\vec{k}_1\right) \left(s_l^{(2)} s_m^{(1)}\right)\left(\vec{k}_2\right)\right\rangle \,,
\end{align}
where $\left(s_l s_m\right)\left(\vec{k}\right)$ indicates the Fourier-transform of the real-space product. The contributions evaluate to
\begin{align}
k_1^l k_2^m \left\langle s_l^{(2)}\left(\vec{k}_1\right)\,s_m^{(2)}\left(\vec{k}_2\right) \right\rangle' 
&= \frac{1}{2} \int\frac{d^3\vec{p}}{(2\pi)^3} P(p)\,P(|\vec{k}_1-\vec{p}|) \left[k_1^i L_i^{(2)}\left(\vec{p}, \vec{k}_1-\vec{p}, \vec{k}_1\right)\right]^2 \nn\\
&\xrightarrow[k\ll p]{} \quad \propto k^4 
\label{eq:1lpterror-term1} \\[.5\baselineskip]
k_1^k k_2^l k_2^m \left\langle s_j^{(2)}\left(\vec{k}_1\right) \left(s_l^{(1)} s_m^{(1)}\right)\left(\vec{k}_2\right)\right\rangle'
&= \int\frac{d^3\vec{p}}{(2\pi)^3}~ P(p) P(|\vec{k}_1-\vec{p}|)~ k_1^j L_j^{(2)}\left(\vec{p}, \vec{k}_1-\vec{p} \right) \nn\\
&\times k_1^l L_l^{(1)}\left(\vec{k}_1 - \vec{p}\right) \times k_1^m
 L_m^{(1)}\left(\vec{p}\right) \nn \\
&\xrightarrow[k\ll p]{} \quad  \propto k^4
\label{eq:1lpterror-term2} \\[.5\baselineskip]
k_1^k k_2^l k_2^m \left\langle s_j^{(1)}\left(\vec{k}_1\right) \left(s_l^{(2)} s_m^{(1)}\right)\left(\vec{k}_2\right)\right\rangle'
&= \frac{i}{2} P(k_2) \int\frac{d^3\vec{p}}{(2\pi)^3}~ P\left(|\vec{k}_2 - \vec{p}|\right) ~ k_2^m L\lin_m\left(\vec{k}_2-\vec{p}\right) \nn\\
&\times k_2^n L_n^{(2)}\left(\vec{k}_2, \vec{p}-\vec{k}_2\right) \nn\\
&\xrightarrow[k\ll p]{} \quad  \propto P(k)\,k^4 
\label{eq:1lpterror-term3}
\end{align}
with the kernels given in eq.~(\ref{eq:analytic-residuals__lpt-kernels}) and $\left\langle \ldots \right\rangle'$ indicating that we have stripped the momentum conserving $(2\pi)^3\delta^\dirac\left(\vec{k}_1 + \vec{k}_2\right)$. The low-k scaling follows from the properties of the LPT kernels, i.e. $k^i L_i^{(1)}\left(\vec{k}-\vec{p}\right) \propto (k/p)^2$ and $k^i L_i^{(2)}\left(\vec{p}, \vec{k}-\vec{p}\right) \propto (k/p)^2$ in the limit of $k\ll p$.

In the case of the second order LPT solution, there is only one leading-order contribution to the residual correlation
\begin{align}
\left\langle \delta\left(\vec{k}_1\right) \Delta\delta_{2\lpt}\left(\vec{k}_2\right) \right\rangle'_\LO 
&=
-k_1^l k_2^m \left\langle s_l^{(1)}\left(\vec{k}_1\right)\, s_m^{(3)}\left(\vec{k}_2\right) \right\rangle' \nn\\
&= \frac{-i}{2} P(k_1) \int\frac{d^3\vec{p}}{(2\pi)^3}~ k_1^m L_m^{(3)}\left(\vec{p}, -\vec{p}, -\vec{k}_1\right)\,P(p) \nn \\
&\xrightarrow[k\ll p]{} \quad  \propto P(k)\,k^2 \,,
\end{align}
where the 3rd order LPT kernel (see e.g. \cite{Rampf:2012up}) behaves as $k^i L_i^{(3)}\left(\vec{p},-\vec{p}, \vec{k}\right)\propto P(k)\, k^2$ for $k\ll p$.

\section{Efficient nLPT computations}
\label{sec:optimizing-ngfwd}
The nLPT solution is found on a grid of size $N_\fwd$ (eq.~\eqref{eq:restframe-accuracy_resolution_ngfwd-rule}), large enough to suppress aliasing below $k_\nyquist\left(N_\eul\right)$ but too small to represent all nonlinearly populated modes up to $n_\lpt\Lambda$. Here, we show that the reduction of $N_\fwd$ does not miss any back-coupling from high-k modes to modes below $k_\nyquist\left(N_\eul\right)$ when higher-order contributions are computed iteratively from the lower-order Lagrangian distortion tensor, $H^{(n)}$. The criterion for $N_\fwd$ is
\begin{equation}
k_\nyquist\left(N_\fwd\right) = \left(\frac{3}{4} + \frac{n_\lpt}{2}\right)\,\Lambda\,,
\label{eq:restframe-accuracy__kny-of-ngfwd}
\end{equation}
and the lowest-order contribution which populates modes above $k_\nyquist\left(N_\fwd\right)$
\begin{equation}
n_\mathrm{cut} = \left\lceil \frac{k_\nyquist\left(N_\fwd\right)}{\Lambda}\right\rceil\,,
\end{equation}
where $\lceil\ldots\rceil$ indicates up-rounding to the next integer. The highest-order, and hence most sensitive terms for mode-coupling in the $n_\lpt$-order solution are of the form
\begin{equation}
\int d^3\vec{k'}~ M^{(n_\lpt-n_\mathrm{cut})}\left(\vec{k'}\right)\,M^{(n_\mathrm{cut})}\left(\vec{k}-\vec{k'}\right)\,.
\end{equation} 
We are only interested in modes that can be represented on a grid of size $N_\eul$, i.e. modes with $k\leq 2/3 \Lambda$. Therefore, the largest mode in $M^{(n_\mathrm{cut})}$ which is relevant for our solution is
\begin{equation}
k_\mathrm{rel,max}\left(M^{(n_\mathrm{cut})}\right) = \frac{3}{2} \Lambda + \left(n_\lpt - n_\mathrm{cut}\right) \Lambda\,.
\end{equation}
Comparing with eq.~(\ref{eq:restframe-accuracy__kny-of-ngfwd}), we see that $k_\mathrm{rel,max} \leq k_\nyquist\left(N_\fwd\right)$, i.e. all modes relevant to the solution can be represented by $N_\fwd$.

We test the aliasing-criterion of eq.~(\ref{eq:restframe-accuracy_resolution_ngfwd-rule}) numerically in figure~\ref{fig:restframe-accuracy_resolution_ngfwd}, where we show the residuals between density grids computed for different $N_\fwd$. While some small residuals remain if $N_\fwd$ is chosen too small, they become suppressed to the level of machine precision by following eq.~(\ref{eq:restframe-accuracy_resolution_ngfwd-rule}). It is important to note that the aliasing criterion for $N_\fwd$ depends on $N_\eul$. For larger $N_\eul$, the residuals are not entirely suppressed, albeit irrelevantly small (right panel).

\begin{figure}
\centering
\includegraphics[width=\figuresize]{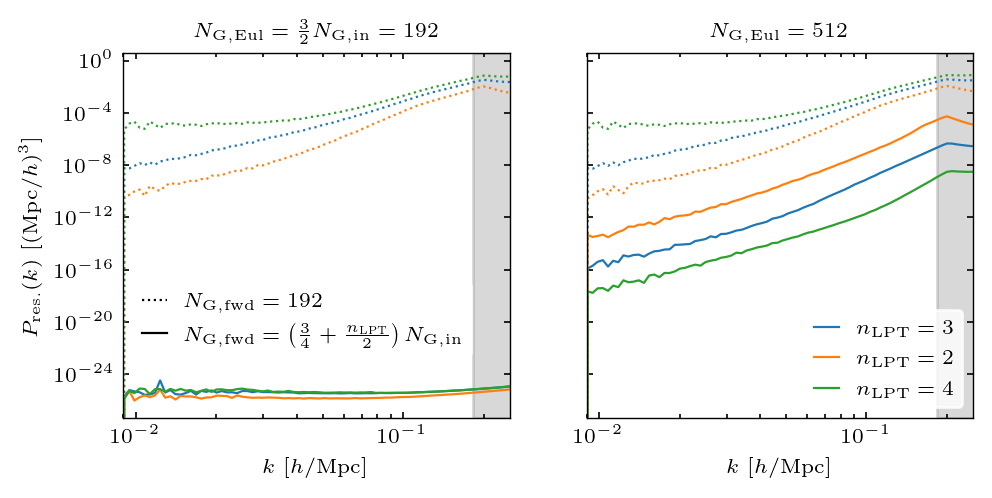}
\caption{Impact of $N_\fwd$, the grid size on which the nLPT solution is computed, on the accuracy of the evolved density field. We consider the 3LPT gravity model without the transverse contribution to the displacement, $\Lambda=0.2\,h/\mpc$ and evolve the density to $z=0$. In the reference model, we choose $N_\fwd = n_\lpt \Lambda$, large enough that all modes populated by nonlinearities in the LPT solution can be represented. We compare this to examples with reduced $N_\fwd$. Indeed, if the aliasing criterion in eq.~(\ref{eq:restframe-accuracy_resolution_ngfwd-rule}) is met, the residuals are at the level of machine precision. Note that the aliasing bound increases with $N_\eul$, hence we still see some residual errors in the right panel, where the displacement is evaluated at a higher resolution.}
\label{fig:restframe-accuracy_resolution_ngfwd}
\end{figure}

\section{Properties of the halo sample}
\label{sec:halo-sample}
To test the accuracy at which cosmological parameters, and in particular $\sigma_8$, can be inferred from non-trivially biased tracers of the fully nonlinear density field, we consider halos in N-body simulations. The setup of these simulations is analogous to the reference simulations introduced in section~\ref{sec:restframe-accuracy_reference-sims}, but without cut-off; i.e. now the initial conditions contain modes down to the smallest resolvable wavenumber. Halos are identified with the ROCKSTAR algorithm \cite{Behroozi:2011ju} in two logarithmically spaced mass bins,
\begin{equation}
\lg M \,\left[h^{-1} M_\odot\right] \in  \left[12.5,\,13.0\right]
\quad\mathrm{and}\quad
\lg M \,\left[h^{-1} M_\odot\right] \in  \left[13.0,\,13.5\right] \,.
\end{equation}
Note that several previous works considered the same simulations but used AMIGA \cite{2009ApJS..182..608K} halos. There are two simulation realizations available and we consider three redshift snapshots at $z=0.0$, $z=0.5$ and $z=1.0$. The average number density over the two simulations is summarized for each mass bin and snapshot in table \ref{tab:halo-number-densities}.

\begin{table}[h]
\centering
\begin{tabular}{l|c|c}
 & $\lg M/\left( h^{-1}\,M_\odot\right) \in \left[12.5, 13.0\right]$ & $\lg M/\left( h^{-1}\,M_\odot\right) \in \left[13.0, 13.5\right]$ \\
\hline
\hline
$z = 0.0$ & $9.6\times 10^{-4} \left(\mpc/h\right)^{-4}$ & $4.0\times 10^{-4} \left(\mpc/h\right)^{-4}$ \\
$z = 0.5$ & $8.7\times 10^{-4} \left(\mpc/h\right)^{-4}$ & $3.2\times 10^{-4} \left(\mpc/h\right)^{-4}$ \\
$z = 1.0$ & $7.0\times 10^{-4} \left(\mpc/h\right)^{-4}$ & $1.7\times 10^{-4} \left(\mpc/h\right)^{-4}$ \\
\end{tabular}
\caption{Number densities of the halo samples considered for the analyses in section~\ref{sec:results}.}
\label{tab:halo-number-densities}
\end{table}

\section{Sampling algorithm and convergence}
\label{sec:sampling}

The free parameters of the analysis in section \ref{sec:results} are $\alpha$ (eq.~\eqref{eq:results-alpha-parametrization}), the linear bias coefficient $b_\delta$, and the noise amplitudes $\sigma_0$ and $\sigma_{\epsilon,2}$ (eq.~\eqref{eq:forward-model__noise-parametrization}); all higher-oder bias coefficients are marginalized over analytically \cite{Elsner:2019rql}. We explore the posterior with a slice sampler \cite{2000physics...9028N} using the ``stepping out'' and the ``shrinkage'' procedure, and in a block sampling approach, we draw one parameter at a time. This sampler has been implemented directly into \leftfield. For all free parameters, we assume uniform priors whose range is given in table~\ref{tab:sampling-priors}, along with the step size of the slice sampler and the initial value the chains are initialized from. 

\begin{table}[h]
\centering
\begin{tabular}{c|c|c|c|c|c}
parameter & lower limit & upper limit & step size & initial value & initial range \\
\hline
$\alpha$ & $-10^{30}$ & $10^{30}$ & $0.01$ & $1.0$ & $0.5 - 2.0$\\
$b_\delta$ & $-10^{30}$ & $10^{30}$ & $0.01$ & $1.0$ & $0.5 - 2.0$\\
$\sigma_0$ & $0.0$ & $10^{30}$ & $0.01$ & $0.3$ & $0.1 - 1.0$ \\
$\sigma_{\epsilon,2}$ & $0.0$ & $10^{30}$ & $3.0$ & $0.0$ & $0 - 50$\\
\end{tabular}
\caption{Free parameters whose posteriors are explored with a slice sampler assuming uniform priors. We list the lower and upper prior limit, the step size for sampling and the initial parameter value. For a subset of analyses, we run ten additional chains for convergence tests (see table~\ref{tab:sampling-gelman-rubin}) with random initial starting points drawn from the ``initial range'' in the last column.}
\label{tab:sampling-priors}
\end{table}

We run each chain for at least 2,000 samples and, after inspection of the trace plots, we discard the first 100 samples of each chain for burn-in. We demand an effective sample size in all free parameters of at least 100.\footnote{\label{fn:numpyro}The effective sample size and the Gelman-Rubin criterion are computed with \texttt{NumPyro} \cite{phan2019composable, bingham2019pyro}} In reality, the effective sample size is higher, and in particular for $\alpha$ and $b_\delta$ it reaches several hundreds to more than one thousand. We pick a representative subset of runs for convergence tests. Those are listed in table~\ref{tab:sampling-gelman-rubin}. For each of them, we run additional 10 chains, starting from random initial values drawn from the ranges listed in table~\ref{tab:sampling-priors}, and we compute the Gelman-Rubin criterion \cite{1992StaSc...7..457G}. For all parameters, we find values very close to one, $\hat{R}-1 < 0.01$, in many cases much less. From these tests we conclude that the chains are well converged.

\newcolumntype{L}[1]{>{\raggedright\let\newline\\\arraybackslash\hspace{0pt}}m{#1}}
\begin{table}
\centering
\begin{tabular}{L{5.48in-5.8cm}||p{1.2cm}|p{1.2cm}|p{1.2cm}|p{1.2cm}}
\multirow[b]{2}{*}{data set and analysis} &
\multicolumn{4}{c}{Gelman-Rubin criterion $\hat{R} - 1$} \\
& $\alpha$ & $b_\delta$ & $\sigma_0$ & $\sigma_{\epsilon,2}$  \\
\hline \hline
3LPT, $o_\bias=3$ (figure~\ref{fig:restframe__results-3lpt-lagrangian__var-bias-order}), \newline
M1, sim. 1, $z=0.0$, $\Lambda = 0.10\,h/\mpc$
& 0.0002 & 0.0001 & 0.006 & 0.006 \\
\hline
3LPT, $o_\bias=3$ (figure~\ref{fig:restframe__results-3lpt-lagrangian__var-bias-order}), \newline
M1, sim. 2, $z=0.0$, $\Lambda = 0.10\,h/\mpc$ 
& 0.0008 & 0.0009 & 0.003 & 0.002 \\
\hline
3LPT, $o_\bias=3$ (figure~\ref{fig:restframe__results-3lpt-lagrangian__var-bias-order}), \newline
M1, sim. 1, $z=0.0$, $\Lambda = 0.18\,h/\mpc$
& $< 10^{-4}$ & $< 10^{-4}$ & 0.002 & 0.003 \\
\hline
3LPT, $o_\bias=3$ (figure~\ref{fig:restframe__results-3lpt-lagrangian__var-bias-order}), \newline
M2, sim. 1, $z=0.0$, $\Lambda = 0.10\,h/\mpc$
& 0.0004 & 0.0004 & 0.002 & 0.002 \\
\hline
3LPT, $o_\bias=3$ (figure~\ref{fig:restframe__results-3lpt-lagrangian__var-bias-order}), \newline
M2, sim. 1, $z=0.0$, $\Lambda = 0.18\,h/\mpc$
& 0.0004 & 0.0007 & 0.005 & 0.005 \\
\hline
3LPT, $o_\bias=3$ (figure~\ref{fig:restframe__results-3lpt-lagrangian__var-bias-order}), \newline
M1, sim. 1, $z=1.0$, $\Lambda = 0.10\,h/\mpc$
& 0.0007 & 0.0006 & 0.002 & 0.002 \\
\hline
3LPT, $o_\bias=3$ (figure~\ref{fig:restframe__results-3lpt-lagrangian__var-bias-order}), \newline
M1, sim. 1, $z=1.0$, $\Lambda = 0.18\,h/\mpc$ 
& 0.0006 & 0.0006 & 0.001 & 0.001 \\
\hline
3LPT, $o_\bias=3$ (figure~\ref{fig:restframe__results-3lpt-lagrangian__var-bias-order}), \newline
M2, sim. 1, $z=1.0$, $\Lambda = 0.10\,h/\mpc$
& 0.002 & 0.002 & 0.002 & 0.002 \\
\hline
3LPT, $o_\bias=3$ (figure~\ref{fig:restframe__results-3lpt-lagrangian__var-bias-order}), \newline
M2, sim. 1, $z=1.0$, $\Lambda = 0.18\,h/\mpc$ 
& 0.0005 & 0.0008 & 0.001 & 0.001 \\
\hline
\hline
2LPT, $o_\bias=3$, $\filterbias=1.25\,\Lambda$ (figure~\ref{fig:restframe-results__eulerianbias__2LPT-varLambda-varKmaxBias}), \newline
M1, sim. 1, $z=0.0$, $\Lambda=k_\mathrm{max}=0.10\,h/\mpc$ 
& 0.0001 & $< 10^{-4}$ & 0.003 & 0.003 \\
\hline
2LPT, $o_\bias=3$, $\filterbias=1.25\,\Lambda$ (figure~\ref{fig:restframe-results__eulerianbias__2LPT-varLambda-varKmaxBias}), \newline
M1, sim. 1, $z=0.0$, $\Lambda=k_\mathrm{max}=0.18\,h/\mpc$ 
& $<10^{-4}$ & $<10^{-4}$ & 0.004 & 0.005 \\
\hline
2LPT, $o_\bias=3$, $\filterbias=1.25\,\Lambda$ (figure~\ref{fig:restframe-results__eulerianbias__2LPT-varLambda-varKmaxBias}), \newline
M2, sim. 1, $z=0.0$, $\Lambda=k_\mathrm{max}=0.10\,h/\mpc$
& $<10^{-4}$ & $<10^{-4}$ & 0.002 & 0.002 \\
\hline
2LPT, $o_\bias=3$, $\filterbias=1.25\,\Lambda$ (figure~\ref{fig:restframe-results__eulerianbias__2LPT-varLambda-varKmaxBias}), \newline
M2, sim. 1, $z=0.0$, $\Lambda=k_\mathrm{max}=0.18\,h/\mpc$ 
& $< 10^{-4}$ & 0.0001 & 0.002 & 0.002 \\
\hline
2LPT, $o_\bias=3$, $\filterbias=1.25\,\Lambda$ (figure~\ref{fig:restframe-results__eulerianbias__2LPT-varLambda-varKmaxBias}), \newline
M1, sim. 1, $z=0.5$, $\Lambda=k_\mathrm{max}=0.10\,h/\mpc$ 
&  0.0003 & 0.0001 & 0.003  & 0.003 \\
\hline
2LPT, $o_\bias=3$, $\filterbias=1.25\,\Lambda$ (figure~\ref{fig:restframe-results__eulerianbias__2LPT-varLambda-varKmaxBias}), \newline
M1, sim. 1, $z=0.5$, $\Lambda=k_\mathrm{max}=0.18\,h/\mpc$ 
& $< 10^{-4}$ & $< 10^{-4}$ & 0.005 & 0.005 \\
\hline
2LPT, $o_\bias=3$, $\filterbias=1.25\,\Lambda$ (figure~\ref{fig:restframe-results__eulerianbias__2LPT-varLambda-varKmaxBias}), \newline
M2, sim. 1, $z=0.5$, $\Lambda=k_\mathrm{max}=0.10\,h/\mpc$
& $<10^{-4}$ & $<10^{-4}$ & 0.002 & 0.002 \\
\hline
2LPT, $o_\bias=3$, $\filterbias=1.25\,\Lambda$ (figure~\ref{fig:restframe-results__eulerianbias__2LPT-varLambda-varKmaxBias}), \newline
M2, sim. 1, $z=0.5$, $\Lambda=k_\mathrm{max}=0.18\,h/\mpc$ 
& $<10^{-4}$ & $<10^{-4}$ & 0.002 & 0.002 \\
\end{tabular}
\caption{Convergence test for selected analyses with Lagrangian (top) and Eulerian (bottom) bias. The lower and higher mass bins are abbreviated as ``M1'' and ``M2'', respectively. We list the Gelman-Rubin statistics from 10 chains with 2,000 samples each after discarding 100 samples for burn in.}
\label{tab:sampling-gelman-rubin}
\end{table}

We report the inferred value of $\alpha$ in terms of the mode-centered 68\% credible interval. We have verified that the 68\% and the 95\% credible interval is very consistent with the mean-centered $1\sigma$ and $2\sigma$ limits, respectively, indicating that the parameter posterior is close to a Gaussian. In figure~\ref{fig:restframe-results__lagrangianbias__contour} and figure~\ref{fig:restframe-results__eulerianbias__contour}, we show two-dimensional contour plots of all free parameters in the inference for some selected chains. We expect a running in $b_\delta$ as a function of the initial cut-off $\Lambda$, because the operator basis (see appendix \ref{sec:analysis-settings}) explicitly refers to the scale $\Lambda$. See Ref. \cite{Rubira:2023vzw} for a detailed discussion of the connection to the commonly used convention for bias coefficients. The number of modes which are integrated out and the inferred noise level decrease towards higher cut-offs. In addition, the figures illustrate the tightening of constraints on $\alpha$ and $b_\delta$ as more modes are included in the analysis and the degeneracy between the two parameters is broken more effectively. Interestingly, at identical cut-off the degeneracy between $\alpha$ and $b_\delta$ is less pronounced in the case of Eulerian bias operators. The cross-correlation coefficient between $\alpha$ and $b_\delta$ is $-0.59$ and $-0.60$ in the lower and higher mass bin at $z=0.0$, $\Lambda=0.14\,h/\mpc$ for the 3LPT, third-order Lagrangian bias model shown in figure~\ref{fig:restframe-results__lagrangianbias__contour}. It becomes $0.01$ and $-0.04$ for the 2LPT third-order Eulerian bias model of figure~\ref{fig:restframe-results__eulerianbias__contour} (again at $z=0.0$, $\Lambda=0.14\,h/\mpc$ in the lower and higher mass bin, respectively).

\begin{figure}
\centering
\includegraphics[width=.5\figuresize]{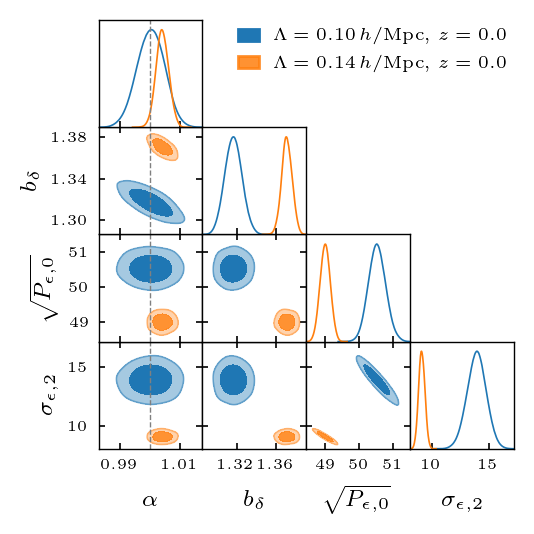}
\includegraphics[width=.5\figuresize]{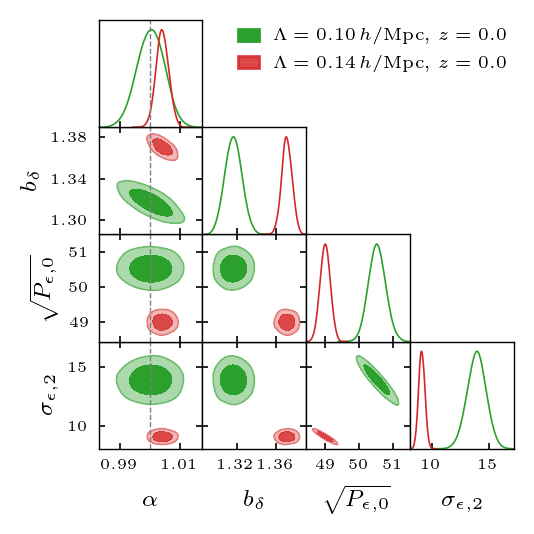}
\caption{Contour plots of free parameters, here for a third order Lagrangian bias expansion with 3LPT gravity at $z=0$ for the lower mass bin (\textbf{left}) and the higher one (\textbf{right}). The noise power spectrum is $P_{\epsilon,0} = \sigma_0^2 \left(L_\mathrm{box}/N_\lh\right)^3$ in units of $\left(\mpc/h\right)^3$. We expect a running in the lowest order bias parameter $b_\delta$ as a function of $\Lambda$, and a reduction of the noise amplitude as the cut-off increases. In addition the tightening of the constraints on $\alpha$ and $b_\delta$ when including more modes is apparent.}
\label{fig:restframe-results__lagrangianbias__contour}
\end{figure}

\begin{figure}
\centering
\includegraphics[width=.5\figuresize]{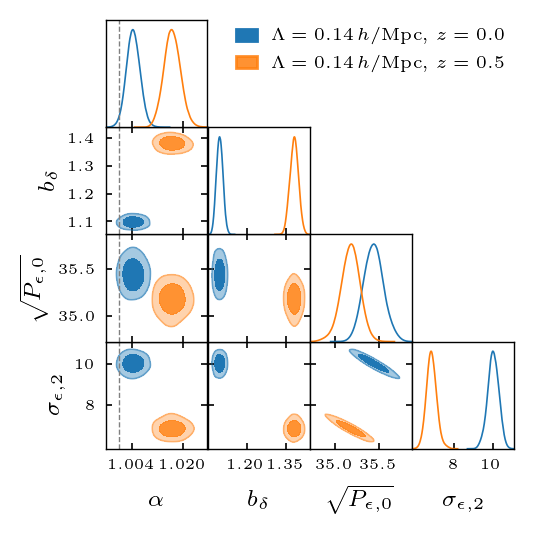}
\includegraphics[width=.5\figuresize]{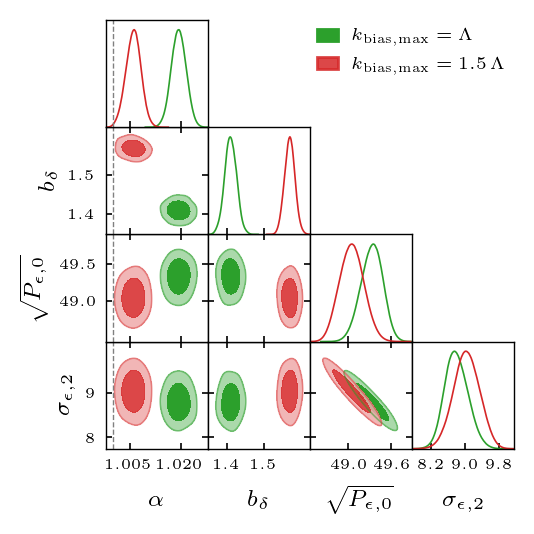}
\caption{Contour plots of all free parameters, here for a third order Eulerian bias expansion with 2LPT gravity. Results on the \textbf{left} are for the lower mass bin at two different redshifts using $\filterbias=\Lambda$. One the \textbf{right} we compare results with different filters in the higher mass bin at $\Lambda=0.14\,h/\mpc$ and $z=0.0$. The noise power spectrum is $P_{\epsilon,0} = \sigma_0^2 \left(L_\mathrm{box}/N_\lh\right)^3$ in units of $\left(\mpc/h\right)^3$.}
\label{fig:restframe-results__eulerianbias__contour}
\end{figure}

\section{Supplementary tests}
\subsection{Forward model accuracy}
Here, we present some additional tests of the convergence of the forward model with numerical resolution, extending those in section~\ref{sec:restframe-accuracy} to different perturbative orders and cut-off scales:
\begin{itemize}
\item We show in figure~\ref{fig:restframe-accuracy_resolution_truncation-lpt} how resolution effects in the forward model evolve between different LPT orders. We find the same qualitative behavior that we expect based on the results of section~\ref{sec:restframe-accuracy__resolution}. The magnitude of resolution effects increases for higher LPT orders, which populate nonlinear modes up to higher wavenumbers.
\item The forward model accuracy for a cut-off of $\Lambda=0.10\,h/\mpc$ is shown in figure~\ref{fig:restframe-accuracy_lpt_lpterror_l010} for $z=0.5$. In this regime, higher LPT-orders $n_\lpt > 2$ achieve a remarkable accuracy, better than $0.2\%$ in the power spectrum and the cross-correlation.
\item In figure \ref{fig:restframe-accuracy_lpt_curl} we investigate the impact of the transverse contributions to the displacement vector (eq.~\eqref{sec:forward-model__shift-decomposition}) on the overall forward model accuracy. These start only at perturbative order $n_\lpt \geq 3$, and they have negligible impact on the residuals of the forward model.
\end{itemize}

\begin{figure}
\centering
\includegraphics[width=\figuresize]{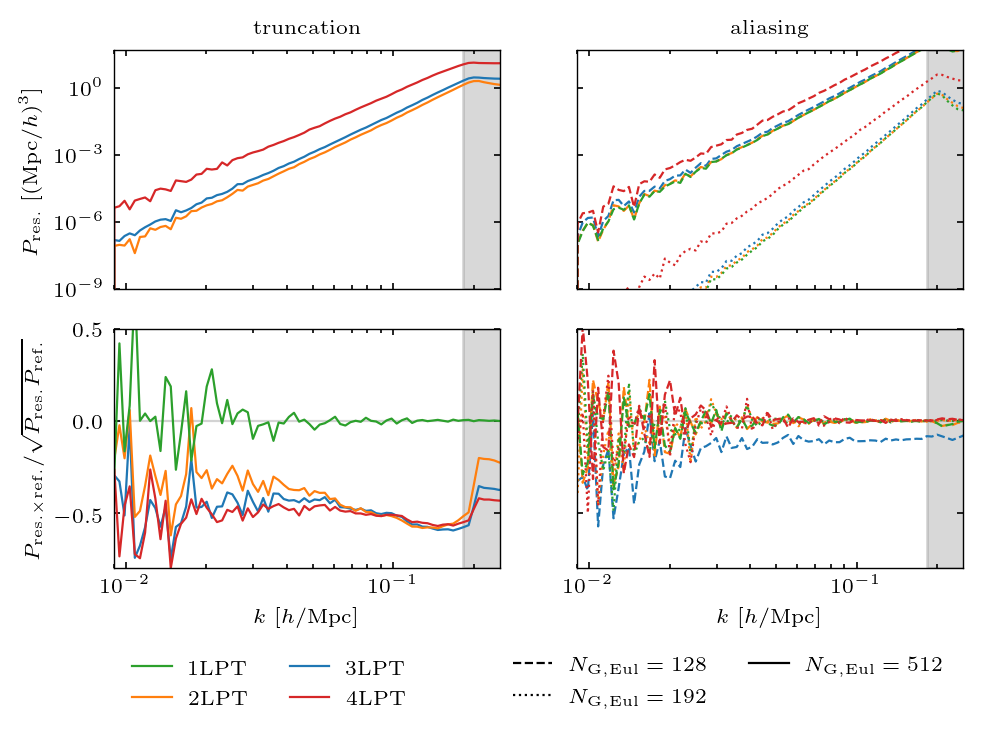}
\caption{Impact of truncation and aliasing effects in the displacement step on the evolved matter density, similar to figure~\ref{fig:restframe-accuracy_resolution_truncation} but considering different LPT orders; the density field is evolved to $z=0$ for a cut-off $\Lambda=0.2\,h/\mpc$. For \textbf{truncation effect  (left)}, we apply a top-hat filter at $\filterfwd=\Lambda$ to the displacement vector before the actual displacement operation and compute the latter at high resolution, $N_\eul=512$. The reference model has no filter. For \textbf{aliasing effects (right)}, we compare densities obtained with identical filters in the forward model at different displacement resolutions. The reference model has the highest resolution, $N_\eul=512$. The numerical results reproduce the expected $k^6$-scaling of the residuals and the white residual correlation towards low-$k$ in the case of truncation effects. At higher LPT orders, small-scale modes become more important and the magnitude of the residuals increases. Since the displacement vector is linear in the initial conditions for 1LPT, there are no truncation effects in this case. }
\label{fig:restframe-accuracy_resolution_truncation-lpt}
\end{figure}

\begin{figure}
\centering
\includegraphics[width=\figuresize]{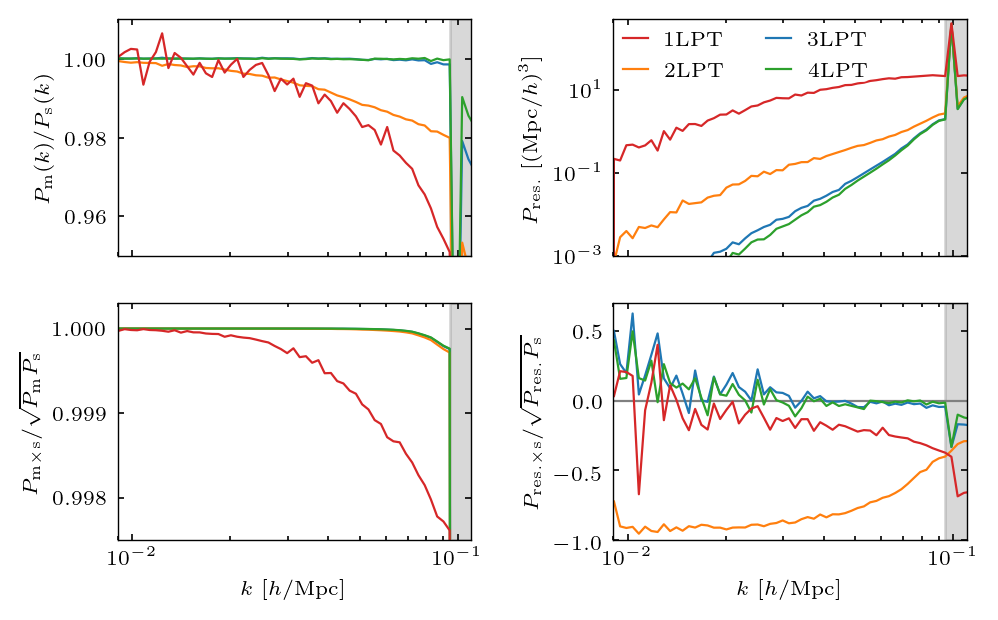}
\caption{Accuracy of the perturbative forward model at $z=0.5$ for different orders in the nLPT expansion. We compare the forward model to a reference N-body simulation which has an identical sharp-k filter applied to its initial conditions at $\Lambda=0.1\,h/\mpc$ (see section~\ref{sec:restframe-accuracy_reference-sims}). This lower cut-off, in comparison to figure~\ref{fig:restframe-accuracy_lpt_lpterror}, further increases the precision to the sub-percent level. Most scenarios considered in this work are in the range spanned by the two cut-offs.}
\label{fig:restframe-accuracy_lpt_lpterror_l010}
\end{figure}

\begin{figure}
\centering
\includegraphics[width=\figuresize]{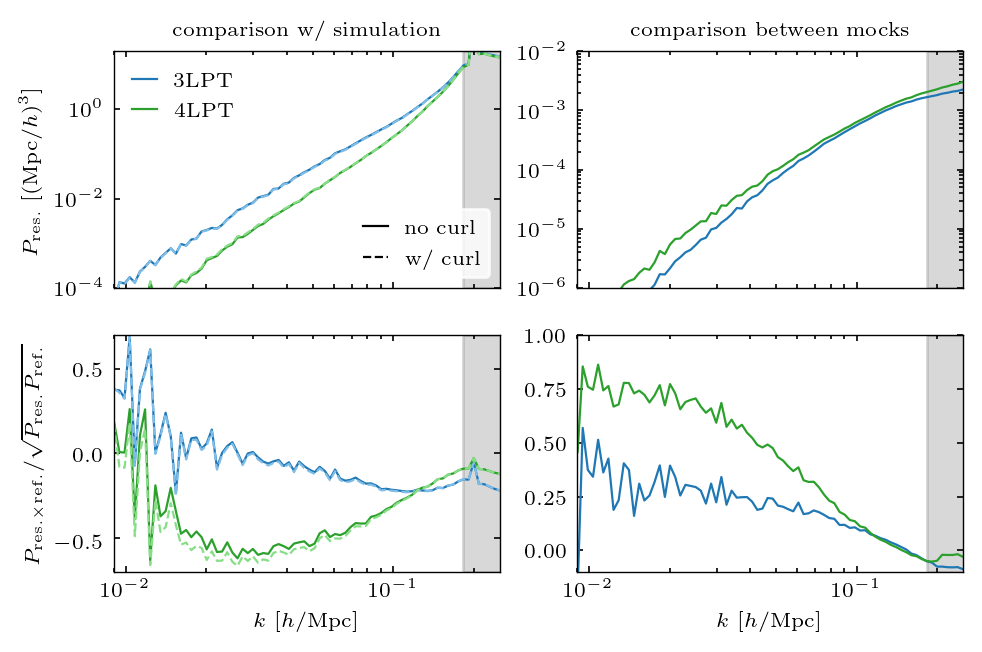}
\caption{The impact of transverse contributions to the displacement on the density field at $z=0$. One the left, we compare the model prediction with (dashed lines) and without (solid lines) the transverse terms to the reference N-body simulations which start from identical initial conditions as the forward model that have been filtered at $\Lambda=0.2\,h/\mpc$. On the right, we compare the model predictions with and without transverse contributions between each other. They contribute only from 3rd order and hence have no impact on the 1LPT and 2LPT prediction, but the impact is negligible also at higher LPT orders.}
\label{fig:restframe-accuracy_lpt_curl}
\end{figure}

\subsection{Parameter inference with Lagrangian bias}
\label{sec:additional-tests__lagrangian-bias-inference}

We show additional tests for the parameter inference from N-body halos with a Lagrangian bias expansion in figures~\ref{fig:restframe__lagrangianbias_var-ngeul} to~\ref{fig:crosscheck-sigma-contours}. These are commented on in section~\ref{sec:results-lagrangina-bias}.

\begin{figure}
\centering
\includegraphics[width=\figuresize]{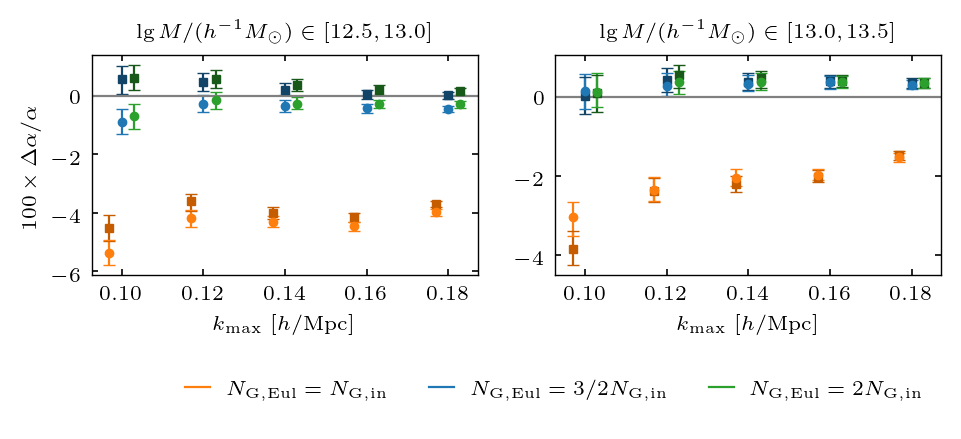}
\caption{Inference of the scaled primordial fluctuation amplitude $\alpha=\sigma_8/\sigma_{8,\mathrm{fid}}$ from N-body halos at $z=0.0$ using the 3LPT gravity model, $\Lambda=k_\mathrm{max}$ and a third order Lagrangian bias expansion. We compare results where the displacement from Lagrangian to Eulerian coordinates is computed at different numerical resolutions, $N_\eul$. While low $N_\eul$ lead to significant shifts in the inferred parameter values, away from the ground truth, an increase beyond the 3/2-rule established in section~\ref{sec:forward-model-numerics} has no impact. This indicates that the 3/2-rule indeed sits a the sweet spot between numerical efficiency and accuracy.}
\label{fig:restframe__lagrangianbias_var-ngeul}
\end{figure}

\begin{figure}
\centering
\includegraphics[width=\figuresize]{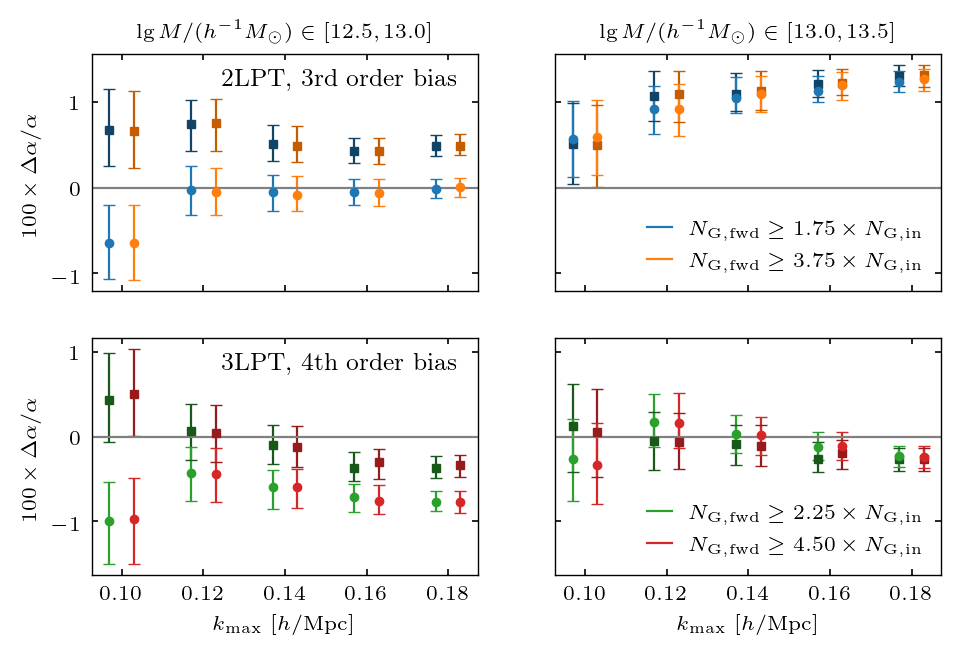}
\caption{Inference of the scaled primordial fluctuation amplitude $\alpha=\sigma_8/\sigma_{8,\mathrm{fid}}$ from N-body halos at $z=0.0$ using the 3LPT gravity model, $\Lambda=k_\mathrm{max}$ and a third order Lagrangian bias expansion. We compare results where $N_\fwd$ is set to avoid aliasing in the LPT solution (eq.~\eqref{eq:restframe-accuracy_resolution_ngfwd-rule}) to those with a higher $N_\fwd$, which also suppresses aliasing in the higher-order bias parameters (see section~\ref{sec:restframe-accuracy__lagrangian-bias}). There is no appreciable difference in the inferred parameters, indicating that the more economic choice of $N_\fwd$ is indeed sufficient.}
\label{fig:restframe-results__lagrangianbias__var-ngfwd}
\end{figure}

\begin{figure}
\centering
\includegraphics[width=\figuresize]{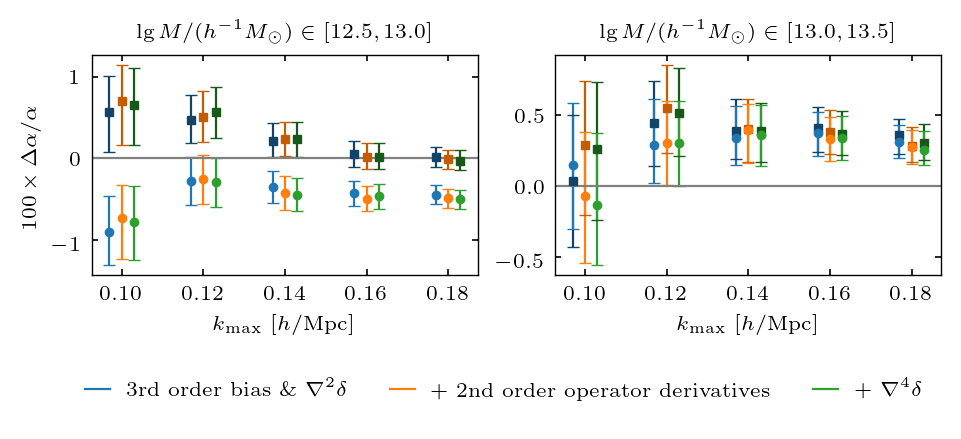}
\caption{Inference of the scaled primordial fluctuation amplitude $\alpha=\sigma_8/\sigma_{8,\mathrm{fid}}$ from N-body halos at $z=0.0$ using the 3LPT gravity model, $\Lambda=k_\mathrm{max}$ and a third order Lagrangian bias expansion. We compare results which include a subset of higher-order derivative operators, whereby ``2nd order operator derivatives'' stands for $\{\nabla^2 \sigma^2,\ \nabla^2 \mathrm{tr}\left(M^{(1)}M^{(1)}\right)\}$. These higher order operators have no appreciable effect on the inferred values for $\sigma_8$, indicating that the back reaction from small scales and tracer non-locality is sufficiently absorbed by the leading-order in derivatives term.}
\label{fig:sampling-results__lagrangianbias__highderiv}
\end{figure}

\begin{figure}
\centering
\includegraphics[width=\figuresize]{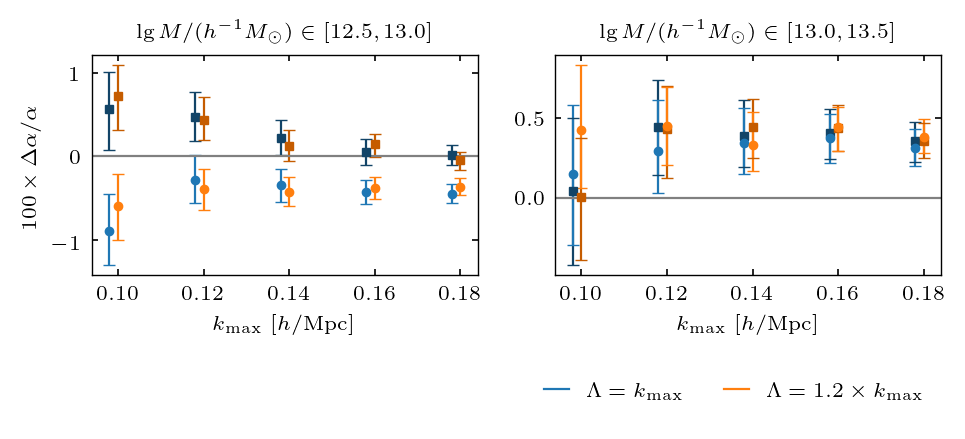}
\caption{Inference of the scaled primordial fluctuation amplitude $\alpha=\sigma_8/\sigma_{8,\mathrm{fid}}$ from N-body halos at $z=0.0$ using the 3LPT gravity model and a third order Lagrangian bias expansion. We investigate the impact of choosing $\Lambda > k_\mathrm{max}$ and find it to be negligible for the inference of $\sigma_8$. As with the higher-order derivative operators (figure~\ref{fig:sampling-results__lagrangianbias__highderiv}), this indicates that the back reaction from small scales and tracer non-locality is sufficiently absorbed by the leading-order in derivatives term.}
\label{fig:sampling-results__lagrangianbias__highlambda}
\end{figure}

\begin{figure}
\centering
\includegraphics[width=\figuresize]{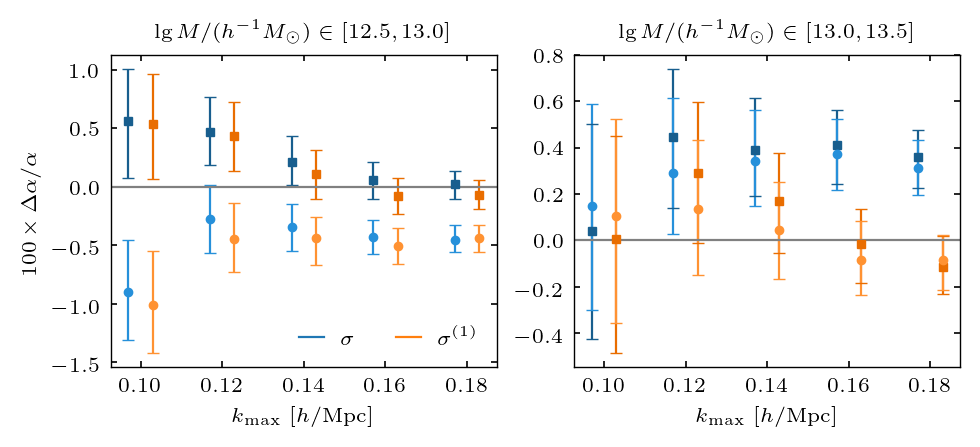}
\caption{Inference of the scaled primordial fluctuation amplitude $\alpha=\sigma_8/\sigma_{8,\mathrm{fid}}$ from N-body halos at $z=0.0$ using the 3LPT gravity model, $\Lambda=k_\mathrm{max}$ and a third order Lagrangian bias expansion. We compare two versions of the bias expansion, where higher-order operators are either constructed from $\sigma = \sum_n \sigma^{(n)}$ or from $\sigma\lin$; blue points are identical to those in the top panel of figure~\ref{fig:restframe__results-3lpt-lagrangian__var-bias-order}. The operators $\sigma^{(n)}$ for $n > 1$ are related to other scalar invariants of the Lagrangian distortion tensor by the LPT recursion relations, hence we expect the two formulations to be equivalent up to higher-order bias terms. Indeed, we find only small deviations in the higher mass bin for large cut-off values $\Lambda$.}
\label{fig:crosscheck-sigma-constraints}
\end{figure}

\begin{figure}
\centering
\includegraphics[width=.5\figuresize]{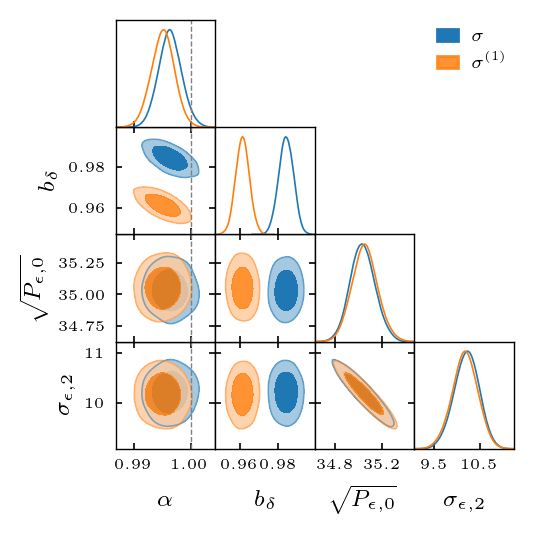}
\includegraphics[width=.5\figuresize]{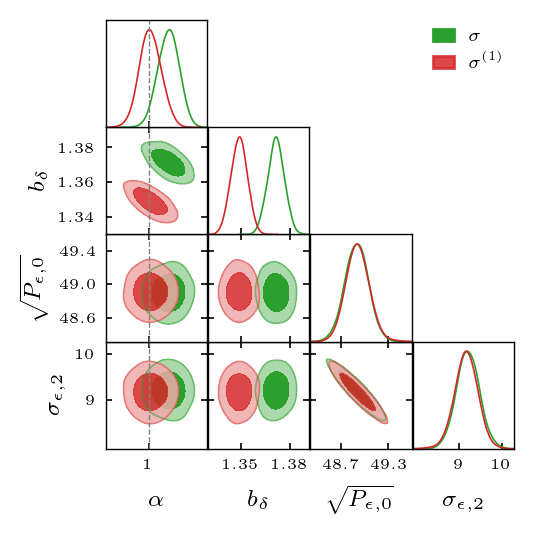}
\caption{Contour plots of free parameters, here for a third order Lagrangian bias expansion with 3LPT gravity at $z=0$ for the lower mass bin (\textbf{left}) and the higher one (\textbf{right}). We compare results where the higher-order bias operators are constructed from $\sigma = \sum_n \sigma^{(n)}$ and $\sigma^{(1)}$. The most noticeable impact is a shift in $b_\delta$, pointing to some degeneracy with the higher-order bias parameters. The qualitative directions of the correlations, however, remain unaffected.}
\label{fig:crosscheck-sigma-contours}
\end{figure}

\subsection{Parameter inference with Eulerian bias}
\label{sec:additional-tests__eulerian-bias-inference}

We show additional tests for the parameter inference from N-body halos with a Eulerian bias expansion in figures~\ref{fig:restframe-results__eulerianbias__3LPT-varLambda-varKmaxBias} to~\ref{fig:restframe-results__eulerianbias__NLOderivatives}. These are commented on in section~\ref{sec:results-eulerian}.

\begin{figure}
\centering
\includegraphics[width=\figuresize]{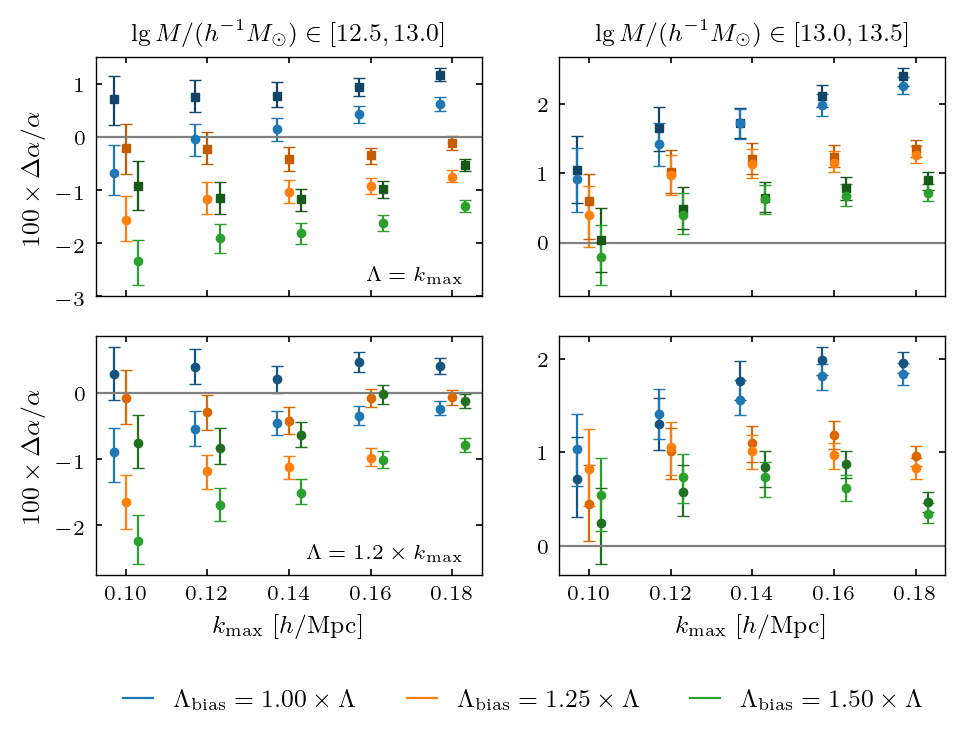}
\caption{Inference of the scaled primordial fluctuation amplitude $\alpha=\sigma_8/\sigma_{8,\mathrm{fid}}$ from N-body halos at $z=0.0$, using third order Eulerian bias and the 3LPT gravity model. We compare different spherical filters, applied to the evolved density before the construction of the bias operators. In the upper panels, we set $\Lambda=k_\mathrm{max}$, i.e. the cut-off in the initial conditions agrees with the largest wavenumber considered in the likelihood, while the lower panels explore increasing $\Lambda$ to $1.2 \, k_\mathrm{max}$. The inferred parameter values change little in comparison to figure~\ref{fig:restframe-results__eulerianbias__2LPT-varLambda-varKmaxBias}, which used the 2LPT solution, and apparently the parameter shifts are not dominated by the gravitational accuracy.}
\label{fig:restframe-results__eulerianbias__3LPT-varLambda-varKmaxBias}
\end{figure}

\begin{figure}
\centering
\includegraphics[width=\figuresize]{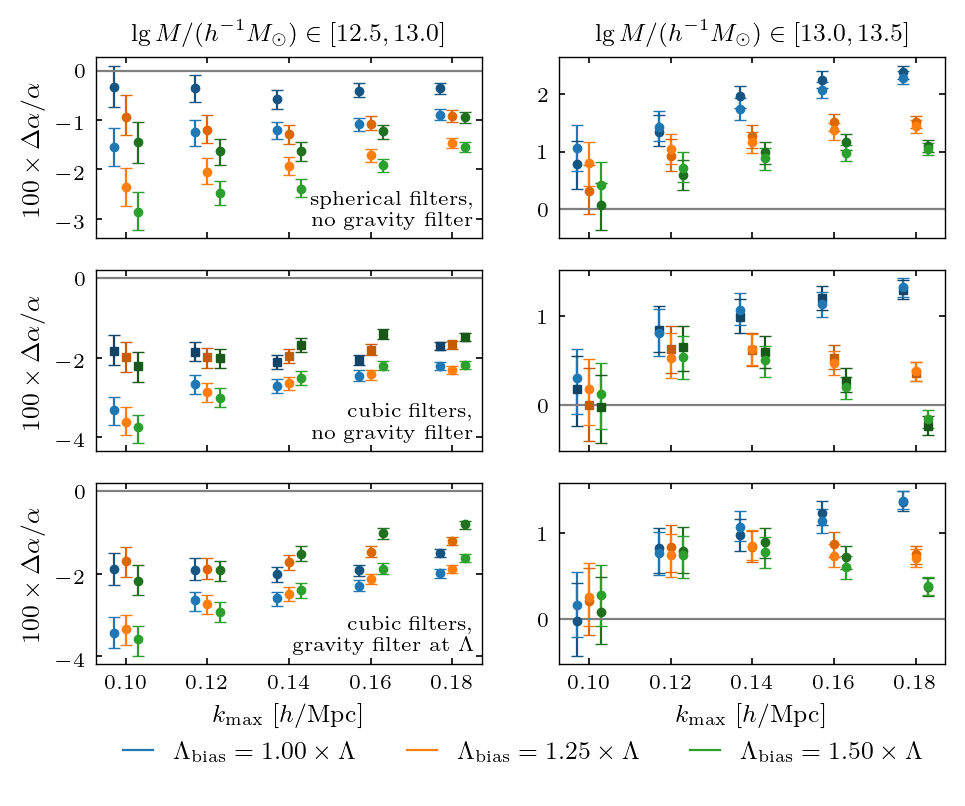}
\caption{Inference of the scaled primordial fluctuation amplitude $\alpha=\sigma_8/\sigma_{8,\mathrm{fid}}$ from N-body halos at $z=0.0$, using the 3rd order Eulerian bias expansion and the 2LPT gravity model at $\Lambda=1.2 \times k_\mathrm{max}$. We test the impact of different filter choices; the top panel uses spherical filters in the initial conditions and before construction of the bias operators; it is identical to the bottom panel of figure~\ref{fig:restframe-results__eulerianbias__2LPT-varLambda-varKmaxBias}. In the middle panel we switch to cubic filters for $\Lambda$ and $\filterbias$ and in the bottom panel we additionally filter the displacement vector at $\filterfwd = \Lambda$ before computing the actual displacement operations. Likelihood filters are spherical in all three cases. The cubic filters pass more modes than spherical ones and tend to shift the inferred $\alpha$ to lower values. There is no clear sign that they would improve the convergence. The impact of the gravity filter, on the other had, is only minor.}
\label{fig:restframe-results__eulerianbias__filtertypes}
\end{figure}

\begin{figure}
\centering
\includegraphics[width=\figuresize]{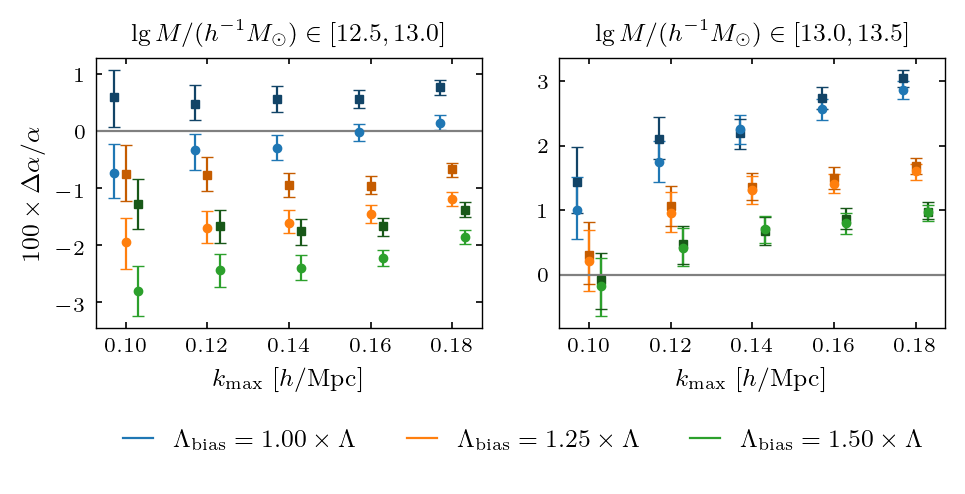}
\caption{Inference of the scaled primordial fluctuation amplitude $\alpha=\sigma_8/\sigma_{8,\mathrm{fid}}$ from N-body halos at $z=0.0$, using the 3rd order Eulerian bias expansion and the 2LPT gravity model at $\Lambda=k_\mathrm{max}$. In contrast to the top panel of figure \ref{fig:restframe-results__eulerianbias__2LPT-varLambda-varKmaxBias}, we allow for the higher-order derivative operators  $\nabla^2(\delta^2)$ and $\nabla^2(K^2)$. The higher-order derivative operators have little impact on the inferred $\sigma_8$ values and apparently cannot absorb the effect of filtering $\delta_\mathrm{fwd}$ before the bias construction.}
\label{fig:restframe-results__eulerianbias__NLOderivatives}
\end{figure}

\clearpage
\bibliographystyle{JHEP} 
\bibliography{forward-model-speed-and-accuracy}

\end{document}

%% file: graphics/flowchart.tex
\begin{tikzpicture}[remember picture]

\newlength{\hmargin}
\setlength{\hmargin}{1in+\hoffset+\oddsidemargin}

\newlength{\nodeshiftV}
\setlength{\nodeshiftV}{.8cm}

\newlength{\nodeheight}
\setlength{\nodeheight}{1cm}

\newlength{\nodewidth}
\setlength{\nodewidth}{2.2cm}

\newlength{\nodeshiftH}
\setlength{\nodeshiftH}{0.22\textwidth}

\newlength{\bbshift}
\setlength{\bbshift}{0.32\textwidth}


\tikzset{leftfieldnode/.style={anchor=north, draw, rounded corners, minimum height=\nodeheight, minimum width=\nodewidth, align=center}}

\tikzset{bbnode/.style={xshift=0.62\textwidth, anchor=center, text width=0.31\textwidth, align=center}}

\node[leftfieldnode] at (current page.north west) (deltaL) {$\delta\lin_\Lambda$};

\node[leftfieldnode, yshift=-\nodeshiftV] at (deltaL.south) (sigman) {$\left\{ \sigma^{(n)}, \vec{t}^{(n)} \right\}$};

\node[leftfieldnode, yshift=-\nodeshiftV] at (sigman.south) (sn) {$\left\{ \vec{s}^{(n)} \right\}$};

\node[leftfieldnode, anchor=north west, xshift=-\nodeshiftH] at (sn.north) (biasL) {$\left\{\op^\lagrangian\right\}_\mathrm{LD}$};

\node[leftfieldnode, yshift=-\nodeshiftV] at (biasL.south) (dispopset) {$\mathbb{1}\,,\left\{\op^\lagrangian\right\}_\mathrm{LD}$};

\node[leftfieldnode, xshift=2\nodeshiftH, anchor=north east] at (dispopset.north west) (unity) {$\mathbb{1}$};

\node[leftfieldnode, yshift=-\nodeshiftV] at (dispopset.south) (biasLevolved) {$\left\{\op^\lagrangian\right\}_\mathrm{LD}$};

\node[leftfieldnode, yshift=-\nodeshiftV] at (unity.south) (deltafwd) {$\delta_\mathrm{fwd}$};

\node[leftfieldnode, yshift=-\nodeshiftV] at (deltafwd.south) (biasELD) {$\left\{\op^\eulerian\right\}_\mathrm{LD}$};

\node[leftfieldnode, yshift=-\nodeshiftV] at (biasELD.south) (biasE) {$\left\{\op^\eulerian\right\}$};

\node[leftfieldnode] at (biasLevolved.south|-biasE.north) (biasLWD) {$\left\{\op^\lagrangian\right\}$};

\node[leftfieldnode, yshift=-\nodeshiftV] at (sn.south|-biasE.south) (deltagdet) {$\dgd$};


\node[anchor=center, xshift=\bbshift, text width=3cm, align=center] at (deltaL.center) (paramtop) {$\Lambda$,\\ $N_\ini$};
\node[anchor=center, xshift=\bbshift, text width=3cm, align=center] at (sigman.center) {$N_\fwd$,\\$n_\lpt$};
\node[anchor=center, xshift=\bbshift, text width=3cm, align=center] at (deltaL.center|-unity.center) {$\filterfwd$,\\ $N_\eul$};
\node[anchor=center, xshift=\bbshift, text width=3cm, align=center] at (deltaL.center|-deltafwd.center) {assignment\\ kernel};
\node[anchor=center, xshift=\bbshift, text width=3cm, align=center] at (deltaL.center|-biasELD.center) (ngfinal) {$\filterbias$,\\ $N_\fin$};
\node[anchor=center, xshift=\bbshift, text width=3cm, align=center] at (deltagdet.center) {$k_\mathrm{max}$,\\ $N_\lh$};

\node[bbnode] at (deltaL.center) (bbini) {$N_\ini = N_{\grid,\nyquist}\left(\Lambda\right)$};
\node[bbnode] at (sigman.center) 
{aliasing limit, \\
$n_\lpt \geq o_\bias-1$};
\node[bbnode] at (sigman.center|-unity.center) {no filter,\\ $N_\eul = 3/2\,N_\ini$};
\node[bbnode] at (sigman.center|-deltafwd.center) {NUFFT};
\node[bbnode] at (sigman.center|-biasELD.center) {$\filterbias\geq \Lambda$,\\
aliasing limit};
\node[bbnode] at (deltagdet.center) {$k_{\max}=\Lambda$,\\$N_\lh = N_{\grid,\nyquist}\left(k_\mathrm{max}\right)$};

\node[anchor=center, yshift=\nodeshiftV+\baselineskip, xshift=-.5\nodeshiftH, align=center, text width=\nodewidth+0.5\nodeshiftH] at (deltaL.north) (headerL) {\bf Lagrangian\\ bias};
\node[anchor=north, xshift=.5\nodeshiftH, align=center, text width=\nodewidth+0.5\nodeshiftH] at (deltaL.north|-headerL.north) {\bf Eulerian\\ bias};
\node[anchor=north, text width=.2\linewidth, align=center] at (paramtop.north|-headerL.north) {\bf accuracy\\ settings};
\node[anchor=north, text width=.3\linewidth, align=center] at (bbini.center|-headerL.north) {\bf best-practice\\ recommendations};

\draw[->] (deltaL.south) -- (sigman.north);
\draw[->] (sigman.south) -- (sn.north);
\draw[->] (sigman.south) -- (sn.north);
\draw[->] (biasL.south) -- (dispopset.north);
\draw[->] (dispopset.south) -- (biasLevolved.north);
\draw[->] (biasLevolved.south) -- (biasLWD.north);
\draw[->] (unity.south) -- (deltafwd.north);
\draw[->] (deltafwd.south) -- (biasELD.north);
\draw[->] (biasELD.south) -- (biasE.north);

\draw[->] (sigman.south) to[out=-90, in=90] (biasL.north);
\draw[->] (sn.south) to[out=-90, in=40] (biasLevolved.north);
\draw[->] (sn.south) to[out=-90, in=140] (deltafwd.north);
\draw[->] (biasE.south) to[out=-90, in=90] (deltagdet.north);
\draw[->] (biasLWD.south) to[out=-90, in=90] (deltagdet.north);
\end{tikzpicture}